\documentclass[12pt]{article}
\usepackage{a4wide,epsfig}
\usepackage{hyperref}
\usepackage{amsmath}

\newcommand{\eq}[1]{Eq.~\eqref{#1}}

\newcommand{\rcite}[1]{Ref.~\cite{#1}}
\newcommand{\rcites}[1]{Refs.~\cite{#1}}

\newcommand{\RG}{{\rm RG}}

\newcommand{\e}{\epsilon}

\def\bfnabla{\mbox{\boldmath $\nabla$}}

\def\bfsigma{\mbox{\boldmath $\sigma$}}

\def\al{\alpha}

\def\siml{{\ \lower-1.2pt\vbox{\hbox{\rlap{$<$}\lower6pt\vbox{\hbox{$\sim$}}}}\ }}

\def\bfnabla{\mbox{\boldmath $\nabla$}}

\def\bfsigma{\mbox{\boldmath $\sigma$}}

\newcommand{\nn}{\nonumber}
\newcommand{\be}{\begin{equation}}
\newcommand{\ee}{\end{equation}}
\newcommand{\bea}{\begin{eqnarray}}
\newcommand{\eea}{\end{eqnarray}}
\def\siml{{\
    \lower-1.2pt\vbox{\hbox{\rlap{$<$}\lower6pt\vbox{\hbox{$\sim$}}}}\ }}

\def\dsl{\,\raise.15ex\hbox{/}\mkern-13.5mu D}

\newcommand{\MS}{\overline{\rm MS}}

\newcommand{\vk}{{\bf k}}

\def\bfnabla{\mbox{\boldmath $\nabla$}}
\def\bfsigma{\mbox{\boldmath $\sigma$}}

\newcommand{\ord}{{\mathcal O}}
\newcommand{\eps}{\epsilon}

\newcommand{\cf}{C_F}

\newcommand{\Appendix}[1]%
    {%
     \section{#1}%
      }

\begin{document}

\title{\vskip-3cm{\baselineskip14pt
}
\vskip1.5cm
S-wave heavy quarkonium spectrum with next-to-next-to-next-to-leading logarithmic accuracy}
\author{C. Anzai, D. Moreno and 
A. Pineda \\[0.5cm]
{\small\it  
Grup de F\'\i sica Te\`orica, Dept. F\'\i sica and IFAE-BIST, Universitat Aut\`onoma de Barcelona,}\\ 
{\small\it E-08193 Bellaterra (Barcelona), Spain}\\}
\date{\today}

\maketitle

\thispagestyle{empty}

\begin{abstract}
We obtain the Potential NRQCD Lagrangian relevant for S-wave states with next-next-to-next-to-leading logarithmic (NNNLL) accuracy. We compute the heavy quarkonium mass of spin-averaged $l= 0$ (angular momentum) states, with otherwise arbitrary quantum numbers, with NNNLL accuracy. These results are complete up to a missing contribution of the two-loop soft running. 
\\[2mm]
\end{abstract}

\newpage

\tableofcontents

\section{Introduction}

High order perturbative computations in heavy quarkonium require the use of effective field theories (EFTs), as they efficiently deal with the different scales of the system. One such EFT is Potential NRQCD (pNRQCD) \cite{Pineda:1997bj,Brambilla:1999xf} (for reviews see \cite{Brambilla:2004jw,Pineda:2011dg}). The key ingredient of the EFT is, obviously, its Lagrangian. At present the pNRQCD 
Lagrangian is known with next-to-next-to-next-to-leading order (NNNLO) accuracy \cite{Kniehl:2002br} (for the nonequal mass case see \cite{Peset:2015vvi}). 

One of the major advantages of using EFTs is that it facilitates the systematic resummation of the large logarithms generated by the ratios of the different scales of the problem. For the case at hand we are talking of
\begin{itemize}
\item
the hard scale ($m$, the heavy quark mass), 
\item
the soft scale ($mv$, the inverse Bohr radius of the problem), 
\item
and the ultrasoft scale ($mv^2$, the typical binding energy of the system). 
\end{itemize}
At present, the pNRQCD Lagrangian is known with next-to-next-to-next-to-leading log (N$^3$LL) precision as far as P-wave states is concerned \cite{Pwave}. For S-wave observables the present precision is NNLL \cite{Pineda:2001ra}. The missing link to 
obtain the complete N$^3$LL pNRQCD Lagrangian is the N$^3$LL running of the delta(-like) potentials\footnote{We use the term ``delta(-like) potentials'' for the delta potential and the potentials generated by the Fourier transform of $\ln^n k$ (in practice only $\ln k$).}. For the spin-dependent case, such precision for the running has already been achieved in \cite{Kniehl:2003ap,Penin:2004xi}. Therefore, what is left is to obtain the N$^3$LL result for the spin-independent delta potential. This is an extremely challenging computation. We undertake this task in this paper. 

The new results we obtain in this paper are the following:
\begin{itemize}
\item
We compute the $\al/m^4$ and the $\al^2/m^3$ spin-independent potentials. These potentials are finite. The expectation value of them produce energy shifts of order $m\al^6$, which contribute to the heavy quarkonium mass at N$^3$LO. Nevertheless, since some expectation values are divergent, 
some of these energy shifts are logarithmic enhanced, i.e. of order ${\cal O}(m\al^6\ln (\frac{\nu}{m\al}))$. Such corrections contribute to the heavy quarkonium mass at N$^3$LL. This divergence, and the associated factorization scale $\nu$, gets canceled by the corresponding divergence in the spin-independent delta potential. By incorporating the HQET Wilson coefficients with LL accuracy\footnote{These are known at ${\cal O}(1/m)$ \cite{EiHi}, 
${\cal O}(1/m^2)$ \cite{Bauer:1997gs,Blok:1996iz} and ${\cal O}(1/m^3)$ \cite{Moreno:2017sgd}.} in the $\al/m^4$ and $\al^2/m^3$ spin-independent potentials, the divergent structure of their expectation value (tantamount to compute potential loops) determines the piece associated to these potentials of the renormalization group (RG) equation of the spin-independent delta potential with N$^3$LL precision. 
\item
We compute the (soft-)$\al^3/m^2$ contribution to the spin-independent delta-like potential proportional to $[c_F^{(1)}]^2$, $[c_F^{(2)}]^2$, $\bar c_1^{(1)hl}$ and $\bar c_1^{(2)hl}$. Unlike before, this potential is divergent. Therefore, for future use, we also give the renormalized expression. The divergent pieces produce corrections of ${\cal O}(m \al^6\ln (\frac{\nu}{m\al}))$ (i.e. of order N$^3$LL). From these divergences we generate the (soft) RG equation of the spin-independent delta potential and resum logarithms with N$^3$LL precision. In order to reach this accuracy, we need the NLL running of the $1/m^2$ HQET Wilson coefficients. For $c_F$ this is known \cite{ABN,Czarnecki:1997dz} but not for $\bar c_1^{hl}$ (the associated missing term is of ${\cal O}(T_fn_f m \al^6\ln(1/\al))$ and is expected to be quite small. Its computation will be carried out elsewhere). The possible mixing between the (soft-)$\al^3/m^2$ and the $\al^2/m^3$ spin-independent potential computed in this paper is also quantified.

 The computation of the (soft-)$\al^3/m^2$ contribution to the spin-independent delta-like potential, proportional to other NRQCD Wilson coefficients, like $[c_k^{(1)}]^2$, $[c_k^{(2)}]^2$, and $c_k^{(1)}c_k^{(2)}$, will be performed in a separated paper. The associated contribution to the running is expected to be small in comparison with the total running of the heavy quarkonium potential. We will estimate its size using the result of the running of the already computed soft contribution.
\item
The N$^3$LL ultrasoft running of the static, $1/m$ and $1/m^2$ potential was originally computed in \cite{Brambilla:2009bi,Pineda:2011db,Pineda:2011aw} (see also \cite{Hoang:2006ht,Hoang:2011gy}). This is enough for P-wave analyses \cite{Pwave}, where such corrections produce a N$^3$LL shift to the energy. Nevertheless, it is not so for S-wave states, as already noted in \cite{Kniehl:2003ap,Penin:2004xi} for the case of the hyperfine splitting. The reason is the generation of singular potentials through divergent ultrasoft loops. We revisit it  in Sec. \ref{sec:ulso} and incorporate the missing contributions needed to have the complete ultrasoft-potential running that produces N$^3$LL shifts to the energy.
\item
Finally, we compute the complete (potential) RG equation of the delta potential with N$^3$LL accuracy (the first nonzero contribution). Solving this equation we obtain the 
complete N$^3$LL running of the delta potential. This allows us to obtain the S-wave mass with N$^3$LL accuracy. It is also one of the missing blocks to obtain the complete NNLL RG improved expression of the Wilson coefficient of the electromagnetic current. This, indeed, 
is what is needed to achieve NNLL precision for non-relativistic sum rules and $t$-$\bar t$ production near threshold. As the spin-dependent (and $l\not=0$) contribution has already been computed in earlier papers \cite{Kniehl:2003ap,Penin:2004xi,Pwave}, we only consider here energy averages of S-wave states where the spin-dependent contributions vanish, and only include terms relevant for the N$^3$LL S-wave spin-average energy. 
\end{itemize}
\medskip

Throughout this paper we work in the $\MS$ renormalization scheme, where bare and renormalized coupling are related as ($D=4+2\epsilon$)
\be
\label{Eq:gB}
g_B^2=g^2\bigg[1+\frac{g^2\bar \nu^{2\eps}}{(4\pi)^2}\beta_0\frac{1}{ \eps} 
+ \left(\frac{g^2\bar \nu^{2\eps}}{(4\pi)^2}\right)^2\left[ \beta_0^2\frac{1}{ \eps^2} +\beta_1\frac{1}{ \eps}\right]
+\ord(g^6)\bigg]\,,
\qquad
\bar\nu^{2\eps}=\nu^{2\eps}\left(\frac{e^{\gamma_E}}{4\pi}\right)^{\eps}
,
\ee
where 
\begin{eqnarray}
  \beta_0&=&{11 \over 3}C_A -{4 \over 3}T_Fn_f\,,
  \nonumber\\
  \beta_1&=&\frac{34}{3}C_A^2-\frac{20}{3}C_AT n_f-4\cf T n_f
  \,.
\end{eqnarray}
$n_f$ is the number of dynamical (active) quarks and $\al=g^2\nu^{2\eps}/(4\pi)$. This definition is slightly different from the one used, for instance, in \cite{Collins:2011zzd}.  

In the following we will only distinguish between the bare coupling $g_B$ and the $\MS$ renormalized coupling $g$ when necessary. The running of $\alpha$ is governed by the $\beta$ function defined through
\be
\frac{1}{2} \nu   {{\rm d} \over {\rm d} \nu}   \frac{\al}{\pi}
=
\nu^2 {{\rm d} \over {\rm d} \nu^2} \frac{\al}{\pi}
=
\beta(\alpha)
=
-\frac{\al}{\pi}
\left\{\beta_0{\al \over 4 \pi}+\beta_1\left({\al \over 4
  \pi}\right)^2 + \cdots\right\}
\,.
\ee

$\al(\nu)$ has 
$n_f$ active light flavours and we define $z=\left[{\al(\nu) \over \al(\nu_h)}\right]^{1 \over
\beta_0}\simeq 1 -1/(2\pi)\al(\nu_h)\ln ({\nu \over \nu_h})$. Note that with the precision achieved in this paper we need in some cases the two-loop running of the coupling when solving the RG equations. 

\section{NRQCD Lagrangian: $1/m^3$ and beyond}

Instrumental in the determination of the Wilson coefficients of the pNRQCD Lagrangian is the determination of the Wilson coefficients of the Lagrangian of the EFT named NRQCD \cite{Caswell:1985ui,Bodwin:1994jh}. We first need to assess which NRQCD operators we have to include in our analysis. We will include light fermions, which we will take to be massless. 

The HQET $1/m^3$ Lagrangian can be found in \cite{Manohar:1997qy}, and including light fermions, though in a different basis, in \cite{Balzereit:1998am}. Here we use the basis and notation from \cite{Moreno:2017sgd}, which also includes light fermions. In \cite{Moreno:2017sgd} one can find the resummed expressions of the Wilson coefficients with LL accuracy for the spin-independent operators. For the spin-dependent $1/m^3$ operators, not relevant for this work, the LL running can be found in \cite{Lobregat:2018tmn,Moreno:2018lbo}. Note that there are no pure gluonic operators of dimension seven.

To obtain the complete $1/m^3$ NRQCD Lagrangian, one also has to consider possible dimension-seven four heavy-fermion operators. There are no such operators, as mentioned in \cite{Brambilla:2008zg}. At ${\cal O}(1/m^4)$, we do not need the complete Lagrangian for the purposes of this paper. For the heavy-quark bilinear sector the complete set of operators was written for the case of QED in \cite{Hill:2012rh} and for QCD in \cite{Gunawardana:2017zix} (in the last case without light fermions). Of those we can neglect most, we do not need the spin-dependent $1/m^4$ operators, nor terms proportional to a single {\bf B}, nor terms with two (either {\bf B} or {\bf E}) terms. The reason is that we only need $1/m^4$ tree level potentials. Therefore, we can take all relevant operators from the QED case. Following the notation of  \cite{Hill:2012rh}, the possible relevant operators are 
\be
\delta {\cal L}_{\psi}^{(4)}=\frac{c^{(1)}_{X1}}{m_1^4}\psi^{\dagger} 
g[{\bf D}^2,{\bf D}\cdot {\bf E} + {\bf E}\cdot {\bf D}]\psi+
\frac{c^{(1)}_{X2}}{m_1^4}\psi^{\dagger} g\{{\bf D}^2,[{\bf \bfnabla}\cdot {\bf E}]\}\psi+
\frac{c^{(1)}_{X3}}{m_1^4}\psi^{\dagger} g[{\bf \bfnabla}^2,{\bf \bfnabla}\cdot {\bf E}]\psi+\cdots
\ee
and similarly for the antiquark. The dots stand for terms that one can trivially see that do not contribute to the S-wave spin-independent spectrum at NNNLL, either 
because involved the emission of two gluons or because they are spin-dependent. 
In principle we need three new coefficients. Nevertheless, we will see later that only $c_{X1}$ contributes to the running of the spin-independent delta potential. Still, we will compute any tree level potential proportional to $c_{X1}$, $c_{X2}$ and $c_{X3}$.

The fact that we need $c_{X1}$, one of the Wilson coefficients of the $1/m^4$ heavy quark bilinear Lagrangian, 
could make it necessary to consider the Wilson coefficients of the $1/m^4$ heavy-light operators as well [light-light operators are subleading for the same reason they are at ${\cal O}(1/m^3)$], as they may enter through RG mixing. Fortunately, $c_{X1}$ can be determined by reparameterization invariance, which gives us the following relation \cite{Hill:2012rh}:
\be
32 c^{(i)}_{X1}=\frac{5Z}{4}-c^{(i)}_F+c^{(i)}_D
\ee
(where one should take $Z=1$ for QCD). 
Note that it depends on $c_D$, so indeed $c^{(i)}_{X1}$ is gauge dependent. Nevertheless, we will see later that it always combines with $c_M$ to produce gauge invariant combinations. This indeed is a nontrivial check of the computation. Note also that the above coefficient has an Abelian term, so it can be checked with QED computations. 

Finally, we consider the heavy four-fermion sector of the $1/m^4$ Lagrangian. They generate local or quasi-local potentials, which do not produce divergent potential loops. The same happens for the potentials generated by $c_{X2}$, $c_{X3}$. Therefore, in both cases, such potentials do not generate contributions to the heavy quarkonium mass at N$^3$LL, and we can neglect them.

Out of this discussion, we conclude that we have the LL running of all necessary Wilson coefficients of the $1/m^4$ NRQCD Lagrangian operators. 

\section{pNRQCD Lagrangian}
\label{Sec:pNRQCD}


Integrating out the soft modes in NRQCD we end up with the EFT named pNRQCD.
The most general pNRQCD Lagrangian 
compatible with the symmetries of QCD that can be constructed
with a singlet and an octet (quarkonium) field, as well as an ultrasoft gluon field to NLO in the 
multipole expansion has the form~\cite{Pineda:1997bj,Brambilla:1999xf}
\bea
& & \!\!\!\!\!
{\cal L}_{\rm pNRQCD} = \!\! \int \!\! d^3{\bf r} \; {\rm Tr} \,  
\Biggl\{ {\rm S}^\dagger \left( i\partial_0 
- h_s({\bf r}, {\bf p}, {\bf P}_{\bf R}, {\bf S}_1,{\bf S}_2) \right) {\rm S} 
+ {\rm O}^\dagger \left( iD_0 
- h_o({\bf r}, {\bf p}, {\bf P}_{\bf R}, {\bf S}_1,{\bf S}_2) \right) {\rm O} \Biggr\}
\nn
\\
& &\qquad\qquad 
+ V_A ( r) {\rm Tr} \left\{  {\rm O}^\dagger {\bf r} \cdot g{\bf E} \,{\rm S}
+ {\rm S}^\dagger {\bf r} \cdot g{\bf E} \,{\rm O} \right\} 
+ \frac{V_B (r)}{ 2} {\rm Tr} \left\{  {\rm O}^\dagger {\bf r} \cdot g{\bf E} \, {\rm O} 
+ {\rm O}^\dagger {\rm O} {\bf r} \cdot g{\bf E}  \right\}  
\nn
\\
& &\qquad\qquad 
- \frac{1}{ 4} G_{\mu \nu}^{a} G^{\mu \nu \, a} 
+  \sum_{i=1}^{n_f} \bar q_i \, i \dsl \, q_i 
\,,
\label{Lpnrqcd}
\\
& &
\nn 
\\
& &
h_s({\bf r}, {\bf p}, {\bf P}_{\bf R}, {\bf S}_1,{\bf S}_2) = 
 \frac{{\bf p}^2 }{ 2\, m_{ r}}
+ 
\frac{{\bf P}_{\bf R}^2 }{ 2\, M} + 
V_s({\bf r}, {\bf p}, {\bf P}_{\bf R}, {\bf S}_1,{\bf S}_2), 
\\
& & 
h_o({\bf r}, {\bf p}, {\bf P}_{\bf R}, {\bf S}_1,{\bf S}_2) = 
\frac{{\bf p}^2 }{ 2\, m_{ r}}
+ 
\frac{{\bf P}_{\bf R}^2 }{ 2\,M}  + 
V_o({\bf r}, {\bf p}, {\bf P}_{\bf R}, {\bf S}_1,{\bf S}_2), 
\eea
where $iD_0 {\rm O} \equiv i \partial_0 {\rm O} - g [A_0({\bf R},t),{\rm O}]$, 
${\bf P}_{\bf R} = -i{\bfnabla}_{\bf R}$ for the singlet,  
${\bf P}_{\bf R} = -i{\bf D}_{\bf R}$ for the octet (where the covariant derivative is in the adjoint representation), 
${\bf p} = -i\bfnabla_{\bf r}$,
\begin{align}
m_{r} = \frac{m_1 m_2}{m_1+m_2}
\end{align}
and $M = m_1+m_2$. 
We adopt the color normalization  
\be
{\rm S} = { S\, 1\!\!{\rm l}_c / \sqrt{N_c}} \,, \quad\quad\quad 
{\rm O} = O^a { {\rm T}^a / \sqrt{T_F}}\,,
\label{SSOO}
\ee 
for the singlet field $S({\bf r}, {\bf R}, t)$ and the octet field $O^a({\bf r}, {\bf R}, t)$.
Here and throughout this paper we denote the quark-antiquark distance vector by ${\bf r}$, the center-of-mass position of the quark-antiquark system by ${\bf R}$, and the time by $t$.

Both $h_s$ and the potential $V_s$ are operators acting on the Hilbert space of a heavy quark-antiquark system in the singlet configuration.\footnote{Therefore, in a more mathematical notation: $h \rightarrow \hat h$, $V_s({\bf r},{\bf p}) \rightarrow \hat V_s(\hat {\bf r},\hat {\bf p})$. We will however avoid this notation in order to facilitate the reading.}
$V_s$ (and $V_o$) can be Taylor expanded in powers of $1/m$ (up to logarithms). At low orders we have
\bea
 V_s &=& 
V^{(0)} +\frac{V^{(1)}}{m_r}+ \frac{V_{{\bf L}^2}^{(2)}}{m_1m_2}\frac{{\bf L}^2 }{ r^2}
+\frac{1}{ 2m_1m_2}
\left\{{\bf p}^2,V^{(2)}_{{\bf p}^2}(r)\right\}+\frac{V_r^{(2)}}{m_1m_2}
\nn
\\
&&+\frac{1}{m_1m_2} V_{S^2}^{(1,1)}(r){\bf S}_1\cdot{\bf S}_2+\frac{1}{m_1m_2} V_{{\bf S}_{12}}^{(1,1)}(r){\bf
  S}_{12}({\bf r})
  \nn
  \\
  &&
  +\frac{1}{m_1m_2}V^{(2)}_{LS_1}(r){\bf L}\cdot{\bf S}_1 +\frac{1}{m_1m_2}V^{(2)}_{LS_2}(r) {\bf L}\cdot{\bf S}_2
+{\cal O}(1/m^3),
\label{V1ovm2}
\eea
where, ${\bf S}_1=\bfsigma_1/2$, ${\bf S}_2=\bfsigma_2/2$, ${\bf L} \equiv {\bf r} \times {\bf p}$, and $\displaystyle{{\bf S}_{12}({\bf r}) \equiv \frac{
3 { {\bf r}}\cdot \bfsigma_1 \,{ {\bf r}}\cdot \bfsigma_2}{r^2} - \bfsigma_1\cdot \bfsigma_2}$.

$V^{(0)}$ is known with N$^3$LL accuracy \cite{Brambilla:2009bi,Pineda:2011db}. The N$^3$LL result for the $1/m$ and $1/m^2$ momentum dependent potential is also known in different matching schemes \cite{Pineda:2011aw,Peset:2017wef,Pwave}: 
on-shell, off-shell (Coulomb, Feynman) and Wilson. In terms of the original definitions used in these papers they read (in four dimensions) 
\be
V^{(1)}=V^{(1,0)}(r)=V^{(0,1)}\equiv - \frac{C_FC_A D^{(1)}}{4 r^2}
\,,
\ee
\be
\frac{V_{{\bf L}^2}^{(2)}}{m_1m_2}\equiv \frac{V_{{\bf L}^2}^{(2,0)}(r)}{m_1^2}+\frac{V_{{\bf L}^2}^{(0,2)}(r)}{m_2^2}+\frac{V_{{\bf L}^2}^{(1,1)}(r)}{m_1m_2}\equiv \frac{C_F D_2^{(2)}}{2 m_1 m_2 r}
\,,
\ee
\be
\frac{V_{{\bf p}^2}^{(2)}}{m_1m_2}\equiv \frac{V_{{\bf p}^2}^{(2,0)}(r)}{m_1^2}+\frac{V_{{\bf p}^2}^{(0,2)}(r)}{m_2^2}+\frac{V_{{\bf p}^2}^{(1,1)}(r)}{m_1m_2}\equiv -\frac{C_F D_1^{(2)}}{ m_1 m_2 r}
\,.
\ee
The spin-dependent and momentum-dependent potentials are also known with N$^3$LL precision \cite{Pwave}. We use the following definitions in this paper (again we refer to \cite{Pwave}):
\bea
\frac{1}{m_1m_2}V^{(2)}_{LS_1}(r)
&\equiv &
\left(\frac{1}{m_1^2}V^{(2,0)}_{LS}(r)+\frac{1}{m_1m_2}V_{L_2S_1}^{(1,1)}(r)
\right) \equiv \frac{3C_F D_{LS_1}^{(2)}}{2m_1m_2}
\,,
\\
\frac{1}{m_1m_2}V^{(2)}_{LS_2}(r)
&\equiv &
\left(\frac{1}{m_2^2}V^{(0,2)}_{LS}(r)+\frac{1}{m_1m_2}
V_{L_1S_2}^{(1,1)}(r) \right) \equiv \frac{3C_F D_{LS_2}^{(2)}}{2m_1m_2}
\,.
\eea

More delicate are $V_{S^2}^{(1,1)}$ and $V_r^{(2)}$, as their running is sensitive to potential loops, which are more efficiently computed in momentum space. 
Therefore, it is more convenient to work with the potential in momentum space, which is defined in the following way:
\be
\tilde V_s \equiv \langle {\bf p}'| V_s | {\bf p} \rangle \,.
\ee
Then the potential reads 
\begin{eqnarray}
 \tilde V_s &=&
 - 4\pi C_F \frac{\alpha_{{\tilde V}}}{{\bf q}^2}
 -\,   {\bf p}^4\left(\frac{c_4^{(1)}}{8m_1^3}+\frac{c_4^{(2)}}{8m_2^3}\right)\, (2\pi)^d\delta^{(d)}({\bf q})
\label{Hpnrqcd}
\\
&&
-\,  C_F C_A \tilde D^{(1)} \frac{\pi^2\, }{2m_r\, |{\bf q}|^{1-2\e}}\, (1+{\cal O}(\epsilon))
\nonumber \\
&&
-\,  \frac{2\pi C_F \tilde D^{(2)}_{1}}{m_1m_2} 
 \frac{{\bf p}^2+{\bf p}^{\prime\, 2}}{{\bf q}^2}
+ \frac{\pi C_F \tilde D^{(2)}_{2}}{m_1m_2} \left(
 \left(\frac{{\bf p}^2-{\bf p}^{\prime\, 2}}{{\bf q}^2}\right)^2 - 1 \right)
\nonumber \\
&& 
+\,  \frac{\pi C_F \tilde D^{(2)}_{d}}{m_1m_2}
- \frac{4\pi C_F \tilde D^{(2)}_{S^2}}{d\, m_1m_2}\, 
       [{\bf S}_1^i,{\bf S}_1^j][{\bf S}_2^i,{\bf S}_2^j]
\nonumber \\
&& 
+\,  \frac{4\pi C_F  \tilde D^{(2)}_{S_{12}}}{d\, m_1m_2}\, 
        [{\bf S}_1^i,{\bf S}_1^r][{\bf S}_2^i,{\bf S}_2^j]
   \left( \delta^{rj}-d\,  \frac{{\bf q}^r {\bf q}^j}{{\bf q}^2} \right)
\nonumber \\
& & 
\nonumber
-\,   \frac{6\pi C_F}{m_1m_2} \,
   \frac{{\bf p}^i {\bf q}^j}{{\bf q}^2}
   \left( \tilde D^{(2)}_{LS_1}[{\bf S}_1^i,{\bf S}_1^j]+ \tilde D^{(2)}_{LS_2}[{\bf S}_2^i,{\bf S}_2^j] \right)
\,,
\end{eqnarray}
where the (Wilson) coefficients $\tilde D$ generically stand for the Fourier transform of the original Wilson coefficients in position space $D$. For them 
(and for $\al_{\tilde V}$) we use the power counting LL/LO for the first nonvanishing correction, and so on. 

$V_{S^2}^{(1,1)}$ is indeed known with the required N$^3$LL accuracy \cite{Kniehl:2003ap,Penin:2004xi} (one should 
be careful when comparing though, as there is a change in the basis of potentials used there, compared with the one we use here). In terms of $\tilde D_{S^2}^{(2)}$ it reads
\bea
\nn
&&
\frac{V_{S^2}^{(1,1)}}{m_1m_2}\equiv
\\
&&
\equiv
\delta^{(3)}({\bf r})\frac{8\pi C_F  \tilde D_{S^2}^{(2)}}{3m_1m_2}
+\frac{8\pi C_F  \tilde D_{S^2}^{(2)}}{3m_1m_2}
\left[- \frac{1}{4\pi} { \rm reg} \frac{1}{r^3}-\ln \nu \delta^{(3)}({\bf r})\right] \left( k {d\over d k} \tilde 
 D_{S^2}^{(2)}\right)\Bigg|_{k=\nu}^{LL}
\,,
\label{eq:Ds2}
\eea
where 
\begin{align}
- \frac{1}{4\pi} { \rm reg} \frac{1}{r^3} 
\equiv  \int \frac{d^3k}{(2\pi)^3} e^{-i{\bf k} \cdot {\bf r}}\ln k 
\, ,
\end{align}
and we neglect higher order logarithms (as they are subleading).

Finally we consider $V_r$. In terms of $\tilde D_{d}^{(2)}$ it reads
\bea
\nn
&&
\frac{V_r^{(2)}}{m_1m_2}\equiv\frac{ V^{(2,0)}_{r}(r)}{m_1^2}+\frac{ V^{(0,2)}_{r}(r)}{m_2^2}
+\frac{ V^{(1,1)}_{r}(r)}{m_1 m_2}
\\
&&
\equiv
\delta^{(3)}({\bf r})\frac{\pi C_F  \tilde D_d^{(2)}}{m_1m_2}
+\frac{\pi C_F}{m_1 m_2}
\left[- \frac{1}{4\pi} { \rm reg} \frac{1}{r^3}-\ln \nu \delta^{(3)}({\bf r})\right] \left( k {d\over d k} \tilde D_d^{(2)}\right)\Bigg|_{k=\nu}^{LL}
\,.
\label{eq:Vr}
\eea
Unlike all the other potentials, we do not know $V_r^{(2)}$ with N$^3$LL expression (though the N$^2$LL expression is known \cite{Pineda:2001ra}). This leads us to the main purpose of this paper: the computation of $V_r$ with  N$^3$LL accuracy. This is equivalent to obtaining the NLL expression of $\tilde D_d^{(2)}$. This will require the use of the other Wilson coefficients to one  order less: LL. Indeed in \eq{Hpnrqcd} we have already approximated the Fourier transform of $V_{{\bf L}^2}^{(2)}$ by its N$^2$LL expression (otherwise the momentum dependence is more complicated).  

At LL the Wilson coefficients are equal in position and momentum space. We only explicitly display those that we will need later. For the static potential we would have at LL that $\al_{V}=\al_{\tilde V}=\al$. For the rest, we show the results in the off-shell Coulomb (which are equal to the Feynman at this order) and on-shell matching schemes, except for $D^{(2)}_{LS_i}$, which we do not need for S-wave:
\be
D^{(1),LL}_{CG}=\tilde D^{(1),LL}_{CG}=\al^2(\nu)+\frac{16}{3\beta_0}\left(\frac{C_A}{2}+C_F\right)\al^2(\nu)\ln\left(\frac{\al(\nu)}{\al(\nu^2/\nu_h)}\right)
\,,
\ee
\be
D^{(1),LL}_{OS}=
\tilde D^{(1),LL}_{OS}=
\al^2(\nu)\left[1-\frac{2 C_F}{C_A}\frac{m_r^2}{m_1m_2}\right]
+\frac{16}{3\beta_0}\left(\frac{C_A}{2}+C_F\right)\al^2(\nu)\ln\left(\frac{\al(\nu)}{\al(\nu^2/\nu_h)}\right)
\,,
\ee
\be
D_{1}^{(2),LL}=\tilde D_{1}^{(2),LL}=\al(\nu)+\frac{(m_1+m_2)^2}{m_1m_2}\frac{2C_A}{3\beta_0}
\al(\nu)\ln\left(\frac{\al(\nu)}{\al(\nu^2/\nu_h)}\right)
\,,
\ee
\be
D_{S_{12}}^{(2),LL}=
\tilde D_{S_{12}}^{(2),LL}=\al(\nu)c_F^2(\nu)
\,,
\ee
\be
D_{S^2}^{(2),LL}=
\tilde D_{S^2}^{(2),LL}=\al(\nu)c_F^2(\nu)-\frac{3}{2\pi C_F}(d_{sv}(\nu)+C_Fd_{vv}(\nu))
\,.\ee

We now turn to $\tilde D_d^{(2)}$. Expanding $\tilde D_d^{(2)}(k,\nu)$ in powers of $\ln k$, we obtain
\be
\tilde D_d^{(2)}(k,\nu)=\tilde D_d^{(2)}(\nu_s,\nu_p,\nu_p^2/\nu_h)\Bigg|_{\nu_s=\nu_p=\nu}+
k{d \over d k} \tilde D_d^{(2)}(k,\nu)\Bigg|_{k=\nu}\ln\left({k\over\nu}\right)+\ldots
\,,
\label{momdepDd}
\ee
where we have made explicit the dependence on the different factorization scales. 

So far we have not made explicit the dependence on $\nu_h \sim m$. Nevertheless, it will play an important role later, when solving the RG equations. Therefore, in the following, we use the notation
$\tilde D_d^{(2)}(\nu_s,\nu_p,\nu_p^2/\nu_h)\Bigg|_{\nu_s=\nu_p=\nu} \equiv \tilde D^{(2)}_d(\nu_h;\nu)$.

$\tilde D^{(2)}_d(\nu_h;\nu)$ can be written in several ways: as a sum of the LL ($\tilde D^{(2),LL}_d(\nu_h;\nu)$) term 
and the NLL ($\delta \tilde D^{(2),NLL}_d(\nu_h;\nu)$) correction, or as the sum of the initial condition ($\tilde D^{(2)}_d(\nu_h;\nu_h)\equiv \tilde D^{(2)}_d(\nu_h)$) at the hard scale and the running contribution ($\delta \tilde D^{(2)}_d(\nu_h;\nu)$ where $\delta \tilde D^{(2)}_d(\nu_h;\nu_h)=0$):
\bea
\tilde D^{(2)}_d(\nu_h;\nu)=\tilde D^{(2)}_d(\nu_h)+\delta \tilde D^{(2)}_d(\nu_h;\nu)= \tilde D^{(2),LL}_d(\nu_h;\nu)+\delta \tilde D^{(2),NLL}_d(\nu_h;\nu)
\,.
\eea
This Wilson coefficient may depend on the matching scheme. Here we mainly consider the off-shell Coulomb gauge matching scheme. Still, for later discussion, we also give expressions in the on-shell matching scheme (see \cite{Peset:2015vvi} for more details).

\begin{figure}[!htb]
	\begin{center}      
	\includegraphics[width=0.74\textwidth]{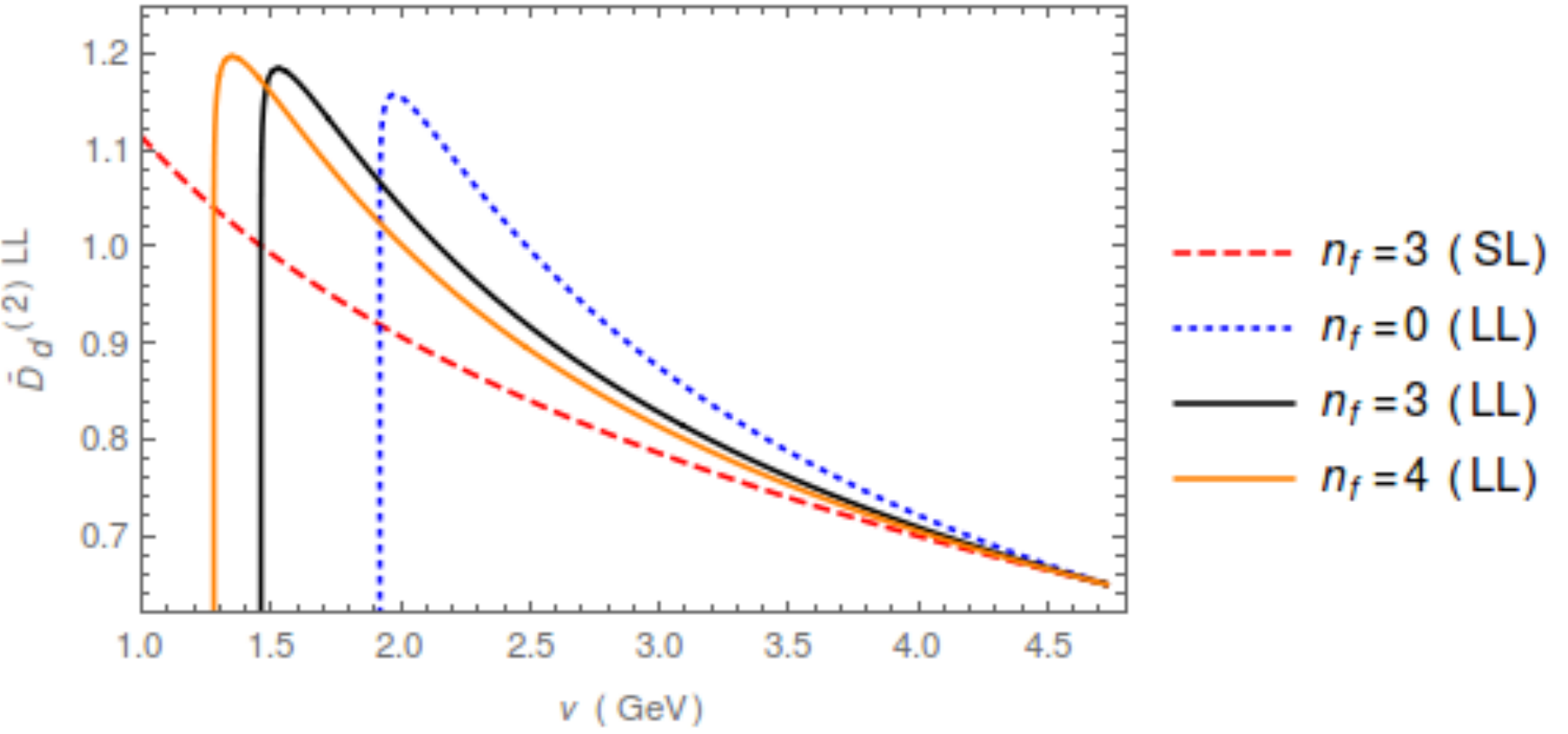}
	\includegraphics[width=0.74\textwidth]{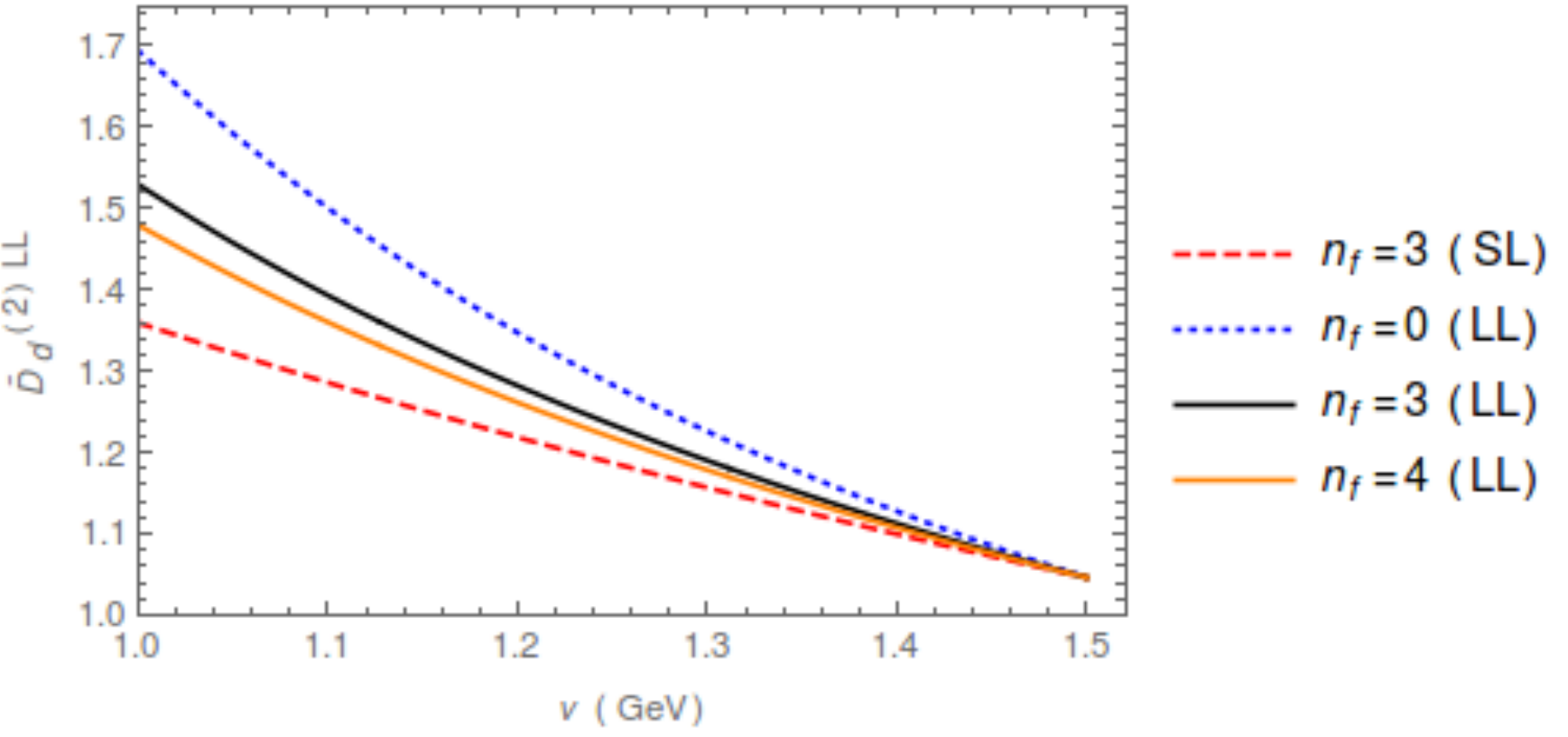}\\
	\includegraphics[width=0.74\textwidth]{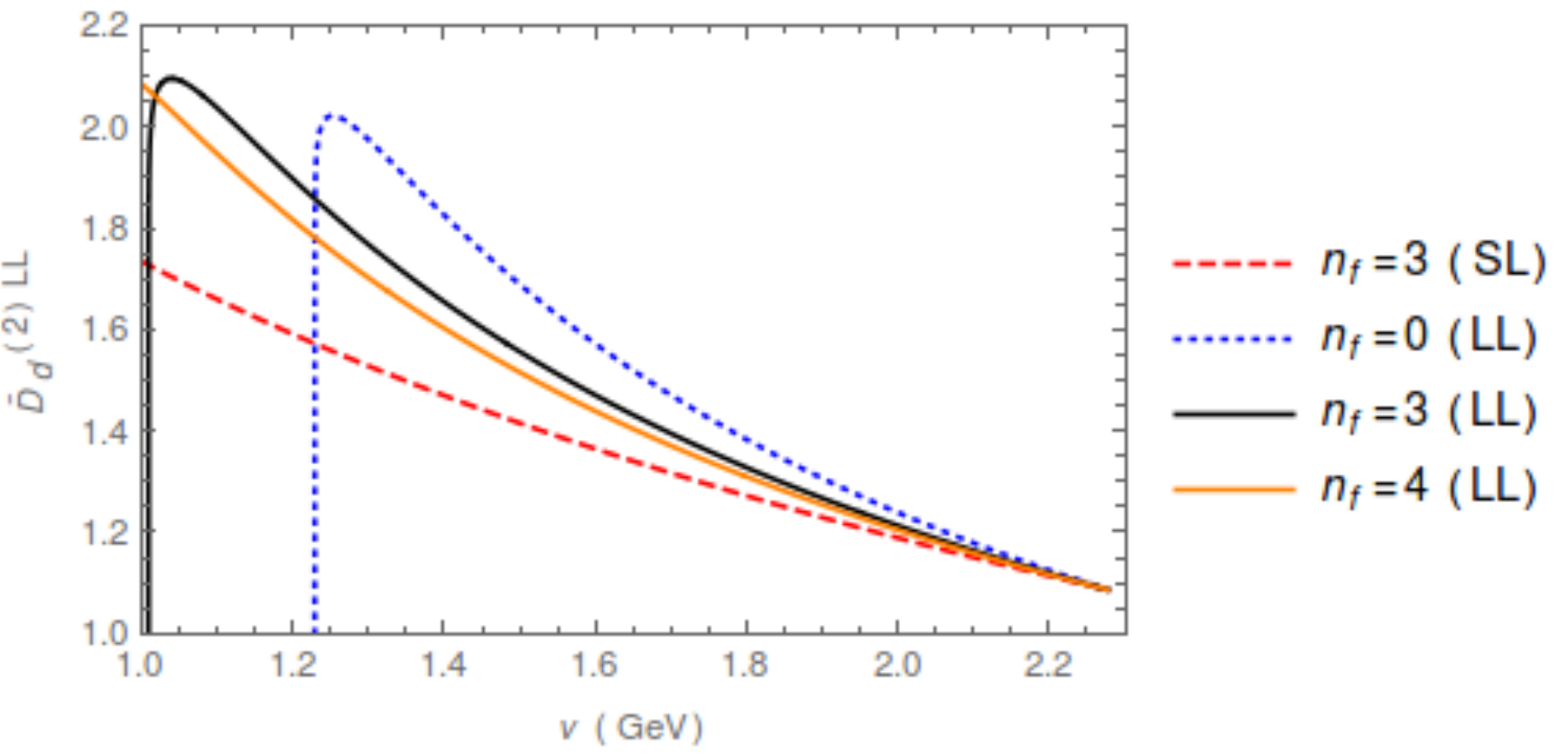}
\caption{Plot of \eq{DdCG}, the LL running in the off-shell (Coulomb/Feynman) matching scheme of $\tilde D_d^{(2)}$ for different values of $n_f$ (0,3,4) and in the single log (SL) approximation (in this case only with $n_f=3$). 
{\bf Upper  panel:} Plot for bottomonium with $\nu_h=m_{b}$. {\bf Middle panel:} Plot for charmonium with 
$\nu_h=m_{c}$. {\bf Lower panel:} Plot for $B_c$ with $\nu_h=2m_{b}m_{c}/(m_{b}+m_{c})$. 
\label{Fig:LL}}   
\end{center}
\end{figure}

The LL running is known \cite{Pineda:2001ra}:
\bea
D_{d,CG}^{(2)LL}(\nu)=\tilde D_{d,CG}^{(2)LL}(\nu)
&=&2\alpha(\nu)
+\frac{1}{\pi C_F}[d_{ss}(\nu)+C_F \bar d_{vs}(\nu)]
\nn
\\
&&
+\frac{(m_1+m_2)^2}{m_1m_2}\frac{8}{3\beta_0}\left(\frac{C_A}{2}-C_F\right)\al(\nu)\ln\left(\frac{\al(\nu)}{\al(\nu^2/\nu_h)}\right)
\,,
\label{DdCG}
\eea
\bea
D_{d,OS}^{(2)LL}(\nu)=\tilde D_{d,OS}^{(2)LL}(\nu)
&=&\alpha(\nu)
+\frac{1}{\pi C_F}[d_{ss}(\nu)+C_F \bar d_{vs}(\nu)]
\nn
\\
&&
+\frac{(m_1+m_2)^2}{m_1m_2}\frac{8}{3\beta_0}\left(\frac{C_A}{2}-C_F\right)\al(\nu)\ln\left(\frac{\al(\nu)}{\al(\nu^2/\nu_h)}\right)
\,,
\label{DdOS}
\eea
where
\be
\bar d_{vs}=\pi \al \frac{c^{(2)}_D}{2}\frac{m_1}{m_2}+
\pi \al \frac{c^{(1)}_D}{2}\frac{m_2}{m_1}+d_{vs}
\ee
is a gauge invariant combination of NRQCD Wilson coefficients, for which its LL running can be found in \cite{Pineda:2001ra}.
In order to visualize the relative importance of the NLL corrections compared with the LL term, we plot the later in Fig. \ref{Fig:LL} in the Coulomb gauge\footnote{Unlike in the other plots, we use here the two-loop running for $\al$. The effect is small.}. For reference, in these and latter figures, we use the following numerical values for the heavy quark masses and $\al$: $m_b = 4.73$ GeV, $\alpha(m_b)$ = 0.216547, $m_c$ = 1.5 GeV, $\alpha(m_c)$ = 0.348536 and 
$\alpha(2m_bm_c/(m_b+m_c))$ = 0.290758. $\nu_h=m_b$ for bottomonium, $\nu_h=m_c$ for charmonium, and $\nu_h=2m_r=2m_bm_c/(m_b+m_c)$ for the $B_c$ system.

\medskip
From the LL result (using the $\nu_s$ independence of the potential at LO) one obtains
\begin{eqnarray}
\label{ddkDd}
  k {d\over d k} \tilde D_{d,\rm CG}^{(2)}\Bigg|_{k=\nu}^{LL}(\nu_h;\nu)
  &=&
  -\beta_0\frac{\al^2}{\pi}+
  \frac{\al^2}{\pi}\left(2C_F-{ C_A \over
2}\right)c_k^{(1)}c_k^{(2)} 
\\
\nn
&&
+
\frac{\al^2}{\pi}\left[
{m_1 \over m_2}\left(\frac{1}{3}T_fn_f \bar c_1^{hl(2)}- { 4 \over 3}(C_A+C_F)[c_k^{(2)}]^2-{ 5 \over 12}C_A[c_F^{(2)}]^2\right)
\right.
\\
\nn
&&
\left.
\qquad
+{m_2 \over m_1}\left(\frac{1}{3}T_fn_f \bar c_1^{hl(1)}- { 4 \over 3}(C_A+C_F)[c_k^{(1)}]^2-{ 5 \over 12}C_A[c_F^{(1)}]^2\right)
 \right]
  \\
  &&
  \nn
  -\frac{(m_1+m_2)^2}{m_1m_2}\frac{4}{3}\left(\frac{C_A}{2}-C_F\right)\frac{\al^2}{\pi}\left[\ln\left(\frac{\al(\nu)}{\al(\nu^2/\nu_h)}\right)+1\right]
\,.
\end{eqnarray}
This term contributes to the N$^3$LL energy shift of the spectrum.

Since we know the NLO expression of $\tilde D_d^{(2)}$, we can determine the initial matching condition. It reads
\bea
\tilde D^{(2)}_{d,OS}(\nu_h)&=&\alpha(\nu_h)
+\frac{\alpha^2(\nu_h)}{4\pi}\left(\frac{28}{9}C_A+\frac{4}{3}C_F-\frac{20}{9}T_F n_f
+\left(\frac{m_1}{m_2}+\frac{m_2}{m_1}\right)\left[\frac{25}{18}C_A-\frac{10}{9}T_F n_f\right]\right) 
\nn
\\
&&+\frac{1}{\pi C_F}\left( d_{ss}(\nu_h)+C_F \bar d_{vs}(\nu_h)\right)\,,
\label{DdhardOS}
\eea

\bea
\tilde D^{(2)}_{d,CG}(\nu_h)&=&2\alpha(\nu_h)
+\frac{\alpha^2(\nu_h)}{4\pi}\left(
\frac{62}{9}C_A+\frac{4}{3}C_F-\frac{32}{3}C_A\ln 2-\frac{28}{9}T_Fn_f
\right.
\nn
\\
&&
\left.+\left(\frac{m_1}{m_2}+\frac{m_2}{m_1}\right)\left[
-\frac{10}{9}T_Fn_f
+ \left(\frac{61}{18}-\frac{16}{3}\ln 2\right)C_A
\right]\right) 
\nn
\\
&&+ \frac{1}{\pi C_F}\left(d_{ss}(\nu_h)+C_F \bar d_{vs}(\nu_h)\right)\,.
\label{DdhardCG}
\eea
$c_D$, and the four-fermion Wilson coefficients $d_{ss}$ and $d_{vs}$, were computed at one loop in 
\cite{Manohar:1997qy} and \cite{Pineda:1998kj} respectively, where one can find the explicit expressions. 

At the order we are working $\delta \tilde D_{d}^{(2)NLL}(\nu_h;\nu)$ can be split into pieces. 
Thus, the NLL approximation for the Wilson coefficient is given by the sum
\bea
\label{Dsum}
 \delta \tilde D_{d}^{(2)NLL}(\nu_h;\nu)
 &=&
 \left(\tilde D_{d}^{(2)}\right)^{~}_{\rm
  1-loop}(\nu_h)\\
  &&
  \nn
  +\delta \tilde D_{d,us}^{(2)NLL}(\nu_h;\nu)
  +\delta \tilde D_{d,s}^{(2)NLL}(\nu_h;\nu)
  +\delta \tilde D_{d,p}^{(2)NLL}(\nu_h;\nu)
\,,
\eea
where  the second line is zero when $\nu=\nu_h$. $\left(\tilde D_{d}^{(2)}\right)^{~}_{\rm
  1-loop}(\nu_h)$ is the ${\cal O}(\al^2)$ term of Eq. (\ref{DdhardOS}) or (\ref{DdhardCG}), depending on the matching scheme. Their numerical values in the Coulomb gauge matching scheme are: for bottomonium 0.042, 0.052 and 0.081 for $n_f$=4, 3, and 0 respectively; for charmonium 0.108, 0.134 and 0.211 for $n_f$=4, 3, and 0 respectively; and
  for $B_c$ 0.048, 0.072 and 0.142 for $n_f$=4, 3, and 0 respectively. We nicely observe that these numbers generate small corrections to the leading order results.
  
\begin{figure}[!htb]
	\begin{center}      
	\includegraphics[width=0.74\textwidth]{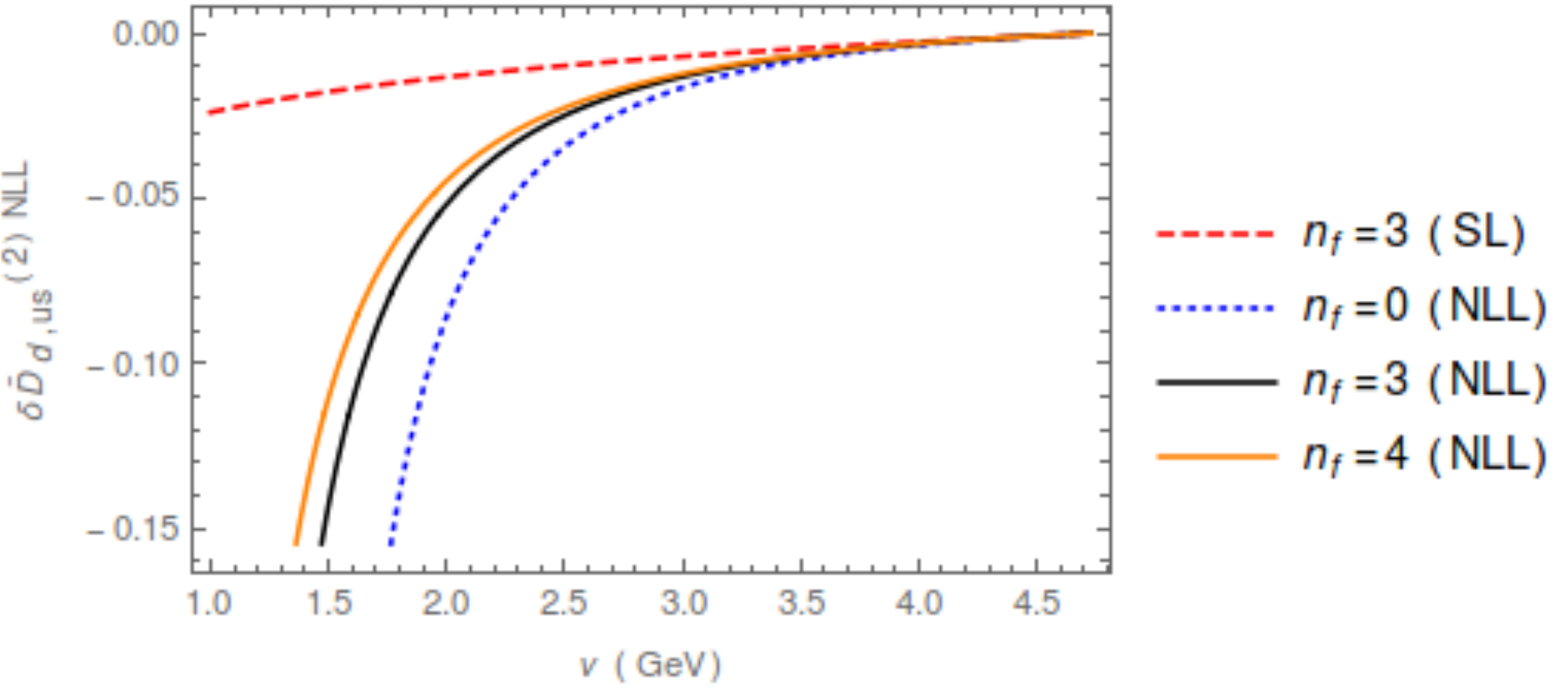}
	\includegraphics[width=0.74\textwidth]{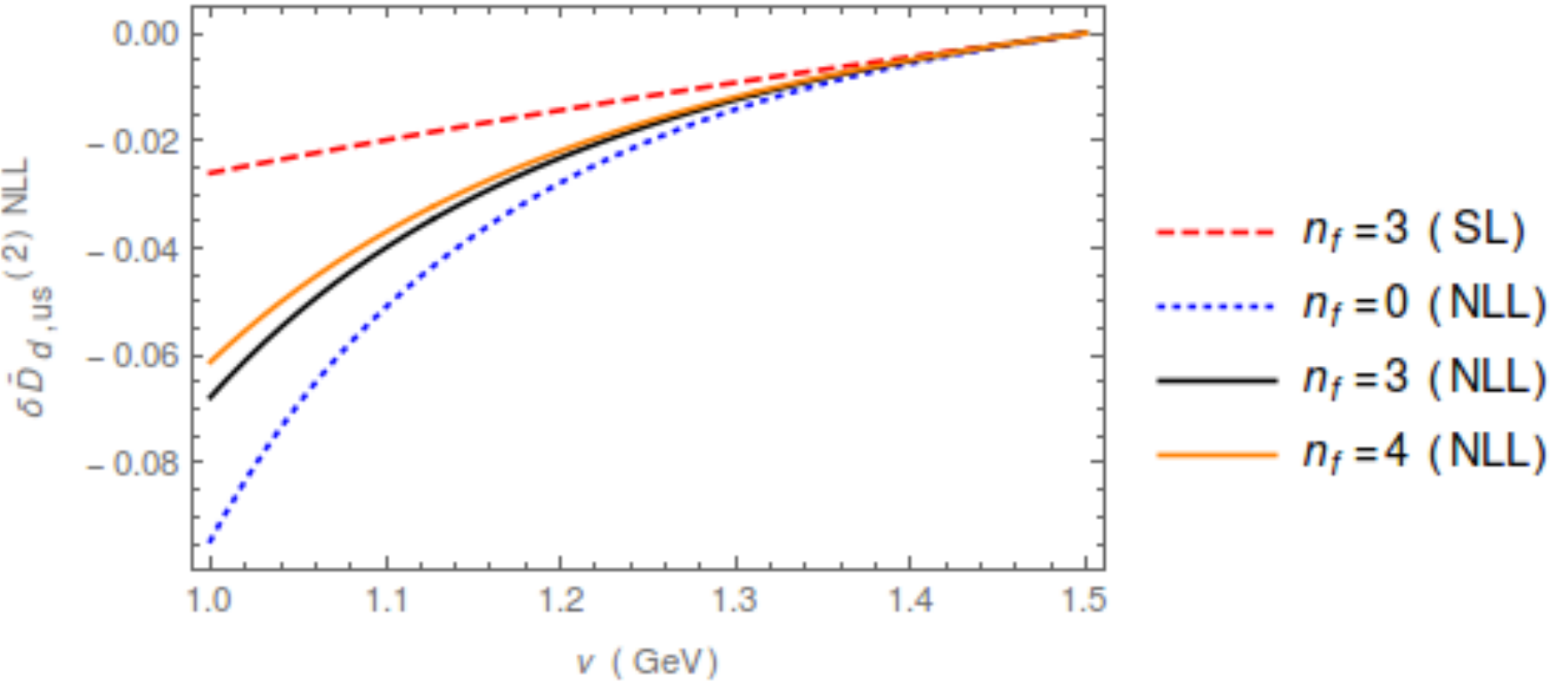}\\
	\includegraphics[width=0.74\textwidth]{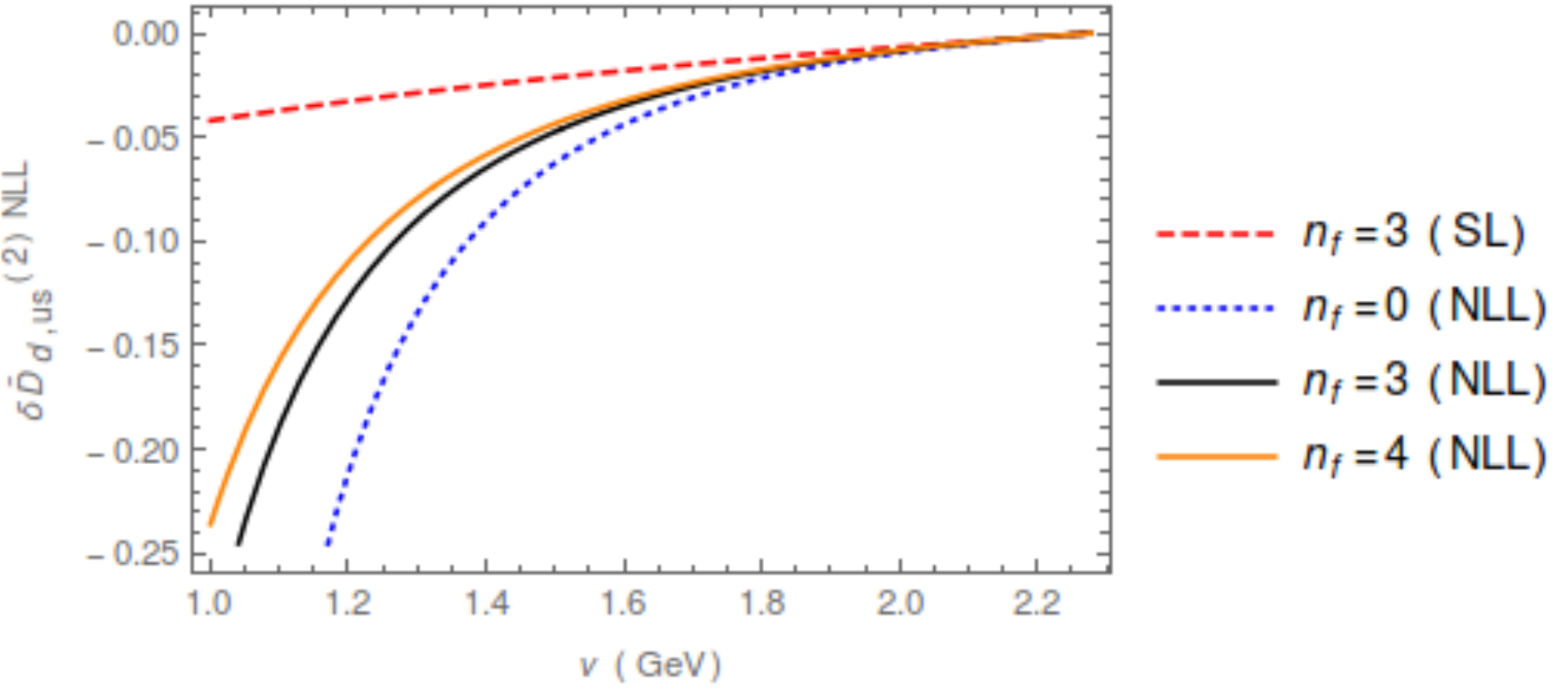}
\caption{Plot of \eq{NNLUS}, the NLL ultrasoft running in the off-shell (Coulomb/Feynman) matching scheme of $\tilde D_d^{(2)}$  for different values of $n_f$ (0,3,4) and in the single log (SL) approximation (in this case only with $n_f=3$). 
{\bf Upper  panel:} Plot for bottomonium with $\nu_h=m_{b}$. {\bf Middle panel:} Plot for charmonium with 
$\nu_h=m_{c}$. {\bf Lower panel:} Plot for $B_c$ with $\nu_h=2m_{b}m_{c}/(m_{b}+m_{c})$. 
\label{Fig:NLLUS}}   
\end{center}
\end{figure}

At present the NLL running is only known for the ultrasoft term \cite{Pineda:2011aw}:
\bea
\nn
&&
\delta \tilde D_{d,us}^{(2),NLL}(\nu_h;\nu)
=
\frac{(m_1+m_2)^2}{m_1m_2}\frac{4\pi}{\beta_0}\left(\frac{C_A}{2}-C_F\right)\al(\nu)
\left\{
\frac{2}{3\pi}\ln\left(\frac{\al(\nu)}{\al(\nu^2/\nu_h)}\right) a_1 \frac{\al(\nu)}{4\pi}
\right.
\\
&&
\qquad
\left.  + (\al(\nu^2/\nu_h)-\al(\nu))
\left(
\frac{8}{3}\frac{\beta_1}{\beta_0}
\frac{1}{(4\pi)^2}-\frac{1}{27\pi^2}\left(C_A\left(47+6\pi^2\right)-10T_Fn_f\right)
\right)
  \right\}
\,,
\label{NNLUS}
\eea
where $a_1=31/9C_A-20T_Fn_f/9$.  
We show the size of this correction in Fig. \ref{Fig:NLLUS}. 
Note that the ultrasoft contribution to the delta potential vanishes in the large $N_c$ limit (it is $1/N_c^2$ suppressed). Nevertheless, it quickly becomes big at relatively small scales because the overall coefficient is large and the ultrasoft scale quickly
 becomes small. Finally, note also that part of the ultrasoft correction (proportional to $\ln k$) is included in Eq. (\ref{ddkDd}).

The missing terms to obtain the complete NLL running of $\tilde D_d^{(2)}$ are then  
$\delta \tilde D_{d,s}^{(2),NLL}(\nu_h;\nu)$ and 
$\delta \tilde D_{d,p}^{(2),NLL}(\nu_h;\nu)$. For $\delta \tilde D_{d,s}^{(2),NLL}(\nu_h;\nu)$ we need the two-loop soft computation of $\tilde D_{d}^{(2)}$, and the associated soft RG equation, which we partially obtain in Secs. \ref{Sec:al2Vr} and \ref{Sec:delta}, respectively. We also discuss the mixing with higher order $1/m$ potentials in Sec. \ref{Sec:EoM}. 
For $\delta \tilde D_{d,p}^{(2),NLL}(\nu_h;\nu)$ we need to determine and solve the potential RG equation. This requires first the matching between NRQCD and pNRQCD to higher orders in $1/m$, which we do in Secs. \ref{Sec:m4} and \ref{Sec:al2m3}, an extra (ultrasoft associated) running, which we obtain in Sec. \ref{sec:ulso}, and obtaining the potential RG equation, which we do in Sec. \ref{Sec:potNLL}. 
 
\section{\label{sec::matchSI}NRQCD--pNRQCD matching, spin-independent}

In this section we compute the potentials for which their expectation values produce corrections to the spectrum of ${\cal O}(m\al^6)$. This means the $\mathcal{O}(\alpha/m^4)$, $\mathcal{O}(\alpha^2/m^3)$ and $\mathcal{O}(\alpha^3/m^2)$ potentials. Of them we mostly care about those that produce logarithmic enhanced contributions to the spectrum. Therefore, in particular, we do not need to consider the $p^6/m^5$ correction to the kinetic term, since it does not give an ultraviolet divergent correction. The $\mathcal{O}(\alpha/m^4)$ and $\mathcal{O}(\alpha^2/m^3)$ potentials are finite. Some of them can be traced back from the QED computation. We mainly compare with \cite{KMY} (but one could also look into  \cite{CMY} for the equal mass case). Logarithmic enhanced corrections are produced by the divergences generated when inserting these potentials in potential loops. On the other hand the logarithmic enhanced contribution to the spectrum due to the $\mathcal{O}(\alpha^3/m^2)$ is not generated by potential loops but by the divergent structure of the potential itself, which we then refer to as soft running. This case will be discussed separately in Sec. \ref{Sec:delta}.

The spin-dependent case was computed in \cite{Kniehl:2003ap,Penin:2004xi}. Explicit expressions for the potentials can be found in the Appendix of \cite{Penin:2004ay}. They produced corrections to the hyperfine splitting (but not to the fine splittings, as shown in \cite{Pwave}). 

\subsection{$\mathcal{O}(\alpha/m^4)$ potential}
\label{Sec:m4}
\begin{figure}[!ht]
\centering
\includegraphics[scale=0.07]{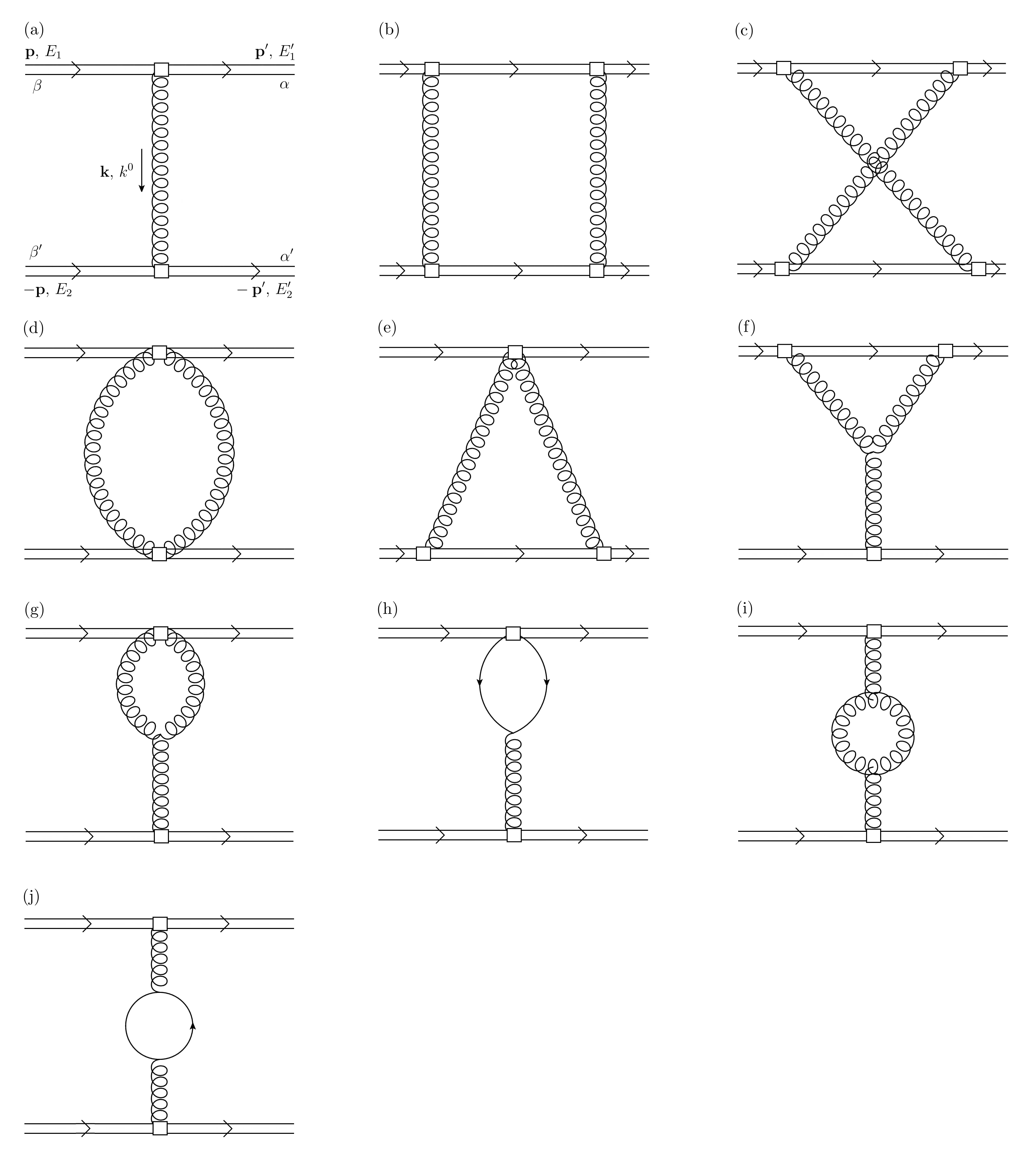}
\caption{The first diagram is the only topology that contributes to the tree level potential. Properly changing the vertex and/or Taylor expanding the denominator of the propagators all potentials are generated. The other diagrams are the general topologies that contribute to the $\al^2/m^3$ potential. Again, properly changing the vertices and/or Taylor expanding the denominator of the propagators, all potentials are generated.}
\label{matchingpNRQCD}
\end{figure}

From a tree level computation (see the first diagram in Fig. \ref{matchingpNRQCD}) we obtain the complete (spin-independent) $\al/m^4$ potentials in momentum space: 
\bea
\tilde {V}_{tree}&=&- c_D^{(1)} c_D^{(2)}C_F\frac{g^2}{64m_1^2 m_2^2}{\bf k}^2
\nn
\\
&&- C_F g^2
\left(\frac{c_{X1}^{(1)}}{m_1^4}+\frac{c_{X1}^{(2)}}{m_2^4}\right)\frac{({\bf p}^2 - {\bf p}'\,^2)^2}{{\bf k}^2}
\nn
\\
&&-C_F g^2\left(\frac{c_{X2}^{(1)}}{m_1^4}+\frac{c_{X2}^{(2)}}{m_2^4}\right)({\bf p}^2+{\bf p}'^2)
\nn
\\
&& - C_F g^2 \left(\frac{c_{X3}^{(1)}}{m_1^4}+\frac{c_{X3}^{(2)}}{m_2^4}\right){\bf k}^2\nn
\\
&&
+C_F\frac{g^2c_k^{(1)2}c_k^{(2)2}}{16m_1^2 m_2^2}\frac{1}{{\bf k}^4}
({\bf p}^2-{\bf p}'^2)^2\left(2({\bf p}^2 + {\bf p}'^2)-{\bf k}^2 -\frac{({\bf p}^2-{\bf p}'^2)^2}{{\bf k}^2}\right)
\nn
\\
&&
+C_F\frac{g^2}{16 m_1 m_2}\left(\frac{c_4^{(1)}c_k^{(2)}}{m_1^2}+\frac{c_4^{(2)}c_k^{(1)}}{m_2^2}\right)
\frac{{\bf p}^2 + {\bf p}'^2}{{\bf k}^2}\left(2({\bf p}^2 + {\bf p}'^2)-{\bf k}^2-\frac{({\bf p}^2-{\bf p}'^2)^2}{{\bf k}^2}\right)
\nn
\\
&&
-C_F\frac{g^2}{16m_1 m_2}\left(\frac{c_M^{(1)}c_k^{(2)}}{m_1^2} + \frac{c_M^{(2)}c_k^{(1)}}{m_2^2}\right)
\left(\frac{({\bf p}^2-{\bf p}'^2)^2}{{\bf k}^2}-({\bf p}+{\bf p}')^2\right)
\,.
\label{eq:tree}
\eea
In this result we have already used the (full) equations of motion replacing \cite{pos}
\be
k_0^2 \rightarrow -\frac{c_k^{(1)}c_k^{(2)}({\bf p}^2-{\bf p'}^2)^2}{4m_1m_2}
\,.
\ee
Such $k_0^2$ terms are generated by Taylor expanding in powers of the energy $k_0$ the denominator of the transverse gluon propagator.

Not all terms in \eq{eq:tree} contribute to the NLL running of the delta potential. The ones that are local (or pseudo-local) do not contribute, as they do not produce potential loop divergences, since the expectation values of these potentials are proportional to 
$|\psi(0)|^2$ and/or (analytic) derivatives of it (kind of $\bfnabla^2|\psi(0)|^2$), which are finite. 
This happens for instance for the potentials proportional to $c_D^2$, $c_{X2}$ and $c_{X3}$. It is also this fact that allows us to neglect 
$1/m^4$ potentials generated by dimension eight four-heavy fermion operators of the 
NRQCD Lagrangian.

As we have incorporated the LL running of the HQET Wilson coefficients, these potentials are already RG improved. 

Note that with trivial modifications these potentials are also valid for QED. 

\subsection{$\mathcal{O}(\alpha^2/m^3)$ potential}
\label{Sec:al2m3}
We now compute the complete set of the $\mathcal{O}(\alpha^2/m^3)$ spin-independent potentials. We show the relevant topologies that contribute to the $\al^2/m^3$ potential in Fig. \ref{matchingpNRQCD}. By properly changing the vertices all potentials are generated.

The (b) type diagrams in Fig. \ref{matchingpNRQCD} do not generate $\mathcal{O}(\alpha^2/m^3)$ potentials (in the Coulomb gauge). 

The (c) type diagrams in Fig. \ref{matchingpNRQCD} do generate $\mathcal{O}(\alpha^2/m^3)$ potentials. They read 
\bea
\label{V1loopq2}
 \tilde V_{1loop}^{(c,1)}&=& -C_F\left(C_F-\frac{C_A}{2}\right)c_k^{(1)}c_k^{(2)}\frac{g^4}{512 m_1 m_2}\frac{E_1+E_2}{|{\bf k}|^{3-2\epsilon}}
\nn
\\
&& \times
\left( 2({\bf p}^2 + {\bf p}'^2)- {\bf k}^2-\frac{({\bf p}^2 -{\bf p}'^2)^2}{{\bf k}^2} 
 - \frac{8({\bf p}\cdot{\bf k})({\bf p}'\cdot{\bf k})}{{\bf k}^2}\right)
\,,
\eea
$$\tilde V^{(c,2)}_{1loop}=-C_F\left(C_F-\frac{C_A}{2}\right)\frac{g^4}{256m_1 m_2}\left(\frac{c_k^{(1)\,2}c_k^{(2)}}{m_1}+\frac{c_k^{(1)}c_k^{(2)\,2}}{m_2}\right)
|{\bf k}|^{1+2\epsilon}
$$
\begin{equation}
\times \bigg( 3({\bf p}^2 + {\bf p}'^2)\frac{({\bf p}\cdot{\bf k})({\bf p}'\cdot{\bf k})}{{\bf k}^6}-\frac{2({\bf p}^2 + {\bf p}'^2)}{{\bf k}^2}+\frac{11}{4}-\frac{1}{4}\frac{({\bf p}^2 - {\bf p}'^2)^2}{{\bf k}^4}
-\frac{1}{2}\frac{({\bf p}^2 + {\bf p}'^2)^2}{{\bf k}^4}\bigg)
\label{Vc2}
\,.\end{equation}

The (d) type diagrams in Fig. \ref{matchingpNRQCD} do not generate $\mathcal{O}(\alpha^2/m^3)$ potentials. 

The (e) type diagrams in Fig. \ref{matchingpNRQCD} do generate $\mathcal{O}(\alpha^2/m^3)$ potentials. They read
\bea
\tilde V^{(e,1)}_{1loop}=C_F\left(2C_F-\frac{C_A}{2}\right)\frac{g^4}{512 m_1 m_2}
\left(\frac{c_k^{(1)\,2}c_k^{(2)}}{m_1}+\frac{c_k^{(1)}c_k^{(2)\,2}}{m_2}\right)|{\bf k}|^{1+2\epsilon}
\nn
\\
\times
\left(\frac{5({\bf p}^2 + {\bf p}'^2)}{{\bf k}^2}-\frac{7}{2}-\frac{3}{2}\frac{({\bf p}^2-{\bf p}'^2)^2}{{\bf k}^4}\right)
\,,
\eea
\bea
\tilde V_{1loop}^{(e,2)}&=&-C_F\left(2C_F-\frac{C_A}{2}\right)\frac{g^4}{256}\left(\frac{c_{A1}^{(1)}}{m_1^3} + \frac{c_{A1}^{(2)}}{m_2^3}\right)
|{\bf k}|^{1+2\epsilon}
\nn
\\
&&-C_F\left(2C_F-\frac{C_A}{2}\right)
\frac{g^4}{512}\left(\frac{c_{A2}^{(1)}}{m_1^3} + \frac{c_{A2}^{(2)}}{m_2^3}\right)|{\bf k}|^{1+2\epsilon}
\nn
\\
&&
-C_F\left(2C_F-\frac{C_A}{2}\right)\frac{g^4}{128 m_1 m_2}
\left(\frac{c^{(1)\,2}_F c_k^{(2)}}{m_1}+\frac{c^{(2)\,2}_F c_k^{(1)}}{m_2}\right)|{\bf k}|^{1+2\epsilon}
\nn
\\
&&
-C_F C_A\frac{g^4}{256m_1 m_2}\left(\frac{c_D^{(1)}c_k^{(2)}}{m_1}+\frac{c_D^{(2)}c_k^{(1)}}{m_2}\right)|{\bf k}|^{1+2\epsilon}
\nn
\\
&&
-\frac{T_F}{N_c} C_F \frac{g^4}{128}\left(\frac{c_{A3}^{(1)}}{m_1^3}+\frac{c_{A3}^{(2)}}{m_2^3}\right)|{\bf k}|^{1+2\epsilon}
\nn
\\
&&
-\frac{T_F}{N_c} C_F \frac{g^4}{256}\left(\frac{c_{A4}^{(1)}}{m_1^3}+\frac{c_{A4}^{(2)}}{m_2^3}\right)|{\bf k}|^{1+2\epsilon}
\,.
\eea

The (f) type diagrams in Fig. \ref{matchingpNRQCD} do generate $\mathcal{O}(\alpha^2/m^3)$ potentials. They read
\begin{equation}
 \tilde V_{1loop}^{(f,1)}= - C_F C_A\frac{g^4}{128}\left(\frac{c_F^{(1)\,2}c_k^{(1)}}{m_1^3}+\frac{c_F^{(2)\,2}c_k^{(2)}}{m_2^3}\right)
 |{\bf k}|^{1+2\epsilon}\frac{{\bf p}\cdot{\bf p}'}{{\bf k}^2}
\,,
\end{equation}

$$\tilde V_{1loop}^{(f,2)}=- C_F C_A\frac{g^4}{512 m_1 m_2}\left(\frac{c_k^{(2)}c_k^{(1)\,2}}{m_1}+\frac{c_k^{(1)}c_k^{(2)\,2}}{m_2}\right)
|{\bf k}|^{1+2\epsilon}\left(1-\frac{3}{2}\frac{{\bf p}^2 +{\bf p}'^2}{{\bf k}^2}\right)$$
$$\times\left(\frac{2({\bf p}^2 + {\bf p}'^2)}{{\bf k}^2}-1-\frac{({\bf p}^2-{\bf p}'^2)^2}{{\bf k}^4}\right)$$
$$- C_F C_A\frac{g^4}{512}\left(\frac{c_k^{(1)\,3}}{m_1^3}+\frac{c_k^{(2)\,3}}{m_2^3}\right)|{\bf k}|^{1+2\epsilon}
 \bigg( \frac{{\bf p}\cdot{\bf p}'}{{\bf k}^2}  + \frac{5({\bf p}\cdot{\bf k})({\bf p}'\cdot{\bf k})}{{\bf k}^4}
-\frac{12({\bf p}\cdot{\bf p}')^2}{{\bf k}^4} + \frac{2{\bf p}^2{\bf p}'^2}{{\bf k}^4}$$
\begin{equation}
 +\frac{6({\bf p}\cdot{\bf p}')({\bf p}\cdot{\bf k})({\bf p}'\cdot{\bf k})}{{\bf k}^6}\bigg)
\,,
\end{equation}

$$ \tilde V^{(f,3)}_{1loop}= -c_k^{(1)}c_k^{(2)} C_F C_A\frac{3g^4}{128m_1 m_2}|{\bf k}|^{-1+2\epsilon}
\left({\bf p}\cdot{\bf p}' - \frac{({\bf p}\cdot{\bf k})({\bf p}'\cdot{\bf k})}{{\bf k}^2}\right) 
\frac{(E_1+E_1')+(E_2+E_2')}{{\bf k}^2}$$
\begin{equation}
 -  C_F C_A\frac{g^4}{256}
 \left(\frac{c_k^{(1)\,2}}{m_1^2}(E_1+E_1')
 +\frac{c_k^{(2)\,2}}{m_2^2}(E_2+E_2') \right) |{\bf k}|^{-1+2\epsilon} 
\left(\frac{5{\bf p}\cdot {\bf p}'}{{\bf k}^2} - \frac{3({\bf p}\cdot{\bf k})({\bf p}'\cdot{\bf k})}{{\bf k}^4}\right)
\,,
\end{equation}

$$\tilde V_{1loop}^{(f,4)}=- C_F C_A\frac{g^4}{256}\left(\frac{c_k^{(1)\,3}}{m_1^3}+\frac{c_k^{(2)\,3}}{m_2^3}\right)|{\bf k}|^{1+2\epsilon}
\bigg(-1+\frac{{\bf p}^2   + {\bf p}'^2}{{\bf k}^2}$$
\begin{equation}
+ \frac{3({\bf p}^4 +{\bf p}'^4) + ({\bf p}^2 + {\bf p}'^2)({\bf p}\cdot{\bf p}') - 6({\bf p}\cdot{\bf p}')^2}{{\bf k}^4}
+\frac{-3({\bf p}^6 + {\bf p}'^6) + 4({\bf p}^4 + {\bf p}'^4)({\bf p}\cdot{\bf p}')-2({\bf p}\cdot{\bf p}')^3}{{\bf k}^6}\bigg)
\,,
\end{equation}

$$\tilde V^{(f,5)}_{1loop}= -C_F C_A\frac{g^4}{128}|{\bf k}|^{1+2\epsilon}
\bigg[\frac{c_k^{(2)\,2}}{m_2^2}\bigg(
\frac{3(E_1+E_1')({\bf p}\cdot{\bf k})({\bf p}'\cdot{\bf k})}{{\bf k}^6} 
+ \frac{(E_1+E_1')({\bf p}\cdot{\bf p}')}{{\bf k}^4}$$
$$+\frac{2(E_1{\bf p}^4 + E_1'{\bf p}'^4)}{{\bf k}^6} 
- \frac{2(E_1{\bf p}^2 +E_1'{\bf p}'^2)({\bf p}\cdot{\bf p}' + {\bf k}^2)}{{\bf k}^6}\bigg)
+ \frac{c_k^{(1)\,2}}{m_1^2}
\bigg( 
\frac{3(E_2+E_2')({\bf p}\cdot{\bf k})({\bf p}'\cdot{\bf k})}{{\bf k}^6}
$$
\begin{equation}
 + \frac{(E_2+E_2')({\bf p}\cdot{\bf p}')}{{\bf k}^4}
+\frac{2(E_2{\bf p}^4 + E_2'{\bf p}'^4)}{{\bf k}^6} 
- \frac{2(E_2{\bf p}^2 +E_2'{\bf p}'^2)({\bf p}\cdot{\bf p}' + {\bf k}^2)}{{\bf k}^6}
\bigg)
\bigg]
\,,
\end{equation}

\begin{equation}
\label{V1loopNAs}
 \tilde V_{1loop}^{(f,6)}=-  C_F C_A\frac{7g^4}{256m_1 m_2}\left(\frac{c_k^{(1)\,2}c_k^{(2)}}{m_1}+\frac{c_k^{(1)}c_k^{(2)\,2}}{m_2}\right)
 |{\bf k}|^{1+2\epsilon}\left(\frac{2({\bf p}^2 + {\bf p}'^2)}{{\bf k}^2}-1-\frac{({\bf p}^2-{\bf p}'^2)^2}{{\bf k}^4}\right)
\,,
\end{equation}

\bea
 \tilde V_{1loop}^{(f,7)}&=&  C_F C_A\frac{g^4}{256m_1 m_2}\left(\frac{c_D^{(1)}c_k^{(2)}}{m_1}
 +\frac{c_D^{(2)}c_k^{(1)}}{m_2}\right)|{\bf k}|^{1+2\epsilon}
\nn
\\
&&
+  C_F C_A\frac{g^4}{1024}\left(\frac{c_4^{(1)}}{m_1^3}+\frac{c_4^{(2)}}{m_2^3}\right)|{\bf k}|^{1+2\epsilon}
\left(\frac{10({\bf p}^2 + {\bf p}'^2)}{{\bf k}^2}-7 + \frac{5({\bf p}^2-{\bf p}'^2)^2}{{\bf k}^4}\right)
\nn
\\
&&
- C_F C_A \frac{g^4}{256}\left(\frac{c_M^{(1)}}{m_1^3}+\frac{c_M^{(2)}}{m_2^3}\right)|{\bf k}|^{1+2\epsilon}
\nn
\\
&&
- C_F C_A\frac{g^4}{512}\left(\frac{c_F^{(1)}c_S^{(1)}}{m_1^3}+\frac{c_F^{(2)}c_S^{(2)}}{m_2^3}\right)|{\bf k}|^{1+2\epsilon}
\,,
\eea

\begin{equation}
\tilde V_{1loop}^{(f,8)}=  C_F C_A\frac{g^4}{64}|{\bf k}|^{-5+2\epsilon}
 \left[\left(c_k^{(1)}\frac{E_1^2}{m_1} + c_k^{(2)}\frac{E_2^2}{m_2}\right)({\bf p}\cdot {\bf k})
 - \left(c_k^{(1)}\frac{E_1'^2}{m_1} + c_k^{(2)}\frac{E_2'^2}{m_2} \right)({\bf p}'\cdot{\bf k})\right]
\,,
\end{equation}

\begin{equation}
\label{eq:Vf9E}
 \tilde V_{1loop}^{(f,9)}= C_F C_A \frac{g^4}{128}
 \left(\frac{c_F^{(1)\,2}}{m_1^2}(E_1+E_1') + \frac{c_F^{(2)\,2}}{m_2^2}(E_2+E_2')\right)
 |{\bf k}|^{-1+2\epsilon}
\,.
\end{equation}

The rest of topologies ((g), (h), (i), (j)) do not contribute. Note that those topologies include, in particular, the one-loop diagrams proportional to $c_i^{hl}$ or $d_i^{hl}$, as they may produce $\sim \al^2/m^3$ potentials. We find that such contributions vanish.

As we have incorporated the LL running of the HQET Wilson coefficients, these potentials are already RG improved. 

Note that with trivial modifications these potentials are also valid for QED. 

\subsection{$\mathcal{O}(\alpha^3/m^2)$ $V_r$ potential}
\label{Sec:al2Vr}

In this section we perform a partial computation of the $\mathcal{O}(\alpha^3/m^2)$ soft contribution to the  $V_r$ potential. 
 The contributions we compute here are those proportional to the 
HQET Wilson coefficients $\bar c^{(i)hl}_1$ and $c_F^{(i)}$. We define
\be
\frac{\tilde D_d^{(2)}}{m_1m_2}=\frac{\tilde D_d^{(2,0)}}{m_1^2}+\frac{\tilde D^{(0,2)}_d}{m_2^2}+
\frac{\tilde D^{(1,1)}_d}{m_1m_2}
\,.
\ee
Using the notation of \cite{Peset:2015vvi},
 \be
\pi C_F \tilde D_{d,B}^{(2,0)}= 
\tilde D_r^{(2,0)}=g_B^2C_F\left\{D_{r,1}^{(2,0)}+\frac{g_B^2k^{2\epsilon}}{16\pi^2}D_{r,2}^{(2,0)}+\frac{g_B^3k^{4\epsilon}}{(4\pi)^3}D_{r,3}^{(2,0)}+\cdots\right\}
\,,
\ee
the bare new result reads 
\bea
\nn
&&
{\tilde D}_{r,3}^{(2,0)}=\bar c_1^{hl}
\left[
T_F n_l \left(C_A \left(-\frac{ 2^{-8 \epsilon -4} \pi ^{\frac{5}{2}-2 \epsilon }3\
   \left(2 \epsilon ^2+7 \epsilon +4\right) \csc (2 \pi  \epsilon ) \csc (\pi  \epsilon
   )}{\epsilon  (2 \epsilon +3) \Gamma \left(2 \epsilon +\frac{5}{2}\right)}
   \right.
   \right.
   \right.
   \\
   \nn
   &&
      -\frac{2^{-6 \epsilon
   -3} \pi ^{\frac{3}{2}-2 \epsilon } \left(40 \epsilon ^4+160 \epsilon ^3+240 \epsilon ^2+167
   \epsilon +44\right) \csc (2 \pi  \epsilon ) \Gamma^2 (\epsilon +1)}{\epsilon  (2 \epsilon +3)
   \Gamma \left(\epsilon +\frac{5}{2}\right) \Gamma (3 \epsilon +3)}
   \\
   \nn
   &&
   \left.
  +\frac{2^{-6 \epsilon -3} \pi
   ^{\frac{3}{2}-2 \epsilon } \left(4 \epsilon ^4+12 \epsilon ^3+12 \epsilon ^2+13 \epsilon
   +6\right) \sin (2 \pi  \epsilon ) \csc ^2(\pi  \epsilon ) \Gamma (-2 \epsilon -3) \Gamma
   (\epsilon +2)}{\epsilon  \Gamma \left(\epsilon +\frac{5}{2}\right)}\right)
   \\
   \nn
   &&
      +C_F
   \left(\frac{2^{-8 \epsilon -4} \pi ^{2-2 \epsilon } (2 \epsilon +1) (2 \epsilon +3)
   \left(\epsilon ^2+2 \epsilon +2\right) \csc (\pi  \epsilon ) \sec (\pi  \epsilon ) \Gamma
   (\epsilon +2) \Gamma (2 \epsilon +2)}{\epsilon ^2 \Gamma^2 \left(\epsilon +\frac{5}{2}\right)
   \Gamma (3 \epsilon +3)}
   \right.
   \\
   \nn
   &&
    \left.\left.\left.
   -\frac{2^{-8 \epsilon -5} \pi ^{3-2 \epsilon } (\epsilon +1) (2
   \epsilon +3) \left(2 \epsilon ^2+\epsilon +2\right) \csc ^2(\pi  \epsilon )}{\epsilon  \Gamma^2
   \left(\epsilon +\frac{5}{2}\right)}\right)\right)+\frac{(T_F n_l)^2 2^{-8 \epsilon -3} \pi
   ^{3-2 \epsilon } (\epsilon +1)^2 \csc ^2(\pi  \epsilon )}{\Gamma^2 \left(\epsilon
   +\frac{5}{2}\right)}
   \right]
   \\
   \nn
   &&
+\left[c_F^{(1)}\right]^2\frac{1}{3} C_A 2^{-8 \epsilon -7} \pi ^{-2 \epsilon } 
\\
\nn
&&
\times
 \Biggl[
 C_A \left(\frac{ 2^{4 \epsilon +5} 3\ (\epsilon  (\epsilon  (\epsilon  (\epsilon  (2 \epsilon  (18 \epsilon  (2 \epsilon +11)+401)+661)+33)-283)-165)-30) \Gamma (1-2 \epsilon ) \Gamma^3 (\epsilon )}{(4 \epsilon  (\epsilon +2)+3)^2 \Gamma (3 \epsilon +3)}
 \right.
 \\
 \nn
 &&
 +\frac{\pi  2^{4 \epsilon +5} \Gamma (1-2 \epsilon ) \Gamma^3 \left(\epsilon +\frac{1}{2}\right)}{\Gamma \left(3 \epsilon +\frac{3}{2}\right)}+\frac{3 \pi ^3 (\epsilon  (\epsilon 
   (22-\epsilon  (12 \epsilon +17))+45)+15) \csc ^2(\pi  \epsilon )}{\epsilon  \Gamma^2 \left(\epsilon +\frac{5}{2}\right)}
   \\
   \nn
   &&
    \left.
   +\frac{24 \pi ^{5/2} (\epsilon  (\epsilon  (\epsilon  (4 \epsilon  (\epsilon +12)+127)+130)+65)+15) \csc (\pi  \epsilon ) \csc (2 \pi  \epsilon )}{\epsilon  (4 \epsilon  (\epsilon +2)+3) \Gamma \left(2 \epsilon +\frac{5}{2}\right)}+\frac{12 \pi ^4 (2 \epsilon -1) \sec ^2(\pi  \epsilon )}{\Gamma^2 (\epsilon +1)}\right)
   \\
   \nn
   &&
   +\frac{24 \pi ^{3/2} n_f T_F }{(2 \epsilon +3)^2}\left(\frac{4^{\epsilon +1} \Gamma (\epsilon
   +1) \left(\epsilon  (4 \epsilon +3) \cot (\pi  \epsilon ) \Gamma (-2 \epsilon -1)-\frac{\left(6 \epsilon ^2+9 \epsilon +4\right) (2 \epsilon  (2 \epsilon +5)+5) \csc (2 \pi  \epsilon ) \Gamma (\epsilon )}{\Gamma (3 \epsilon +3)}\right)}{\Gamma \left(\epsilon +\frac{3}{2}\right)}
   \right.
   \\
   &&
   \left.
   -\frac{\pi  (2 \epsilon +1)^2 (2 \epsilon +3) \csc (\pi  \epsilon ) \csc (2 \pi  \epsilon )}{\Gamma \left(2 \epsilon +\frac{5}{2}\right)}\right)
   \Biggr]
\,.
\eea
With obvious changes the same result is obtained for ${\tilde D}_{r,3}^{(0,2)}$. It is worth emphasizing that this expression vanishes in pure QED. A non trivial check of this result is that $c_D$ and $c_1^{hl}$ appear in the gauge invariant combination $\bar c_1^{hl}=c_D+c_1^{hl}$. Another nontrivial check is that the counterterm is independent of $k$ and that the $1/\epsilon^2$ terms comply with the constraints from RG. This computation has been done in the Feynman gauge (with a general gauge parameter $\xi$) in the kinematic configuration ${\bf p}={\bf k}$ and ${\bf p}'=0$. We also set the external energy to zero. Not setting it to zero produces subleading corrections (we recall that the one-loop computation of this contribution has no energy dependence \cite{Peset:2015vvi}). The result is shown to be independent of the gauge fixing parameter $\xi$.

For future computations, it is useful to explain the convention we have taken for the $D$-dimensional spin matrices. For the $c_F^{(i)}$ vertex we typically take a covariant notation $\sim \sigma^{\mu\nu}$ (see for instance \cite{Finkemeier:1996uu}) and project to the particle to single out the spin-independent part: $ \sim Tr[(I+\gamma_0)/2(\cdots \cdots)(I+\gamma_0)/2]$. At one loop this procedure gives the same result than using Pauli matrices with the conventions used in \cite{Peset:2015vvi}.

Though not directly relevant for this work, we also give the $\MS$ renormalized expression of the bare potential computed above. It will be of relevance for future computations of the spectrum (and decays) at N$^4$LO. The result reads ($\al=\al(\nu)$) 
\bea
\nn
\tilde D_{r,\MS}^{(2,0)}(\vk)&=&\frac{C_F \alpha^2}{2} \left[\frac{13}{36}c_F^{(1)2}
C_A-\frac{5}{9}\bar c_1^{hl\,(1)}T_F n_f
+
\left(-C_A\frac{5}{6}c_F^{(1)2}+\frac{2}{3}\bar c_1^{hl\,(1)}T_F n_f\right)\ln\left(k/\nu\right)\right]
\\
\nn
&&
+
c_F^{(1)2}C_FC_A^2\frac{\al^3}{2\pi} 
\left(
\frac{ 1080 \zeta (3)+706-900 \gamma +432 \pi ^2-81 \pi
   ^4+900 \ln (4\pi )}{5184 }
\right.
\\
&&
\left.
\quad
-\frac{179 
 }{108}\ln(k/ \nu)+\frac{10  }{9}\ln^2(k/ \nu)
\right)
\nn
\\
&&
+c_F^{(1)2}C_FC_An_fT_F\frac{\al^3}{2\pi}   
\left(
\frac{  -3581+750 \gamma -750
   \ln (4\pi )}{2592 }
+\frac{  91
     }{54 }\ln(k/ \nu)-\frac{7  }{12}\ln^2(k/ \nu)
\right)
\nn
\\
&&
+
{\bar c}_1^{hl}C_Fn_fT_FC_A\frac{\al^3}{2\pi}
\left(
\frac{   -1008 \zeta (3)+627-130 \gamma +130 \ln (4\pi )}{864}
\right.
\nn
\\
&&
\left.
\quad
+\frac{5  
  }{6 }\ln(k/ \nu) -\frac{31  }{36 }\ln^2(k/ \nu) 
\right)
\nn
\\
&&
+ {\bar c}_1^{hl}C_F^2 n_fT_F  \frac{\al^3}{2 \pi }  
\left(\frac{ 48 \zeta (3)-55+6 \gamma -6 \ln (4\pi)}{48}+\frac{1}{2}
\ln(k/ \nu) 
\right)
\nn
\\
&&
+
{\bar c}_1^{hl}C_Fn^2_fT^2_F\frac{\al^3}{2\pi}\left(
\frac{25  }{81 }-\frac{20  }{27 }\ln(k/ \nu) 
+\frac{4}{9 }\ln^2(k/ \nu) \right)
\,,
\eea
where we have also included the ${\cal O}(\al^2)$ term. Note that this contribution does not mix with 
$V_{\bf L}^{(2)}$. Therefore, it really corresponds to the contributions proportional to $c_F^{(1)}$ and $\bar c_1^{hl(1)}$ of $D_r^{(2,0)}$, as defined in \cite{Peset:2015vvi}. With obvious changes a similar expression is obtained for $\tilde D_{r,\MS}^{(0,2)}(\vk)$.

Finally, note that the missing part of the soft term should carefully be computed in a way consistent with the scheme we have used for the rest of the computation, in particular of the $\al^2/m^3$ potential, as a strong mixing (if using field redefinitions) of the terms proportional to $c_k^2$ is expected.

\subsection{Equations of motion}
\label{Sec:EoM}

Some of the potentials we have obtained in Sec. \ref{Sec:al2m3} are energy dependent. If we want to eliminate such energy dependence, and write an energy independent potential, this could be achieved by using field redefinitions. At the order we are working it is enough to use the full equation of motion (at leading order), which includes the Coulomb potential. Let us see how it works. We first consider \eq{V1loopq2}. It depends on the total energy of the heavy quarkonium and does not contribute to the running of the delta potential. We next consider \eq{eq:Vf9E}, which is the only energy dependent potential proportional to $c_F^{(i)2}$. Such potential is generated by the following interaction Lagrangian
$$L_{\tilde V_{1loop}^{(f,9)}} =
-C_F C_A\frac{g^4}{128} \frac{c_F^{(1)\,2}}{m_1^2}\int d^3 x_1 d^3 x_2 (\psi^\dagger (i\partial_0\psi(t,{\bf x}_1)) - (i\partial_0\psi^\dagger) \psi(t,{\bf x}_1))
 \int\frac{d^3 k}{(2\pi)^3}
 \frac{e^{i{\bf k}\cdot{\bf x}}}
 {|{\bf k}|^{1-2\epsilon}} \chi_c^\dagger \chi_c (t,{\bf x}_2)$$
\begin{equation}
 - C_F C_A\frac{g^4}{128}\frac{c_F^{(2)\,2}}{m_2^2}\int d^3 x_1 d^3 x_2 \psi^\dagger \psi(t,{\bf x}_1)
\int\frac{d^3 k}{(2\pi)^3}
 \frac{e^{i{\bf k}\cdot{\bf x}}}
 {|{\bf k}|^{1-2\epsilon}}
(\chi_c^\dagger i\partial_0\chi_c (t,{\bf x}_2) - (i\partial_0\chi_c^\dagger) \chi_c (t,{\bf x}_2))
\,.
\end{equation}
For this Lagrangian one can use the equations of motion ($V_C({\bf x})=-C_F\al/|{\bf x}|$):
\be
\left(i\partial_0 +\frac{\bfnabla^2}{2m_1}\right)\psi(t,{\bf x})-\int d^3x_2 \psi(t,{\bf x})V_C({\bf x}-{\bf x}_2)\chi_c^\dagger \chi_c (t,{\bf x}_2)=0
\ee
and similarly for the other fields. We then obtain
\bea
&&
L_{\tilde V_{1loop}^{(f,9)}} =
-C_F C_A\frac{g^4}{128} \frac{c_F^{(1)\,2}}{m_1^2}\int d^3 x_1 d^3 x_2 \left[\psi^\dagger \left(-\frac{\bfnabla^2}{2m_1}\psi(t,{\bf x}_1)\right)
+\left(-\frac{\bfnabla^2}{2m_1}\psi^\dagger
\right) \psi(t,{\bf x}_1)\right]
\nn
\\
\nn
&&
\qquad\qquad\qquad\qquad
\times
 \int\frac{d^3 k}{(2\pi)^3}
 \frac{e^{i{\bf k}\cdot{\bf x}}}
 {|{\bf k}|^{1-2\epsilon}} \chi_c^\dagger \chi_c (t,{\bf x}_2)
\\
&&
\qquad\qquad
-C_F C_A\frac{g^4}{64} \frac{c_F^{(1)\,2}}{m_1^2}\int d^3 x_1 d^3 x_2 d^3x_3\psi^\dagger\psi(t,{\bf x}_1) V_C({\bf x}_1-{\bf x}_3)
\nn
\\
&&
\qquad\qquad\qquad\qquad
\times
\int\frac{d^3 k}{(2\pi)^3}
 \frac{e^{i{\bf k}\cdot({\bf x}_1-{\bf x}_2)}}
 {|{\bf k}|^{1-2\epsilon}}
 \chi_c^\dagger \chi_c (t,{\bf x}_2)\chi_c^\dagger \chi_c (t,{\bf x}_3)
+ \cdots
\,,
\label{eq:Vf9EoM}
\eea
where the dots stand for the analogous contribution for the antiparticle. 

The first term in \eq{eq:Vf9EoM} yields the potential we had obtained after using the free on-shell equations of motion in \eq{eq:Vf9E}. It reads
\begin{equation}
\label{eq:f9free}
 \tilde V_{1loop}^{(f,9)}= C_F C_A \frac{g^4}{256}\left(\frac{c_F^{(1)\,2}c_k^{(1)}}{m_1^3}+\frac{c_F^{(2)\,2}c_k^{(2)}}{m_2^3}\right)
 |{\bf k}|^{1+2\epsilon}\frac{{\bf p}^2 +{\bf p}'^2}{{\bf k}^2}
\,.
\end{equation}
The second term is a six-fermion field term. After contracting two of them, a new $\al^3/m^2$ potential is generated (here we only care about the divergent part). It reads 
\be
\label{eq:f9delta}
\delta \tilde V_{1loop}^{(f,9)}=\frac{1}{32 \epsilon} C_F^2C_A \frac{g^6k^{4\epsilon}}{(4\pi)^2}
\left[\frac{c_F^{(1)2}}{m_1^2}+\frac{c_F^{(2)2}}{m_2^2}\right]
\,.
\ee
 It is worth mentioning that this contribution has a different color structure as those (purely soft) computed in Sec. \ref{Sec:al2Vr}, and that they are $\pi^2$ enhanced compared to those also. Therefore, one could expect them to be more important than the strict pure-soft contribution. 

Remarkably enough, we will see later that the contributions from \eq{eq:f9free} and \eq{eq:f9delta} to the running of the delta potential cancel each other in the equal mass case (but not for different masses). This was to be expected, since in the equal mass case, the potential can be written in terms of the total energy of the heavy quarkonium, which does not produce divergences that should be absorbed in the delta potential. 

It is worth mentioning that this exhausts all possible $c_F^{(i)2}$ structures that can be generated. To be sure of this statement, we have to check that the result does not depend on the gauge. Therefore, we have redone the diagrams proportional to $c_F^{(i)2}$ (i.e. the associated contributions to $\tilde V_{1loop}^{(e,2)}$, $\tilde V_{1loop}^{(f,1)}$ and $\tilde V_{1loop}^{(f,9)}$) in the Feynman gauge and found the same result.

\medskip

The other potentials that are dependent on the energy are proportional to $c_k^2$. As before, these contributions will mix with the $\al^3/m^2$ pure-soft contribution proportional to $c_k^2$, which we have not computed anyhow. Therefore, in this paper, we only include the explicit contribution generated 
using the free equations of motion and postpone the incorporation of the other contribution to have the full result. The contributions we explicitly include in this paper then read:

$$\tilde V_{1loop}^{(f,3)}=- C_F C_A\frac{3g^4}{1024m_1 m_2}\left(\frac{c_k^{(1)\,2}c_k^{(2)}}{m_1}+\frac{c_k^{(1)}c_k^{(2)\,2}}{m_2}\right)
|{\bf k}|^{1+2\epsilon}
\frac{{\bf p}^2 + {\bf p}'^2}{{\bf k}^2}
\left(\frac{2( {\bf p}^2 + {\bf p}'^2)}{{\bf k}^2}-1-\frac{({\bf p}^2 - {\bf p}'^2)^2}{{\bf k}^4}\right)$$
\begin{equation}
- C_F C_A\frac{g^4}{512}\left(\frac{c_k^{(1)\,3}}{m_1^3}+\frac{c_k^{(2)\,3}}{m_2^3}\right) |{\bf k}|^{1+2\epsilon} 
\frac{{\bf p}^2 + {\bf p}'^2}{{\bf k}^2}
\left(\frac{5{\bf p}\cdot {\bf p}'}{{\bf k}^2} - \frac{3({\bf p}\cdot{\bf k})({\bf p}'\cdot{\bf k})}{{\bf k}^4}\right)
\,,
\end{equation}

$$\tilde V_{1loop}^{(f,5)}=-C_F C_A\frac{g^4}{256 m_1 m_2}
|{\bf k}|^{1+2\epsilon}\left(\frac{c_k^{(1)\,2}c_k^{(2)}}{m_1}+\frac{c_k^{(1)}c_k^{(2)\,2}}{m_2}\right)
\bigg( \frac{3({\bf p}^2 + {\bf p}'^2)({\bf p}\cdot{\bf k})({\bf p}'\cdot{\bf k})}{{\bf k}^6}$$
\begin{equation}
-\frac{2({\bf p}^4 +{\bf p}'^4)}{{\bf k}^4} 
+\frac{({\bf p}^2 +{\bf p}'^2)({\bf p}\cdot{\bf p}')}{{\bf k}^4}
+\frac{2({\bf p}^6 + {\bf p}'^6)}{{\bf k}^6}
-\frac{2({\bf p}^4 + {\bf p}'^4)({\bf p}\cdot{\bf p}')}{{\bf k}^6}\bigg)
\,,
\end{equation}

\begin{equation}
\tilde V_{1loop}^{(f,8)}=  C_F C_A\frac{g^4}{512}\left(\frac{c_k^{(1)\,3}}{m_1^3} + \frac{c_k^{(2)\,3}}{m_2^3}\right)|{\bf k}|^{1+2\epsilon}
\left(\frac{2({\bf p}^6 + {\bf p}'^6)}{{\bf k}^6}-\frac{({\bf p}^4 + {\bf p}'^4)({\bf p}^2 + {\bf p}'^2)}{{\bf k}^6}
+ \frac{{\bf p}^4 + {\bf p}'^4}{{\bf k}^4}\right)
\,.
\end{equation}

\section{$\tilde D_d^{(2)}$ NLL running}
\label{Sec:RGpot}

We now compute the NLL soft and potential running of $\tilde D_d^{(2)}$.

\subsection{Soft running}
\label{Sec:delta}

From the results obtained in Sec. \ref{Sec:al2Vr} we can obtain the ${\cal O}(\al^3)$ RG soft equation of $\tilde D_d$ (the ${\cal O}(\al^2)$ RG soft equation can be found in \cite{Pineda:2001ra}) proportional to $c_F^{(i)2}$ and $c_1^{hl(i)}$. In practice, such computation can be understood as getting the NLL soft running of $d_{ss}+C_F \bar d_{vs}$ (see \eq{DdCG} or \eq{DdOS}). It reads
\begin{eqnarray}
&&
  \nu_s {d\over d \nu_s}(d_{ss}+C_F\bar d_{vs})\Bigg|_{soft}
  =
  C_F\al^2\left(2C_F-{ C_A \over
2}\right)c_k^{(1)}c_k^{(2)} 
\\
\nn
&&
+C_F\al^2\left[
{m_1 \over m_2}\left(\frac{1}{3}T_fn_f \bar c_1^{hl(2)}- { 4 \over 3}(C_A+C_F)[c_k^{(2)}]^2-{ 5 \over 12}C_A[c_F^{(2)}]^2\right)
\right.
\\
\nn
&&
\left.
\qquad
+{m_2 \over m_1}\left(\frac{1}{3}T_fn_f \bar c_1^{hl(1)}- { 4 \over 3}(C_A+C_F)[c_k^{(1)}]^2-{ 5 \over 12}C_A[c_F^{(1)}]^2\right)
 \right]
\\
\nn
&&
+C_F\frac{\al^3}{4\pi}\left[
{m_1 \over m_2}\left(-\frac{T_Fn_f}{54}(65C_A-54C_F) \bar c_1^{hl(2)}-\frac{C_A}{18}(25C_A-\frac{125}{3}T_Fn_f)[c_F^{(2)}]^2\right)
\right.
\\
\nn
&&
\left.
\qquad
+{m_2 \over m_1}\left(-\frac{T_Fn_f}{54}(65C_A-54C_F)  \bar c_1^{hl(1)}-\frac{C_A}{18}(25C_A-\frac{125}{3}T_Fn_f)[c_F^{(1)}]^2\right)
 \right]
 \\
\nn
&&
+{\cal O}(\al^3)
\,.
  \end{eqnarray}
The ${\cal O}(\al^3)$ stands for terms proportional to NRQCD Wilson coefficients different from $c_F^{(i)2}$ and $c_1^{hl(i)}$. 
This equation is meant to represent the pure soft running of the NRQCD Wilson coefficients. It does not give the full running of $\tilde D^{(2)}_d$, as one should also include the potential and ultrasoft running. We fix the initial matching condition to zero, since we only need the initial matching condition of the total potential, which can be determined in the final step, when combining the different contributions. 

The strict NLL contribution to the solution of this equation reads (the LL is already included in \eq{DdCG})
\begin{eqnarray}
&&\pi C_F \delta D_{d,s}^{(2),NLL}=[d_{ss} + C_F \bar d_{vs}]^{NLL}= 
- \alpha^2(\nu_h) C_F \bigg[  \bigg(465 C_A^6 (757 m_1^2 - 306 m_1 m_2 + 757 m_2^2)
\nonumber
\\
&&
-  13824 C_F^2 (2 m_1^2 - 3 m_1 m_2 + 2 m_2^2) n_f^4 T_F^4 + 
      C_A^5 \Big(5580 C_F (53 m_1^2 + 102 m_1 m_2 + 53 m_2^2) 
\nonumber
\\
&&      
      + (-590218 m_1^2 + 
            342117 m_1 m_2 - 590218 m_2^2) n_f T_F\Big) - 
      C_A^4 n_f T_F \Big(34 C_F (8347 m_1^2 + 38772 m_1 m_2 
\nonumber
\\
&&      
      + 8347 m_2^2) - 
         3 (115117 m_1^2 - 101466 m_1 m_2 + 115117 m_2^2) n_f T_F\Big) + 
      32 C_A n_f^3 T_F^3 \Big(81 C_F^2 (70 m_1^2
\nonumber
\\
&&      
      - 83 m_1 m_2 + 70 m_2^2) - 
         4 C_F (5 m_1^2 - 459 m_1 m_2 + 5 m_2^2) n_f T_F + 
         120 (m_1^2 + m_2^2) n_f^2 T_F^2\Big)
\nonumber
\\
&&         
         -    8 C_A^2 n_f^2 T_F^2 \Big(81 C_F^2 (566 m_1^2 - 563 m_1 m_2 + 566 m_2^2) - 
         3 C_F (193 m_1^2 - 17595 m_1 m_2 + 193 m_2^2) n_f T_F
\nonumber
\\
&&         
         +          2 (739 m_1^2 + 1080 m_1 m_2 + 739 m_2^2) n_f^2 T_F^2\Big) + 
      6 C_A^3 n_f T_F \Big(360 C_F^2 (106 m_1^2 - 93 m_1 m_2 + 106 m_2^2)
\nonumber
\\
&&      
      +  C_F (10129 m_1^2 + 187731 m_1 m_2 + 10129 m_2^2) n_f T_F - 
         4 (2536 m_1^2 - 4959 m_1 m_2 +  2536 m_2^2) n_f^2 T_F^2\Big)\bigg)
\nonumber
\\
&&         
         \times \frac{1}{36 m_1 m_2 (31 C_A - 16 n_f T_F) (5 C_A - 4 n_f T_F) (11 C_A - 4 n_f T_F)^2 (2 C_A - n_f T_F) } 
\nonumber
\\
&&
+  5 C_A (m_1^2 + m_2^2) \bigg( 397 C_A^3 + 48 C_F n_f^2 T_F^2 + 
    11 C_A^2 (33 C_F - 35 n_f T_F) + 10 C_A n_f T_F (-21 C_F 
\nonumber
\\
&&    
    + 10 n_f T_F)\bigg) z^{\frac{1}{3}(5 C_A - 4 n_f T_F)} 
    \frac{1}{6 m_1 m_2 (5 C_A - 4 n_f T_F) (11 C_A - 4 n_f T_F)^2}
\nonumber
\\
&& 
 - \frac{1}{468 m_1 m_2 (11 C_A - 4 n_f T_F)^2} 
    \bigg( 1989 C_A^3 (8 m_1^2 + 3 m_1 m_2 + 8 m_2^2) + 
     8 C_F n_f T_F \Big(81 C_F (6 m_1^2 
\nonumber
\\
&&      
     + 13 m_1 m_2 + 6 m_2^2) + 
        1240 (m_1^2 + m_2^2) n_f T_F\Big) + 
     2 C_A n_f T_F \Big(C_F (-15134 m_1^2 + 5967 m_1 m_2
\nonumber
\\
&&      
     - 15134 m_2^2) + 
        3100 (m_1^2 + m_2^2) n_f T_F\Big) + 
     2 C_A^2 \Big(3978 C_F (2 m_1^2 - 3 m_1 m_2 + 2 m_2^2) - 
        5 (2263 m_1^2
\nonumber
\\
&&         
        + 351 m_1 m_2 + 2263 m_2^2) n_f T_F\Big)\bigg) z^{\frac{2}{3} (11 C_A - 4 n_f T_F)}  
\nonumber
\\
&&         
        + \frac{2 (5 C_A + 8 C_F) (m_1^2 + m_2^2) n_f T_F (-1327 C_A + 594 C_F + 
     620 n_f T_F) z^{\frac{31}{6}C_A - \frac{8}{3}n_f T_F} }{117 m_1 m_2 (31 C_A - 16 n_f T_F) (11 C_A - 4 n_f T_F)}
\nonumber
\\
&&   
  -  C_A (m_1^2 + m_2^2) \bigg(15 C_A^3 - 188 C_A^2 n_f T_F - 
     2 n_f^2 T_F^2 (27 C_F + 10 n_f T_F) + 
     C_A n_f T_F (216 C_F 
\nonumber
\\
&&      
     + 137 n_f T_F)\bigg) z^{\frac{8}{3}(2 C_A - n_f T_F)} 
     \frac{1}{12 m_1 m_2 (11 C_A - 4 n_f T_F)^2 (2 C_A - n_f T_F)}
\nonumber
\\
&&     
    -  \frac{5 C_A^2 \Big(1-z^{\frac{1}{3} (5 C_A - 4 n_f T_F)}\Big)  \bigg(m_2^2 \ln\left(\frac{\nu_h}{m_1}\right) + 
    m_1^2 \ln\left(\frac{\nu_h}{m_2}\right)\bigg)}{2 m_1 m_2 (5 C_A - 4 n_f T_F)}\bigg]
\,.
\label{eq:DsNLL}
\end{eqnarray}

\begin{figure}[!htb]
	\begin{center}      
	\includegraphics[width=0.75\textwidth]{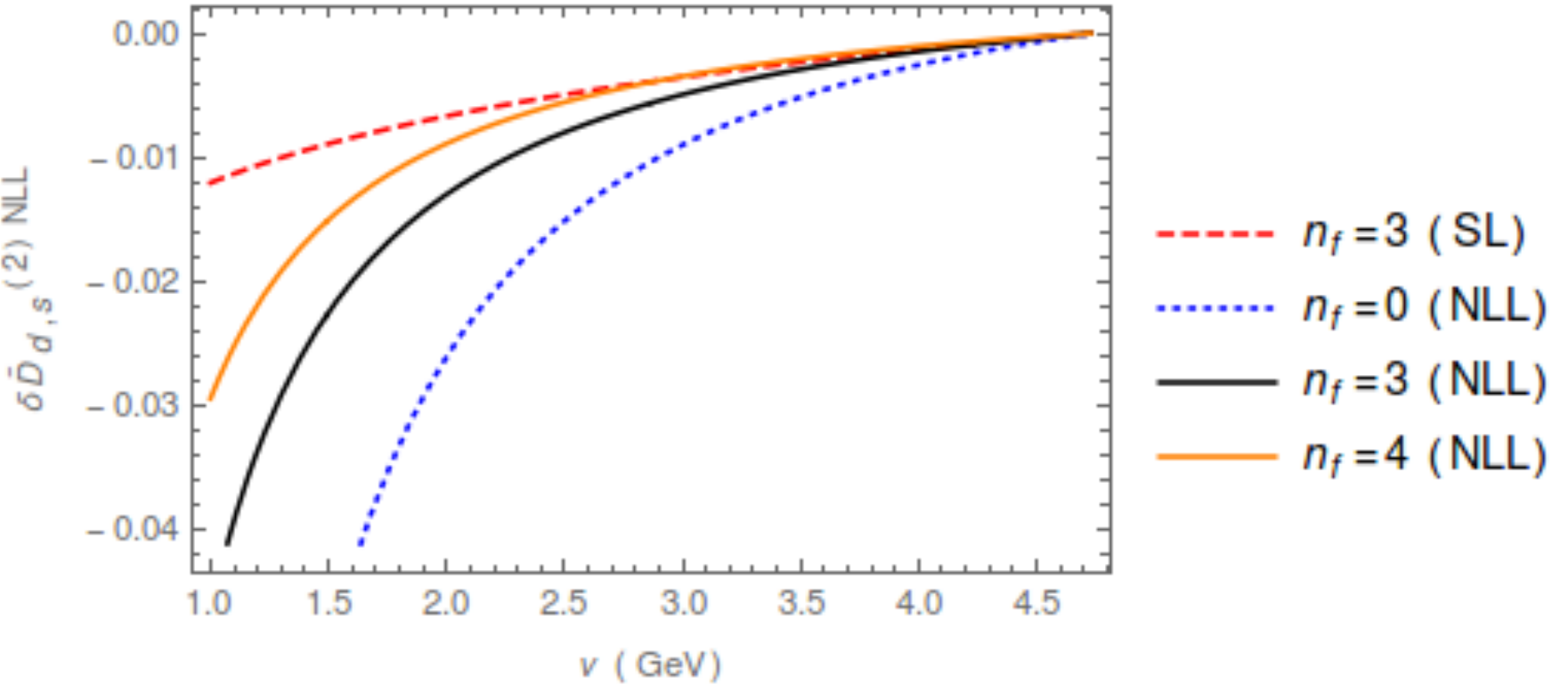}
	\includegraphics[width=0.75\textwidth]{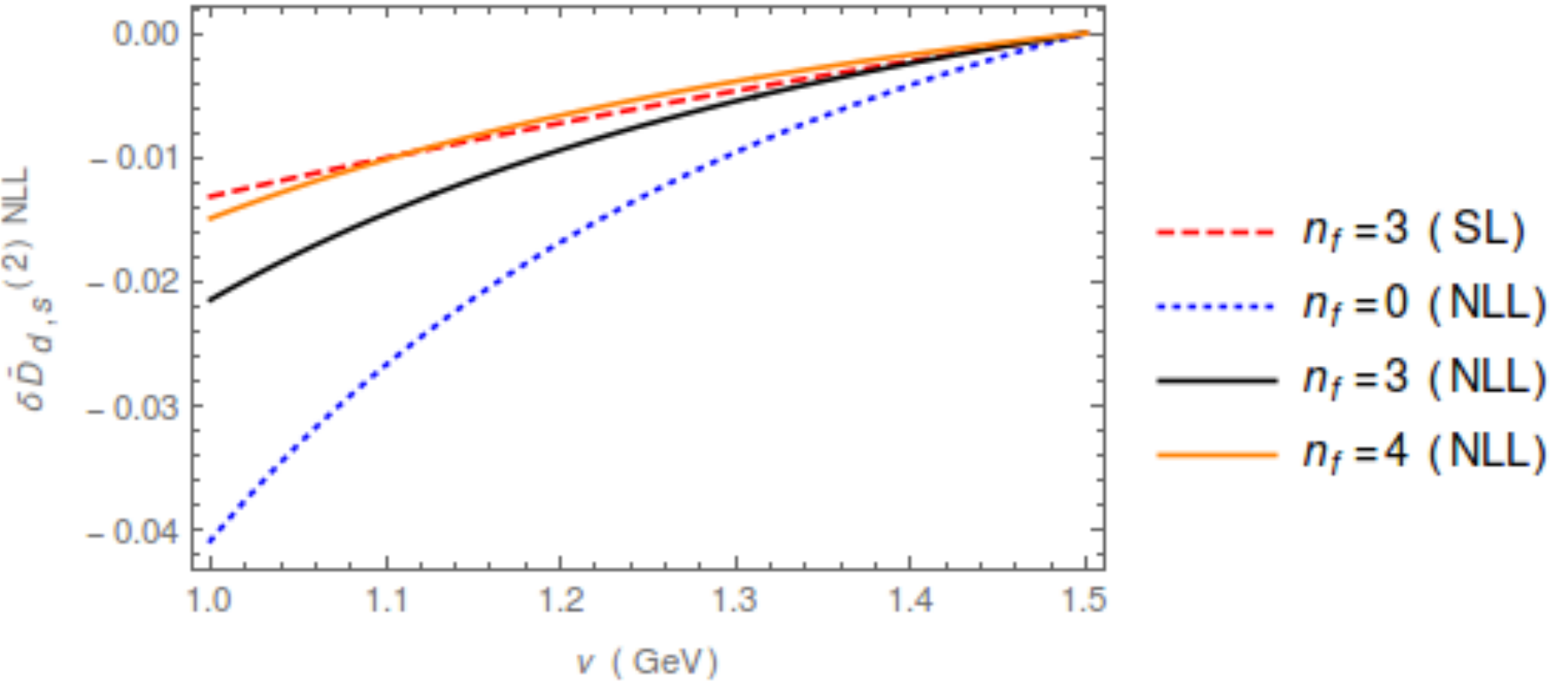}\\
	\includegraphics[width=0.75\textwidth]{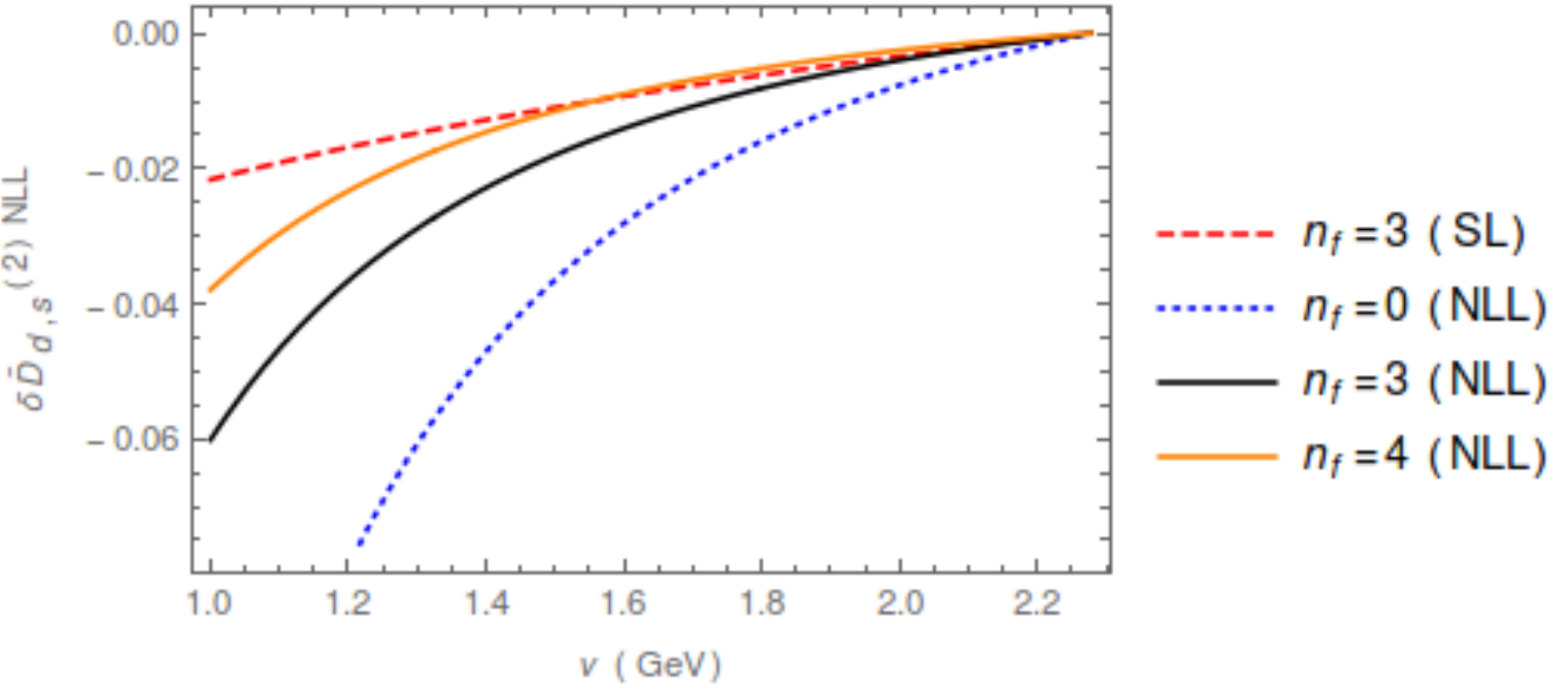}
\caption{Plot of the NLL soft running due to \eq{eq:DsNLL} to $\delta \tilde D_{d,s}^{(2)NLL}$ for different values of $n_f$ (0,3,4) and in the single log (SL) approximation (in this case only with $n_f=3$). 
{\bf Upper  panel:} Plot for bottomonium with $\nu_h=m_{b}$. {\bf Middle panel:} Plot for charmonium with 
$\nu_h=m_{c}$. {\bf Lower panel:} Plot for $B_c$ with $\nu_h=2m_{b}m_{c}/(m_{b}+m_{c})$. 
\label{Fig:soft}}   
\end{center}
\end{figure}
We do not aim in this paper to give a full fledged phenomenological analysis. Still, we compute numerically the running of $\delta \tilde D_{d,s}^{(2),NLL}$ to see its size. 
We show the result in Fig. \ref{Fig:soft}. The contribution is small.

\begin{figure}[!htb]
	\begin{center}      
	\includegraphics[width=0.75\textwidth]{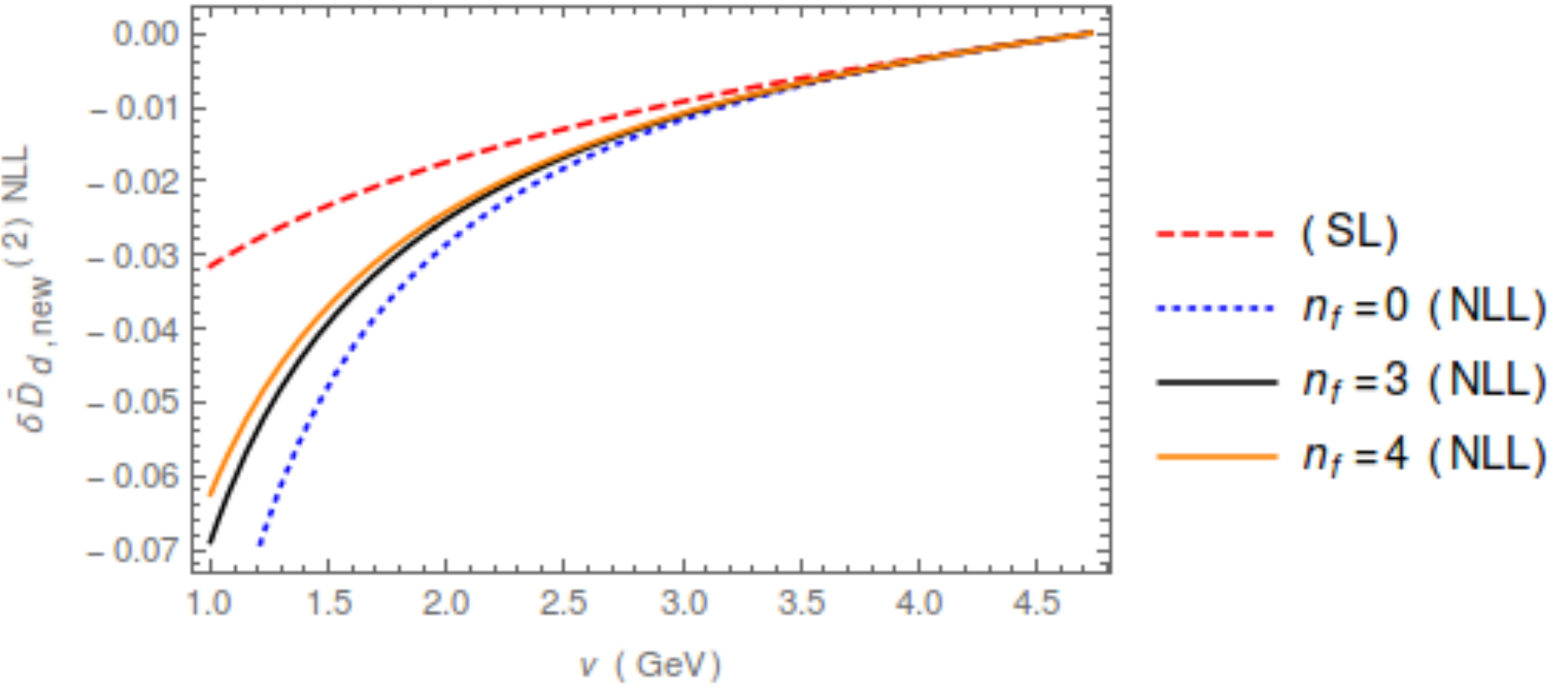}
	\includegraphics[width=0.75\textwidth]{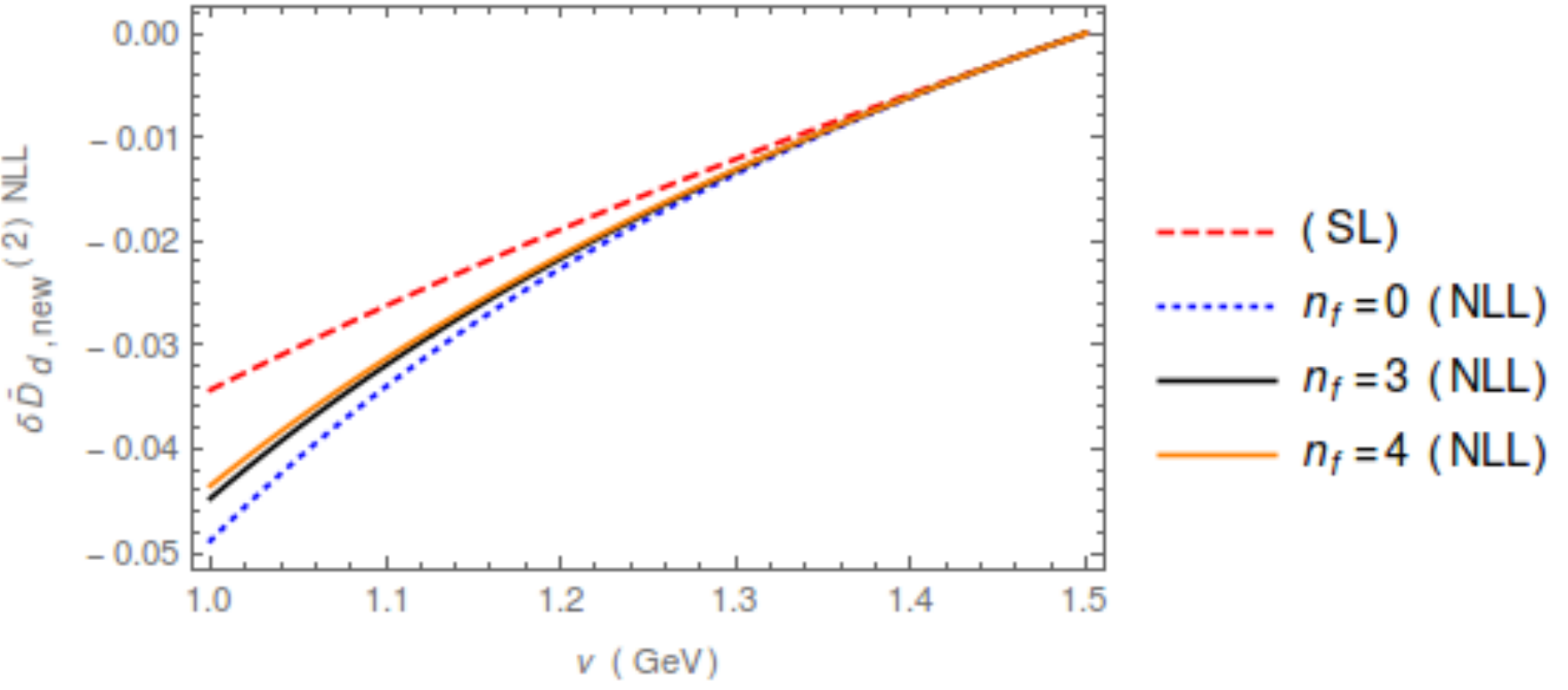}\\
	\includegraphics[width=0.75\textwidth]{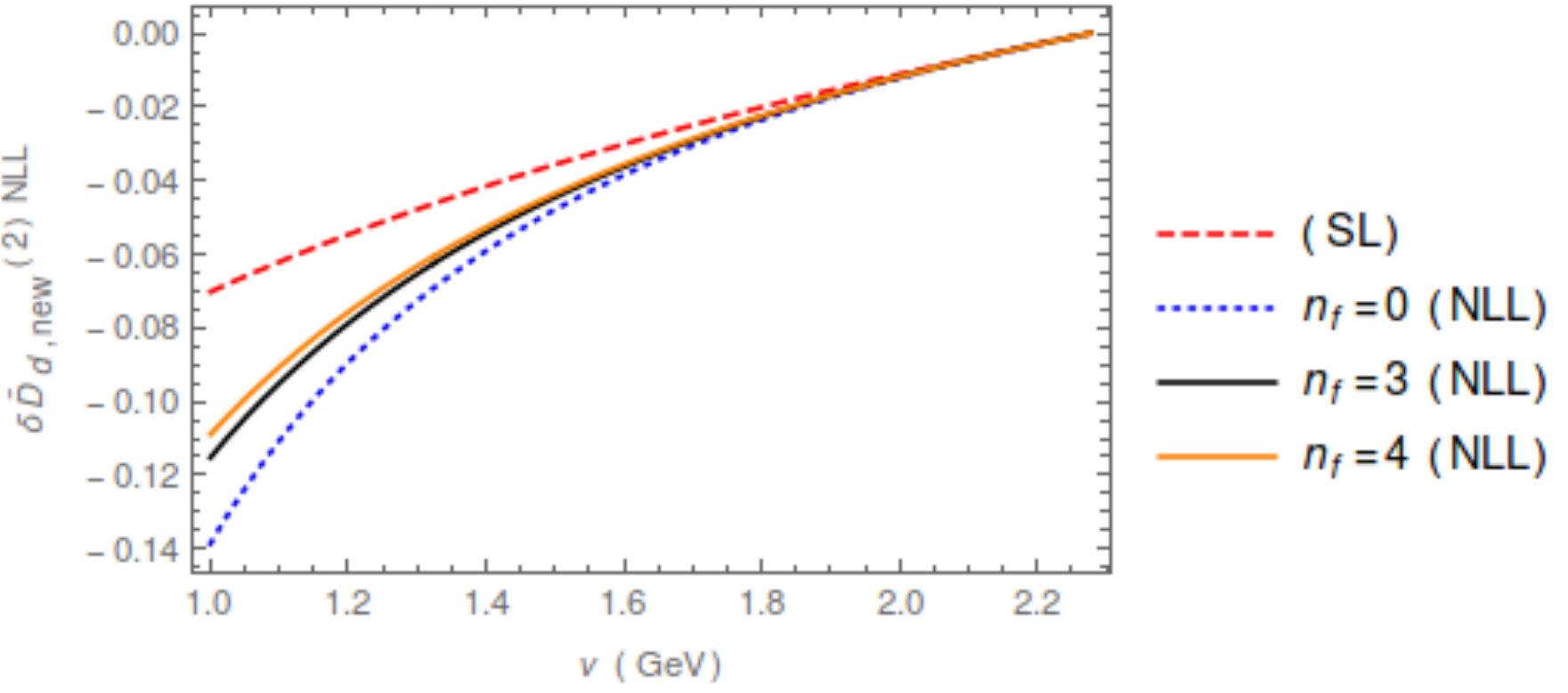}
\caption{Plot of the extra contribution to the NLL soft running, $\delta \tilde D_{d,s}^{(2)NLL}$, due to \eq{eq:DsNLLnew}, for different values of $n_f$ (0,3,4) and in the single log (SL) approximation (in this case only with $n_f=3$). 
{\bf Upper  panel:} Plot for bottomonium with $\nu_h=m_{b}$. {\bf Middle panel:} Plot for charmonium with 
$\nu_h=m_{c}$. {\bf Lower panel:} Plot for $B_c$ with $\nu_h=2m_{b}m_{c}/(m_{b}+m_{c})$. 
\label{Fig:softnew}}   
\end{center}
\end{figure}

To this contribution one should also add the contributions generated by the new $\al^3/m^2$ potentials that appear after using the full equations of motion. Of those we only computed the contributions proportional to $c_F^{(i)2}$ and $c_1^{hl}$ (the latter happened to be zero). This generates a new contribution to the soft RG equation:
\be
\nu_s {d\over d \nu_s}(d_{ss}+C_F\bar d_{vs})\Bigg|_{soft}=\cdots+
\frac{1}{16}C_F^2C_Ag^2\al^2\left[\frac{m_2}{m_1}c_F^{(1)2}+\frac{m_1}{m_2}c_F^{(2)2}
\right]
\,.
\ee 
Its solution reads
\begin{equation}
\label{eq:DsNLLnew}
 \delta \tilde D_{d,new}^{(2)NLL} = \frac{1}{\pi C_F}(d_{ss}+C_F\bar d_{vs})=
 - \frac{\pi C_A C_F (m_1^2 + m_2^2)  (1 -  z^{-2(C_A - \beta_0)}) \alpha^2(\nu_h)}{4 m_1 m_2 (C_A - \beta_0)}
\,.
\end{equation}
We then show the size of this new contribution in Fig. \ref{Fig:softnew}. 

Finally, let us note that the $c_k^2$ terms can also mix with $\al^2/m^3$ potentials through field redefinitions, see the discussion in the Appendix. Therefore, this contribution could be different for other matching schemes.

\subsection{\label{sec:ulso}Ultrasoft running}

To obtain the complete potential RG equation, we also need an extra potential divergence that is generated by ultrasoft divergences. This term was already computed in \cite{Penin:2004ay}, and applied to the spin-dependent case. Here, we give the full term, which contributes to both, the spin-dependent and spin-independent term. It is generated by the following diagram
\vspace{-82mm}
\begin{figure}[!htb]
\makebox[1.0cm]{\phantom b}
\includegraphics[width=0.85\textwidth]{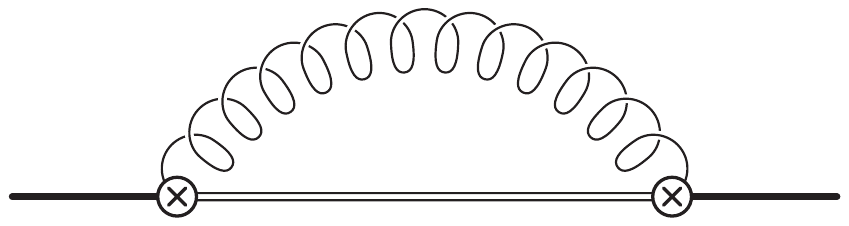}
\put(-250,215){$\frac{c_F^{(i)}}{m_i}$}
\put(-140,215){$\frac{c_F^{(j)}}{m_j}$}
\put(-237,219){$\underbrace{\hbox{~~~~~~~~~~~~~~~~~~~~~}}$}
\put(-220,195){$1/(E-V_o^{(0)}-{\bf p}^2/(2m_r))$}
\end{figure}
\vspace{-65mm}

which produces the following ultrasoft RG equation
\bea
\nn
\nu_{us} {{\rm d} V_{{\bf S}^2,1/r^3}\over {\rm d}\nu_{us}}
&=&
\frac{4C_F}{3}
\left[
{2{\bf S_1}\cdot {\bf S}_2 c_F^{(1)}(\nu_{us})c_F^{(2)}(\nu_{us})\over m_1m_2}-\frac{3}{4}\left(\frac{c_F^{(1)2}(\nu_{us})}{m_1^2}+\frac{c_F^{(2)2}(\nu_{us})}{m_2^2}\right)
\right]
\\
&&
\times
\left(
(V_o-V_s)^3+\left(\frac{1}{m_1}+\frac{1}{m_2}\right)\frac{(V_o-V_s)^2}{2r^2}
\right)
\left[{\al(\nu_{us}) \over 2\pi}\right]
\,,
\eea 
or alternatively (but equivalent at this order)
\bea
\nn
\nu_{us} {{\rm d} V_{{\bf S}^2,1/r^3}\over {\rm d}\nu_{us}}
&=&
\frac{4C_F}{3}
\left[
{2{\bf S_1}\cdot {\bf S}_2 c_F^{(1)}(\nu_{us})c_F^{(2)}(\nu_{us})\over m_1m_2}-\frac{3}{4}\left(\frac{c_F^{(1)2}(\nu_{us})}{m_1^2}+\frac{c_F^{(2)2}(\nu_{us})}{m_2^2}\right)
\right]
\\
&&
\times
V_o
(V_o-V_s)^2
\left[{\al(\nu_{us}) \over 2\pi}\right]
\,.
\eea 
Using that the LL running of $c_F$ is independent of the masses (we take the initial matching condition to be $\nu_h$ for both heavy quarks), its solution reads
\be
V_{{\bf S}^2,1/r^3}=\frac{4C_F}{3}
\left[
{2{\bf S}_1\cdot {\bf S}_2\over m_1m_2}-\frac{3}{4}\left(\frac{1}{m_1^2}+\frac{1}{m_2^2}\right)
\right]V_o
(V_o-V_s)^2
D_{1/r^3,{\bf S}^2}
\,,
\ee 
or
\bea
\nn
V_{{\bf S}^2,1/r^3}&=&\frac{4C_F}{3}
\left[
{2{\bf S}_1\cdot {\bf S}_2 \over m_1m_2}-\frac{3}{4}\left(\frac{1}{m_1^2}+\frac{1}{m_2^2}\right)
\right]
D_{1/r^3,{\bf S}^2}
\\
&&
\times
\left(
(V_o-V_s)^3+\left(\frac{1}{m_1}+\frac{1}{m_2}\right)\frac{(V_o-V_s)^2}{2r^2}
\right)
\,,
\eea 
where (we use the same notation as in \cite{Penin:2004ay})
\be
\label{Dr3}
D_{1/r^3,{\bf S}^2}={1 \over 2C_A}
\left[
\left(\frac{\al(\nu_h)}{\al(\nu_{us})}\right)^{2C_A/\beta_0}-
\left(\frac{\al(\nu_h)}{\al(1/r)}\right)^{2C_A/\beta_0}
\right]
\,.
\ee
$V_{{\bf S}^2,1/r^3}$ is singular and will contribute to the potential running of $\tilde D_d^{(2)}$. 

\subsection{Potential running}
\label{Sec:potNLL}
\begin{figure}[!ht]
\centering
\includegraphics[scale=0.5]{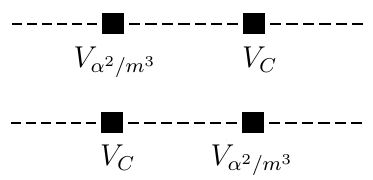}\\
\includegraphics[scale=0.5]{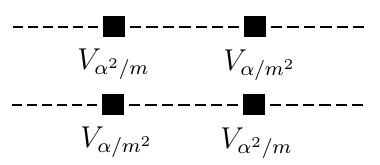}
\caption{Divergent diagrams with one potential loop that contribute to the running of $\tilde D_d^{(2)}$ at ${\cal O}(\al^3)$.}
\label{pl1}
\end{figure}
\begin{figure}[!ht]
\centering
\includegraphics[scale=0.5]{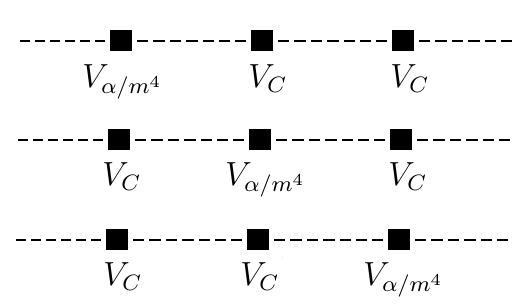}\\
\includegraphics[scale=0.5]{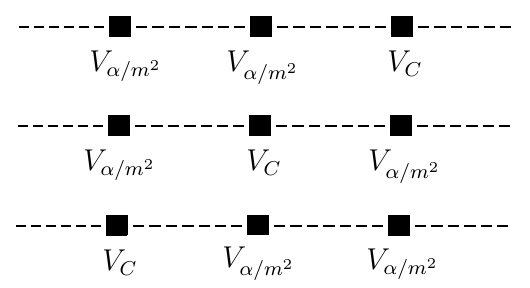}\\
\includegraphics[scale=0.5]{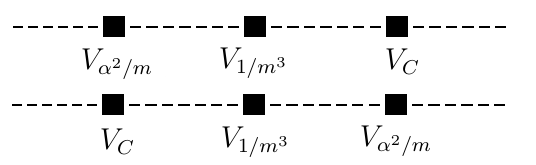}
\caption{Divergent diagrams with two potential loops that contribute to the running of $\tilde D_d^{(2)}$ at ${\cal O}(\al^3)$.}
\label{pl2}
\end{figure}
\begin{figure}[!ht]
\centering
\includegraphics[scale=0.5]{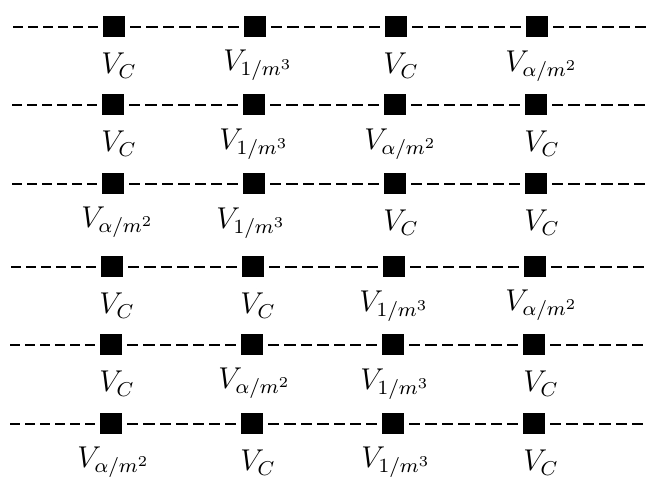}
\caption{Divergent diagrams with three potential loops that contribute to the running of $\tilde D_d^{(2)}$ at ${\cal O}(\al^3)$.}
\label{pl3}
\end{figure}
\begin{figure}[!ht]
\centering
\includegraphics[scale=0.5]{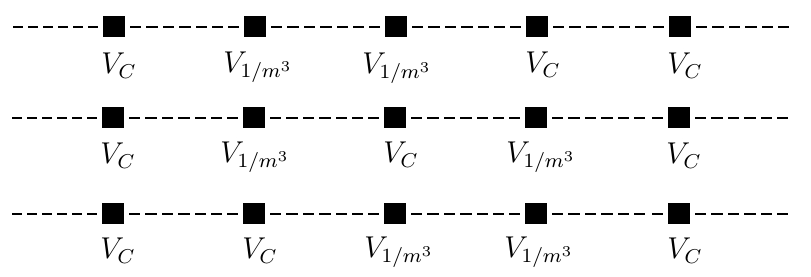}
\caption{Divergent diagrams with four potential loops that contribute to the running of $\tilde D_d^{(2)}$ at ${\cal O}(\al^3)$.}
\label{pl4}
\end{figure}

We now have all the necessary preliminary ingredients to obtain the complete potential RG equation. The next step is to compute all potential loops that produce ultraviolet divergences that get absorbed in $\tilde D_d^{(2)}$ and are at most of ${\cal O}(\al^3)$. Since the delta-like
potential is of $\mathcal{O}(1/m^2)$ we must construct potential
loop diagrams of $\mathcal{O}(\alpha^n/m^2)$ with $n\leq 3$ describing
the interaction between the two heavy quarks in the bound state
through several potentials. The first
non-vanishing contribution to the
potential running is indeed of $\mathcal{O}(\alpha^3/m^2)$. To construct such potential loop diagrams, we must consider the
power of $\alpha$ and $m$ of each potential and take into account that
each
propagator adds an extra power of the mass $m$ in the numerator. We summarize all kind of diagrams that contribute to the NLL potential running of $\tilde D_d^{(2)}$, in Figs. \ref{pl1}, \ref{pl2}, \ref{pl3} and \ref{pl4}. The ultraviolet divergences
arising in such diagrams must be absorbed in the $1/m^2$ potentials.
However,
after the computation, we observe that all divergences are only
absorbed by the delta-like potential. It is important to mention that
the iteration of
two or more spin-dependent potentials can give a contribution to
$\tilde D_d^{(2)}$, associated to a spin-independent potential. The relevant
diagrams are
shown in Figs. [\ref{pl1}-\ref{pl4}], where $ V_C$ is the tree
level, $\mathcal{O}(\alpha)$, Coulomb potential, $
V_{\alpha^r/m^s}$ is the
$\mathcal{O}(\alpha^r/m^s)$ potential and $ V_{1/m^3}$ corresponds
to the first relativistic correction to the kinetic energy, and it is
proportional
to $c_4$. 

It is interesting to discuss in more detail which, of the novel $\al^2/m^3$ potentials computed in Secs. \ref{Sec:al2m3} and \ref{Sec:EoM} (we remind that here we use the potentials after using the (free) equations of motion, i.e. the expressions in Sec. \ref{Sec:EoM} for the energy dependent potentials), contribute to the running of $\tilde D_d^{(2)}$. The potentials in  Eqs. (\ref{V1loopq2}-\ref{Vc2}) do not contribute to the running of $\tilde D_d^{(2)}$. Equation (\ref{V1loopq2}) does not because it is proportional to a total 
derivative, whereas Eq. (\ref{Vc2}) does not because of the following argument: the only possible potential loop that can be constructed with an 
$\mathcal{O}(\alpha^2/m^3)$ potential is the iteration of it with a Coulomb potential. As a consequence, the $\alpha^2/m^3$ potential is always applied to 
an external momentum. When the high loop momentum limit is taken in the integral in order to find the ultraviolet pole, all these external momenta vanish and 
all the terms become proportional to $|{\bf k}|^{1+2\epsilon}$. After doing so and summing all the terms the overall coefficient is zero, explaining the fact that they do not contribute. This argument also applies to $\tilde V^{(e,1)}$ and $\tilde V^{(f,i)}$ (with $i=1$ to 6). On the other hand $\tilde V^{(e,2)}$ and $\tilde V^{(f,7/8/9)}$  do contribute to the running. Note that $\tilde V^{(f,8)}$ and $\tilde V^{(f,9)}$ were originally dependent on the energy. 

Diagrams with $V_{1/m^3}$ in the extremes of a
potential loop, i.e. acting over a external momentum have not been drawn
because they
do not produce any ultraviolet divergence. Similarly, diagrams with $ V_{1/m^5}$ do not produce ultraviolet divergences. One can then easily convince himself that there are no diagrams with five potential loops or more that can contribute to the ${\cal O}(\al^3)$ anomalous dimension of $\tilde D_d$. Therefore, the above discussion exhausts all possible contributions to the ${\cal O}(\al^3)$ anomalous dimension of $\tilde D_d$, and the potential RG equation finally reads
\bea
\label{Ddpotprel}
 &&
 \nu\frac{d \tilde D_d^{(2)}}{d\nu}= - 2 C_F^2 \alpha_{V}^2 m_r^3\left(\frac{c_4^{(1)}}{m_1^3} +\frac{c_4^{(2)}}{m_2^3}\right) D_d^{(2)}
\\
\nn
&&
+ C_F^2 \alpha_{V}\frac{m_r^2}{m_1 m_2}\left( D_{d}^{(2)\,2} -8 D_d^{(2)} D_1^{(2)} +12  D_1^{(2)\,2}
 -\frac{5}{6}  D_{S_{12}}^{(2)\,2} +\frac{4}{3}  D_{S^2}^{(2)\,2}\right)
 \\
 \nn
 &&
 + 2 C_F^2 \alpha_{V}^2 m_r^3\left(\frac{c_4^{(1)}}{m_1^3} +\frac{c_4^{(2)}}{m_2^3}\right) \left(4 D_1^{(2)} \right)
\\
&&
 + C_F C_A\left[ 2  D_1^{(2)}D^{(1)} 
 -  D_d^{(2)}D^{(1)} 
 + D^{(1)}\alpha_{V} m_r m_1 m_2 \left(\frac{c_4^{(1)}}{m_1^3}+\frac{c_4^{(2)}}{m_2^3}\right)\right]
 \nn
 \\
 &&
 + C_F^2\alpha_{V}^3 m_r^4 m_1 m_2\left(\frac{c_4^{(1)}}{m_1^3}+\frac{c_4^{(2)}}{m_2^3}\right)^2
 \nn
 \\
 &&
 +\left(\frac{m_1}{m_2}+\frac{m_2}{m_1}\right)
 {C_A^2(C_A-2C_F)\alpha_s^3 \over 2}D^{(2)}_{S^2,1/r^3}
 \nn
 \\
 \nn
 &&-2 C_F^2 \alpha_{V}^2\alpha m_r^2
 \left[ 16 m_1 m_2 \left(\frac{c_{X1}^{(1)}}{m_1^4}+\frac{c_{X1}^{(2)}}{m_2^4}\right)
 + 2\left(\frac{c_4^{(1)}}{m_1^2}+\frac{c_4^{(2)}}{m_2^2}\right)
 +\frac{1}{m_1 m_2}
 + \left(\frac{c_M^{(1)}}{m_1^2}+\frac{c_M^{(2)}}{m_2^2}\right)\right]
 \eea
$$ +C_F\left(2C_F-\frac{C_A}{2}\right) \alpha^2 \alpha_{V} m_r m_1 m_2\left[\frac{1}{2}\left(\frac{c_{A1}^{(1)}}{m_1^3} + \frac{c_{A1}^{(2)}}{m_2^3}\right)
+ \frac{1}{4}\left(\frac{c_{A2}^{(1)}}{m_1^3} + \frac{c_{A2}^{(2)}}{m_2^3}\right)
+\frac{1}{m_1 m_2}\left(\frac{c^{(1)\,2}_F}{m_1}+\frac{c^{(2)\,2}_F}{m_2}\right)\right]$$
\begin{eqnarray}\nonumber
&& 
- \frac{1}{2} C_F C_A \alpha^2 \alpha_{V} m_r m_1 m_2\left[2\left(\frac{c_4^{(1)}}{m_1^3} + \frac{c_4^{(2)}}{m_2^3}\right) + \left(\frac{1}{m_1^3} + \frac{1}{m_2^3}\right)
-\left(\frac{c_M^{(1)}}{m_1^3}+\frac{c_M^{(2)}}{m_2^3}\right)
+\left(\frac{c_F^{(1)\,2}}{m_1^3}+\frac{c_F^{(2)\,2}}{m_2^3}\right)
\right.\\ \nonumber&& \left.
- \left(\frac{c_F^{(1)}c_S^{(1)}}{m_1^3}+\frac{c_F^{(2)}c_S^{(2)}}{m_2^3}\right)\right]
+  \frac{T_F}{N_c} C_F \alpha^2 \alpha_{V}m_r m_1 m_2\left[\left(\frac{c_{A3}^{(1)}}{m_1^3}+\frac{c_{A3}^{(2)}}{m_2^3}\right)
+ \frac{1}{2}\left(\frac{c_{A4}^{(1)}}{m_1^3}+\frac{c_{A4}^{(2)}}{m_2^3}\right)\right]
\,.
\end{eqnarray}
The first five lines are generated by potential loops with $\al^2/m$, $\al/m^2$ and the ${\bf p}^4/m^3$ correction to the kinetic energy (besides the iteration of the Coulomb potential, accounted for by $\al_V$). The 6th line is the term generated by the potential computed in Sec. \ref{sec:ulso}. The last four lines are generated by potential loops with the $\al^2/m^3$ and $\al/m^4$ potentials (besides the iteration of the Coulomb potential, accounted for by $\al_V$). Note that, for simplicity, we have already used $c_k^{(i)}=1$ \cite{Luke:1992cs} in the terms that do not have NRQCD Wilson coefficients in the above expression. A part of this equation was already computed in \cite{Manohar:2000rz}. Also, several of these terms (for QED) can be checked with the computations in \cite{KMY}. 

It is interesting to see that there is a matching scheme dependence of the individual  $\al^2/m^3$ and $\al/m^4$ potentials that cancels out in the sum. In the above expression the coefficients $c_{A_2}$, $c_D$, $c_M$, $c_{X1}$ appear (note that the last two coefficients are dependent on $c_D$ due to reparameterization invariance). They are gauge dependent quantities. Such gauge dependence should vanish in the final result. Indeed it does. This is actually a strong check of the computation. In \eq{Ddpotprel} we can approximate $\al_V=\al$ (everything is needed with LL accuracy). Then we can show that everything can be written in terms of $\bar c_{A_2}$, which is gauge independent (it is an observable in the low energy limit of the Compton scattering, see the discussion in \cite{Moreno:2017sgd}), and the explicit dependence in $c_D$, $c_M$, $c_{X1}$, $c_{A2}$ disappears. The resulting expression reads
\bea
\label{eq:RGpot}
 &&
 \nu\frac{d \tilde D_d^{(2)}}{d\nu}= - 2c_4 C_F^2 \alpha_{V}^2 m_r^3\left(\frac{1}{m_1^3} +\frac{1}{m_2^3}\right) D_d^{(2)}
\\
\nn
&&
+ C_F^2 \alpha_{V_s}\frac{m_r^2}{m_1 m_2}\left( D_{d}^{(2)\,2} -8 D_d^{(2)} D_1^{(2)} +12  D_1^{(2)\,2}
 -\frac{5}{6}  D_{S_{12}}^{(2)\,2} +\frac{4}{3}  D_{S^2}^{(2)\,2}\right)
 \\
\nn
 &&
 + 2c_4 C_F^2 \alpha_{V}^2 m_r^3\left(\frac{1}{m_1^3} +\frac{1}{m_2^3}\right) \left(4 D_1^{(2)} \right)
\\
&&
 + C_F C_A\left[ 2  D_1^{(2)}D^{(1)} 
 -  D_d^{(2)}D^{(1)} 
 + c_4 D^{(1)}\alpha_{V} m_r m_1 m_2 \left(\frac{1}{m_1^3}+\frac{1}{m_2^3}\right)\right]
 \nn
 \\
 &&
 + c_4^2 C_F^2\alpha_{V}^3 m_r^4 m_1 m_2\left(\frac{1}{m_1^3}+\frac{1}{m_2^3}\right)^2
 \nn
 \\
 &&
 +\left(\frac{m_1}{m_2}+\frac{m_2}{m_1}\right)
 {C_A^2(C_A-2C_F)\alpha^3 \over 2}D^{(2)}_{S^2,1/r^3}
 \nn
 \\
\nn
 &&-2 C_F^2 \alpha^3 m_r^2
 \left[ \frac{5}{8} m_1 m_2 \left(\frac{1}{m_1^4}+\frac{1}{m_2^4}\right)
 + 2\left(\frac{c_4^{(1)}}{m_1^2}+\frac{c_4^{(2)}}{m_2^2}\right)
 +\frac{1}{m_1 m_2}
 \right]
 \eea
$$ +C_F\left(2C_F-\frac{C_A}{2}\right) \alpha^3  m_r m_1 m_2\left[\frac{1}{2}\left(\frac{c_{A1}^{(1)}}{m_1^3} + \frac{c_{A1}^{(2)}}{m_2^3}\right)
+ \frac{1}{4}\left(\frac{\bar c_{A2}^{(1)}}{m_1^3} + \frac{\bar c_{A2}^{(2)}}{m_2^3}\right)
+\frac{1}{m_1 m_2}\left(\frac{c^{(1)\,2}_F}{m_1}+\frac{c^{(2)\,2}_F}{m_2}\right)\right]$$
\begin{eqnarray}\nonumber
&& 
- \frac{1}{2} C_F C_A \alpha^3 m_r m_1 m_2\left[2\left(\frac{c_4^{(1)}}{m_1^3} + \frac{c_4^{(2)}}{m_2^3}\right) + \left(\frac{1}{m_1^3} + \frac{1}{m_2^3}\right)
+\left(\frac{c_F^{(1)\,2}}{m_1^3}+\frac{c_F^{(2)\,2}}{m_2^3}\right)
\right.\\ \nonumber&& \left.
- \left(\frac{c_F^{(1)}c_S^{(1)}}{m_1^3}+\frac{c_F^{(2)}c_S^{(2)}}{m_2^3}\right)\right]
-  \left(C_F-\frac{C_A}{2}\right) C_F \alpha^3 m_r m_1 m_2\left[\left(\frac{c_{A3}^{(1)}}{m_1^3}+\frac{c_{A3}^{(2)}}{m_2^3}\right)
+ \frac{1}{2}\left(\frac{c_{A4}^{(1)}}{m_1^3}+\frac{c_{A4}^{(2)}}{m_2^3}\right)\right]
\,.
\end{eqnarray}

From this result one may think that $c_{A3}$ and $c_{A4}$ contribute to the Abelian case. Nevertheless, the LO matching condition is zero for these Wilson coefficients, and all the running vanishes in the Abelian limit. Therefore, there is no contradiction with the pure QED case.

\begin{figure}[!htb]
	\begin{center}      
	\includegraphics[width=0.75\textwidth]{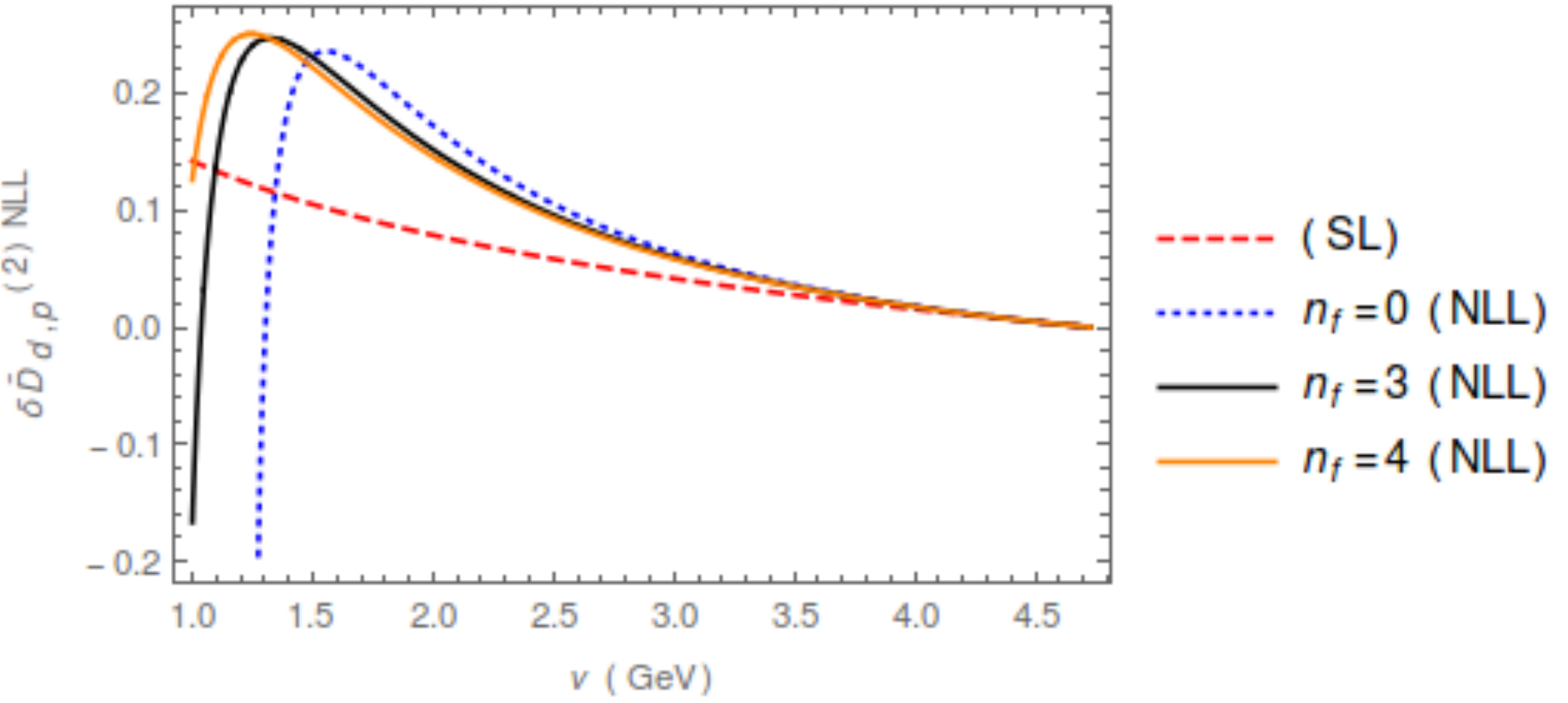}
	\includegraphics[width=0.75\textwidth]{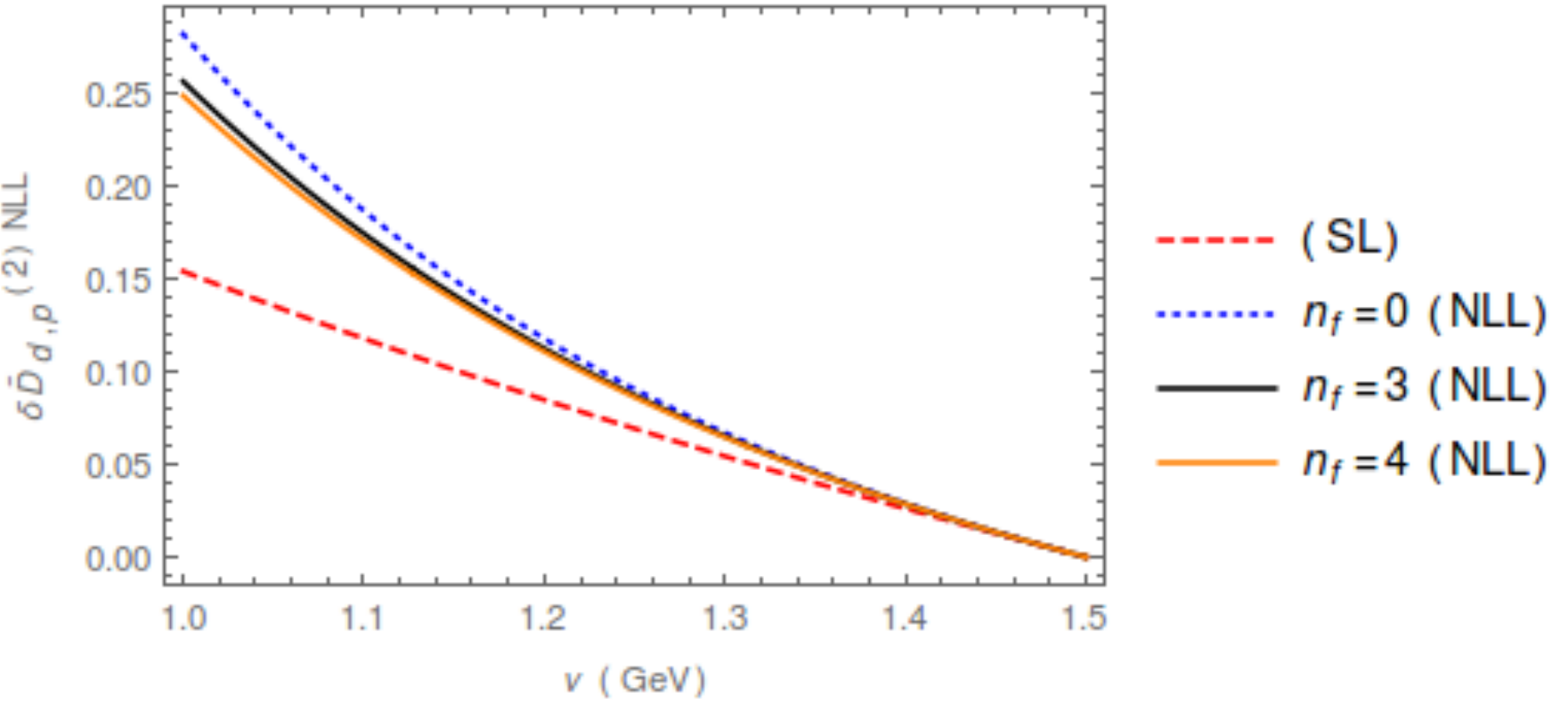}\\
	\includegraphics[width=0.75\textwidth]{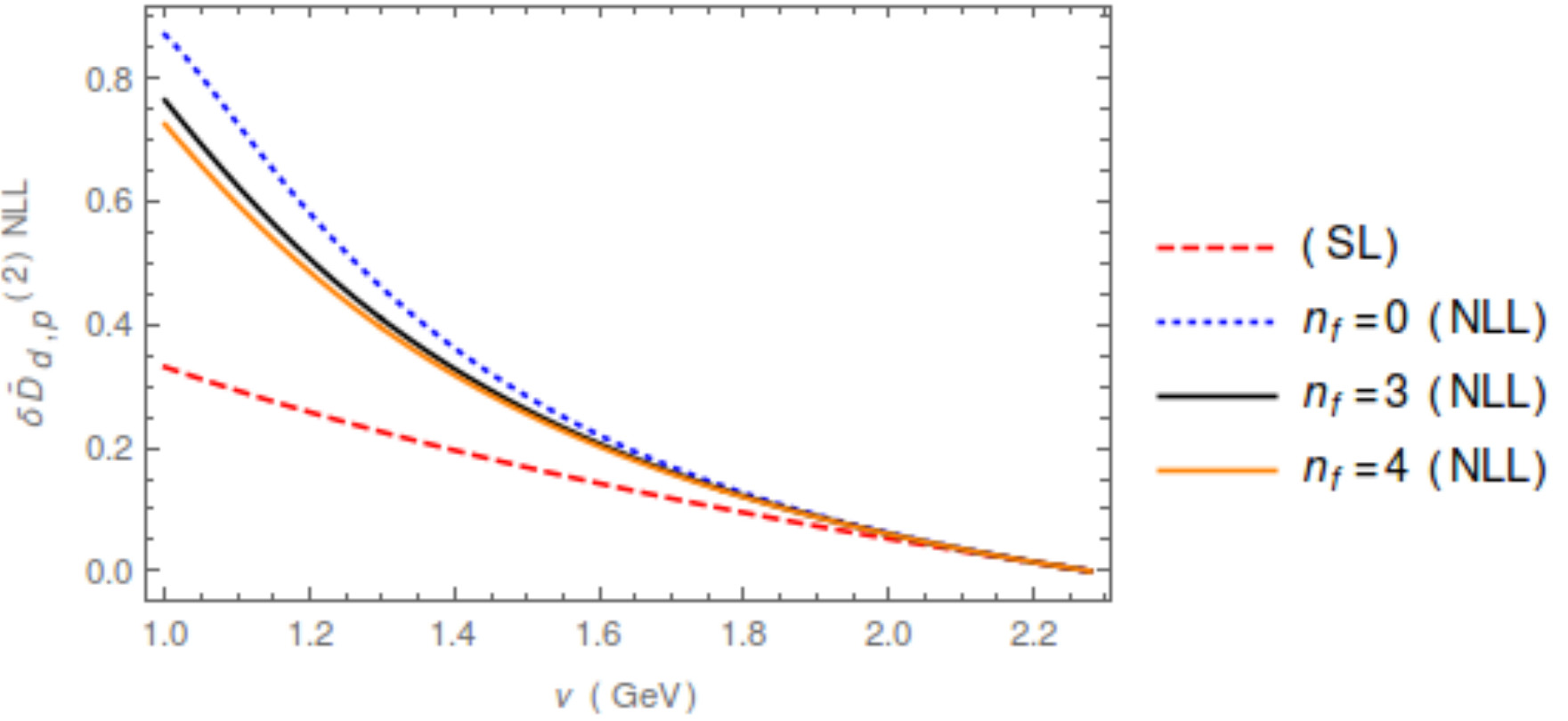}
\caption{Plot of $\delta \tilde D_{d,p}^{(2)NLL}$ for different values of $n_f$ (0,3,4) and in the single log (SL) approximation (in this case only with $n_f=3$). 
{\bf Upper  panel:} Plot for bottomonium with $\nu_h=m_{b}$. {\bf Middle panel:} Plot for charmonium with 
$\nu_h=m_{c}$. {\bf Lower panel:} Plot for $B_c$ with $\nu_h=2m_{b}m_{c}/(m_{b}+m_{c})$. 
\label{Fig:pot}}   
\end{center}
\end{figure}

In order to solve \eq{eq:RGpot}, we need to introduce the $D's$, the Wilson coefficients of the potentials. The necessary expressions can be found in Sec. \ref{Sec:pNRQCD}. Note that in those expressions we have already correlated the ultrasoft factorization scale $\nu_{us}$ with $\nu$ and $\nu_h$ 
using $\nu_{us}=\nu^2/\nu_h$. We also do so in \eq{Dr3} (where we also set $1/r=\nu$, consistent with the precision of our computation). This correlation of scales was first introduced and motivated in \cite{Luke:1999kz}.

For $n_f=3$ or 4 it is not possible to get an analytic result for the solution of the RG equation, more specifically for the coefficients multiplying the different $z$ functions (note that this comes back to the fact that the polarizability Wilson coefficients $c_{A1}$, $c_{A2}$, ..., cannot be computed analytically). On top of that the resulting expressions are too long. Therefore, we only explicitly show the analytic result with $n_f=0$. 
It reads
\begin{eqnarray}
&&\delta\tilde D_{d,p}^{(2) NLL} = 
 \pi \alpha (\nu_h)^2\bigg[
-\frac{1}{48656036000 C_A^3 m_1^5 m_2^5}
    \bigg( 4561503375 C_A^4 m_1^4 m_2^4 (m_1^2 + m_2^2) 
\nonumber
\\
&&
- 1754688000 C_F^4 m_1^2 m_2^2 (18 m_1^4 + 142 m_1^3 m_2
+ 101 m_1^2 m_2^2 + 142 m_1 m_2^3 + 18 m_2^4) m_r^2
\nonumber
\\
&&
- 14250 C_A^3 C_F m_1^3 m_2^3 \Big(480 m_1^3 (1924 m_2 - 5885 m_r) 
+ 481 m_1^2 m_2 (720 m_2 - 1331 m_r)
\nonumber
\\
&&
+        481 m_1 m_2^2 (1920 m_2 - 1331 m_r) - 2824800 m_2^3 m_r\Big) 
-        98800 C_A C_F^3 m_1 m_2 \Big(-293040 m_2^5 m_r^3  
\nonumber
\\
&&
      - 222 m_1 m_2^4 m_r^2 (691 m_2 + 5720 m_r) + 
         m_1^2 m_2^3 m_r (-144320 m_2^2 + 2370183 m_2 m_r - 293040 m_r^2)
\nonumber
\\
&&
         + 2 m_1^5 (497280 m_2^3 - 72160 m_2^2 m_r - 76701 m_2 m_r^2 -146520m_r^3)
         + 111m_1^4 m_2 (15680 m_2^3 
\nonumber
\\
&&          
         + 21353 m_2 m_r^2        
         - 11440 m_r^3)         
         +  444 m_1^3 m_2^2 (2240 m_2^3 - 1831 m_2 m_r^2 - 660 m_r^3)\Big)
\nonumber
\\
&& 
         - 4 C_A^2 C_F^2 \Big(-2741700 m_1^2 m_2^5 (5109 m_2 
         - 1100 m_r) m_r^2 
         - 5767851375 m_1 m_2^6 m_r^3 
\nonumber
\\
&&           
         + 3317457000 m_2^6 m_r^4 - 25 m_1^3 m_2^3 m_r ( 298447930 m_2^3       
         + 693732351 m_2^2 m_r + 
            230714055 m_2 m_r^2 
\nonumber
\\
&&            
            - 265396560 m_r^3) + 
         27417 m_1^4 m_2^3 (380000 m_2^3 + 332750 m_2^2 m_r 
         - 922808 m_2 m_r^2 - 210375 m_r^3) 
\nonumber
\\
&&
         + 685425 m_1^5 m_2^2 (9440 m_2^3 + 13310 m_2^2 m_r - 25303 m_2 m_r^2 + 
            4400 m_r^3) 
\nonumber
\\
&&
            + 25 m_1^6 (416738400 m_2^4 - 298447930 m_2^3 m_r - 
            560293812 m_2^2 m_r^2 - 230714055 m_2 m_r^3
\nonumber
\\
&&             
            + 132698280 m_r^4)\Big) \bigg)   
            + \frac{1}{1375 C_A^2 m_1^3 m_2^3}\bigg(6 C_F^2 \Big(100 C_F (2 m_1^4 - 3 m_1^3 m_2 
            + 4 m_1^2 m_2^2 - 3 m_1 m_2^3 
\nonumber
\\
&& 
            + 2 m_2^4) 
            + C_A (200 m_1^4 
	    + 75 m_1^3 m_2 + 334 m_1^2 m_2^2
            + 75 m_1 m_2^3
            + 200 m_2^4)\Big) m_r^2 z^{5 C_A/3}\bigg)
\nonumber
\\
&&
             - \frac{3 C_F^2 (25 m_1^4 - 32 m_1^2 m_2^2 + 25 m_2^4) m_r^2 z^{10 C_A/3}}{500 C_A m_1^3 m_2^3}
\nonumber
\\
&&
             + \frac{24C_F(43C_A^2 - 66C_A C_F - 40C_F^2)(m_1^3 + m_2^3)m_r z^{11C_A/3}}{605C_A^2 m_1^2 m_2^2}
\nonumber
\\
&&
             - \frac{1}{1331 C_A^3 m_1^4 m_2^4} 6 C_F \bigg( 32 C_F^3 m_1 m_2 (2 m_1^4     
     + 36 m_1^3 m_2 + 19 m_1^2 m_2^2 + 
          36 m_1 m_2^3 + 2 m_2^4) m_r^2 
\nonumber
\\
&&          
          + 11 C_A^3 m_1^2 m_2^2 \Big(3 m_1^2 m_2^2 + 8 m_1 m_2^3 +  8 m_1^3 ( m_2 - m_r) 
        - 8 m_2^3 m_r \Big)
        + 8 C_A C_F^2 m_1 m_2 \Big(132 m_1 m_2^3 m_r^2 
\nonumber
\\
&&
        - m_2^3 m_r^2 (8 m_2 + 77 m_r)
          +  8m_1^4 (6 m_2^2 -  m_r^2) 
          + 8 m_1^2 (6 m_2^4
           - 7 m_2^2 m_r^2) +  m_1^3 (96 m_2^3 + 132 m_2 m_r^2
\nonumber
\\
&&           
           - 77 m_r^3)\Big) 
          + 2 C_A^2 C_F \Big(44 m_2^5 m_r^3         
          -  m_1 m_2^4 m_r^2 (64 m_2 + 55 m_r) 
          + 4 m_1^2 m_2^3 m_r (-22 m_2^2 + 32 m_2 m_r 
\nonumber
\\
&&          
          + 11 m_r^2) 
          +  m_1^4 m_2 (38 m_2^3 + 128 m_2 m_r^2 - 55 m_r^3) + 
           m_1^3 m_2^2 (96 m_2^3            
           - 71 m_2 m_r^2 + 44 m_r^3)
\nonumber
\\
&&           
           + m_1^5 (96 m_2^3 
          - 88 m_2^2 m_r - 64 m_2 m_r^2 + 44 m_r^3)\Big)\bigg) z^{11 C_A/3}
\nonumber
\\
&&         
         - \frac{99 (3C_A-8C_F)C_F (m_1^3 + m_2^3) m_r z^{13 C_A/3}}{442 C_A m_1^2 m_2^2}
          + \frac{(C_A -2C_F)C_F(m_1^3 + m_2^3)m_r z^{16C_A/3}}{10C_A m_1^2 m_2^2}
\nonumber
\\
&&
          + \frac{(C_A - 4C_F)C_F(11m_1^3 + 3m_1^2 m_2 + 3m_1 m_2^2 + 11 m_2^3)m_r z^{16C_A/3}}{16C_A m_1^2 m_2^2}
\nonumber
\end{eqnarray}

\begin{eqnarray}
&&     
 + \frac{1}{352 C_A^2 m_1^4 m_2^4} 3\bigg(11 C_A^3 m_1^3 m_2^3 (m_1^2 + m_2^2) 
          - 22 C_A^2 C_F m_1^2 m_2^2 ( m_1^3 + m_2^3 ) m_r
        - 16 C_F^3 m_1 m_2 (2 m_1^4   
\nonumber
\\
&&
 - 3 m_1^3 m_2 + 4 m_1^2 m_2^2 - 3 m_1 m_2^3 + 
          2 m_2^4) m_r^2 
          - 4 C_A C_F^2 (m_1^2 + m_2^2)m_r^2 \Big( -19 m_1^2 m_2^2 + 8m_1 m_2^3 
\nonumber
\\
&&
          + m_1^3(8m_2-11m_r)-11m_2^3 m_r \Big)\bigg) z^{16 C_A/3}
          + \frac{6C_F^2 (m_1^3 + m_2^3)m_r z^{19C_A/3}}{19C_A m_1^2 m_2^2}
\nonumber
\\
&&
          + \frac{3C_F ( 51C_A^2 - 54C_A C_F - 224 C_F^2)(m_1^3 + m_2^3)m_r z^{22C_A/3}}{121 C_A^2 m_1^2 m_2^2}
\nonumber
\\
&&
+ \frac{1}{5324 C_A^3 m_1^5 m_2^5} 
  3 C_F \bigg(-64 C_F^3 m_1^2 m_2^2 (10 m_1^4 - 2 m_1^3 m_2 + 25 m_1^2 m_2^2  - 2 m_1 m_2^3 + 10 m_2^4) m_r^2
\nonumber
\\
&&
          + 2 C_A^3 m_1^3 m_2^3 \Big(42 m_1^2 m_2^2 + 112 m_1 m_2^3 + 209 m_2^3 m_r
          + m_1^3 (112 m_2 - 209 m_r)\Big) 
\nonumber
\\
&&
          - 32 C_A C_F^2 m_1 m_2 \Big(-33 m_2^5 m_r^3 + 
          m_1 m_2^4 m_r^2 (32 m_2 + 11 m_r) - 
          m_1^2 m_2^3 m_r^2 (47 m_2 + 33 m_r)
\nonumber
\\
&&          
          + m_1^5 (16 m_2^3 + 32 m_2 m_r^2 - 33 m_r^3) + 
          m_1^3 m_2^2 (16 m_2^3 + 87 m_2 m_r^2 - 33 m_r^3) 
\nonumber
\\
&&
          + m_1^4 m_2 (4 m_2^3 - 47 m_2 m_r^2 + 11 m_r^3)\Big)
          + C_A^2 C_F \Big(m_1^2 m_2^5 (221 m_2 - 1320 m_r) m_r^2 + 
          880 m_1 m_2^6 m_r^3 
\nonumber
\\
&&
          - 484 m_2^6 m_r^4 + 
          8 m_1^3 m_2^3 m_r (44 m_2^3 + 356 m_2^2 m_r + 110 m_2 m_r^2 - 
             121 m_r^3)
             - 8 m_1^5 m_2^2 (42 m_2^3 
\nonumber
\\
&&
             - 356 m_2 m_r^2 + 165 m_r^3) + 
          4 m_1^4 m_2^3 (4 m_2^3 + 149 m_2 m_r^2 + 220 m_r^3) + 
          m_1^6 (16 m_2^4 + 352m_2^3 m_r 
\nonumber
\\
&&
          + 221 m_2^2 m_r^2
          + 880 m_2 m_r^3 - 484 m_r^4) \Big)\bigg) z^{22 C_A/3}
          - \frac{1}{11949113 C_A^2 m_1^2 m_2^2}\bigg( 4(3C_A - 10C_F)C_F 
\nonumber
\\
&&
     \times \Big( ( 2691765
      + 51647 \sqrt{157}) C_A 
      + 272 (10205 - 269 \sqrt{157}) C_F\Big) (m_1^3 + m_2^3) m_r z^{-\frac{1}{12} (-71 + \sqrt{157}) C_A}\bigg)      
\nonumber
\\
&&
     + \frac{1}{11949113 C_A^2 m_1^2 m_2^2}\bigg( 4(3C_A - 10C_F)C_F\Big((-2691765      
     + 51647\sqrt{157})C_A - 272(10205 
\nonumber
\\
&&
     + 269\sqrt{157})C_F\Big)
     (m_1^3 + m_2^3)m_r z^{\frac{1}{12}(71 + \sqrt{157})C_A}\bigg)
\nonumber
\\
&&     
      + \frac{1}{32 m_1 m_2}
      \bigg(3 (C_A - 2 C_F) (m_1^2  + m_2^2)
      \mbox{Hypergeometric2F1}(-6/11, 1, 27/11, -1)\bigg)
\nonumber
\\
&&
   - \frac{1}{16\cdot 2^{5/11} m_1 m_2 (2 - z^{11 C_A/3})^{6/11}}
   \bigg(3 (C_A - 2 C_F) (m_1^2 + m_2^2) z^{22 C_A/3} (-1 + 2 z^{-11 C_A/3})^{6/11}
\nonumber
\\
&&
 \times\mbox{Hypergeometric2F1}(-6/11, 16/11, 27/11, z^{11 C_A/3}/2 )\bigg)
 - \frac{1}{ 18 C_A^2 m_1^3 m_2^3}\bigg(C_F (m_1^2 + m_2^2)
\nonumber
\\
&&
 \times \Big(C_A^2 m_1^2 m_2^2 + 2 C_F^2 (m_1 + m_2)^2 m_r^2 + 
      C_A C_F \big(2 m_1 m_2 m_r^2 + m_2^2 m_r^2 
      + m_1^2 (2 m_2^2 + m_r^2)\big)\Big) 
\nonumber
\\
&&
      \times\mbox{Hypergeometric2F1}(1, 1, 38/11, -1)\bigg) 
      + \frac{1}{ 36 C_A^2 m_1^3 m_2^3}
     \bigg(C_F (C_A + 2 C_F) (m_1^2 + m_2^2)
\nonumber
\\
&&
     \times\Big(C_A m_1^2 m_2^2 + 
      C_F (m_1 + m_2)^2 m_r^2\Big) z^{9 C_A}
     \mbox{Hypergeometric2F1}(1, 27/11, 38/11, z^{11 C_A/3}/2 )\bigg)
\nonumber
\end{eqnarray}
          
\begin{eqnarray}
&&    - \frac{1}{3025 C_A^2 m_1^3 m_2^3}
     8 C_F^2 \bigg(100 C_F^2 (2 m_1^2 - 3 m_1 m_2 + 2 m_2^2)^2 + 
     50 C_A C_F (16 m_1^4 - 18 m_1^3 m_2 
\nonumber
\\
&&
      + 23 m_1^2 m_2^2 
       - 18 m_1 m_2^3 + 16 m_2^4) + 
       C_A^2 (400 m_1^4 + 300 m_1^3 m_2 + 1673 m_1^2 m_2^2 + 300 m_1 m_2^3 
\nonumber
\\
&&
    + 400 m_2^4) \bigg) m_r^2 \ln(z)
     -  \frac{1}{1331 C_A^3 m_1^4 m_2^4}
     96 C_F (m_1 + m_2) \bigg(-11 C_A^3 m_1^2 m_2^2 (m_1^2 - m_1 m_2 
\nonumber
\\
&&     
     + m_2^2) m_r 
     + 24 C_F^3 m_1 m_2 (m_1 + m_2)^3 m_r^2
     + 2 C_A C_F^2 (m_1 + m_2) \Big(2 m_1^2 m_2^2 m_r^2 + 12 m_1 m_2^3 m_r^2 
\nonumber
\\
&&
          - 11 m_2^3 m_r^3 + m_1^3 (24 m_2^3 + 12 m_2 m_r^2 - 11 m_r^3)\Big)
    + C_A^2 C_F \Big(-22 m_1^2 m_2^4 m_r - 11 m_1 m_2^3 m_r^3 
\nonumber
\\
&&   
    - 11 m_2^4 m_r^3 
    + m_1^4 (13 m_2^3 - 22 m_2^2 m_r - 11 m_r^3)
    + m_1^3 (13 m_2^4 + 22 m_2^3 m_r
\nonumber
\\
&&    
    - 11 m_2 m_r^3)\Big) \bigg) \ln(2 - z^{11 C_A/3})
    - \frac{1}{1331 C_A^3 m_1^3 m_2^3} 
     48 C_F \bigg(C_A^3 m_1^2 m_2^2 (8 m_1^2 + 3 m_1 m_2 + 8 m_2^2) 
\nonumber
\\
&&
     - 56 C_F^3 m_1 m_2 (m_1 + m_2)^2 m_r^2 
+ C_A^2 C_F \Big(19 m_1 m_2^3 m_r^2 + 8 m_2^4 m_r^2 + 
          19 m_1^3 m_2 (-2 m_2^2 + m_r^2)
\nonumber
\\
&&          
          + 8 m_1^4 (m_2^2 + m_r^2) + 
          m_1^2 (8 m_2^4 
          + 22 m_2^2 m_r^2)\Big)
- 2 C_A C_F^2 \Big(11 m_1 m_2^3 m_r^2 - 4 m_2^4 m_r^2 + 
          m_1^4 (8 m_2^2 
\nonumber
\\
&&          
          - 4 m_r^2) + 11 m_1^3 m_2 (4 m_2^2 + m_r^2) 
          + m_1^2 (8 m_2^4 + 30 m_2^2 m_r^2)\Big) \bigg) z^{11 C_A/3}
      \ln(2 - z^{11 C_A/3}) 
\nonumber
\\
&&      
      + \frac{1}{22 C_A^2 m_1^3 m_2^3}
      \bigg(3 C_F (C_A + 2 C_F) 
(m_1^2 + m_2^2) \Big(C_A m_1^2 m_2^2 + 
      C_F (m_1 + m_2)^2 m_r^2\Big) z^{16 C_A/3}
\nonumber
\\
&&
      \times\ln(2 - z^{11 C_A/3})\bigg)    
     + \frac{24 C_F}{1331 C_A^3 m_1^4 m_2^4}
      \bigg(8 C_F^3 m_1 m_2 (m_1 + m_2)^2 (3 m_1^2 - m_1 m_2 + 3 m_2^2) m_r^2 
\nonumber
\\
&&      
      + C_A^3 m_1^2 m_2^2 \Big(3 m_1^2 m_2^2 
+ 8 m_1 m_2^3
+ m_1^3 (8 m_2 - 11 m_r) - 11 m_2^3 m_r\Big) + 
       2 C_A C_F^2 \Big(2 m_1 m_2^4 (8 m_2 
\nonumber
\\
&&       
       - 11 m_r) m_r^2 + 
          m_1^2 m_2^3 (15 m_2 
          - 11 m_r) m_r^2
          - 11 m_2^5 m_r^3 + 
          m_1^4 m_2 (4 m_2^3 + 15 m_2 m_r^2 - 22 m_r^3) 
\nonumber
\\
&&          
          + m_1^3 m_2^2 (16 m_2^3 - 2 m_2 m_r^2 - 11 m_r^3)           
          + m_1^5 (16 m_2^3
          + 16 m_2 m_r^2 - 11 m_r^3)\Big) 
          + C_A^2 C_F \Big(2 m_1 m_2^4 (4 m_2
\nonumber
\\
&&          
          - 11 m_r) m_r^2 - 11 m_2^5 m_r^3
          +  m_1^2 m_2^3 m_r (-22 m_2^2 + 19 m_2 m_r - 11 m_r^2) 
          + m_1^4 m_2 (-12 m_2^3 + 19 m_2 m_r^2
\nonumber
\\
&&          
          - 22 m_r^3) + 
          m_1^5 (21 m_2^3 
          - 22 m_2^2 m_r + 8 m_2 m_r^2 - 11 m_r^3)
   + m_1^3 m_2^2 (21 m_2^3 + 22 m_2 m_r^2 - 11 m_r^3)\Big)\bigg)
\nonumber
\\
&&   
  \times z^{22 C_A/3}
      \ln(2 - z^{11 C_A/3})
 + \frac{1}{1331 C_A^3 m_1^3 m_2^3}
     \bigg(768 C_F^2 (m_1 + m_2)^2 \Big(C_A^2 m_1^2 m_2^2  
\nonumber
\\
&&      
     + C_F^2 (m_1 + m_2)^2 m_r^2 
     + C_A C_F \big(2 m_1 m_2 m_r^2 + m_2^2 m_r^2 
     + m_1^2 (2 m_2^2 + m_r^2)\big)\Big) \ln^2(2 - z^{11 C_A/3})\bigg)
\nonumber
\\
&& 
      - \frac{1}{1331 C_A^3 m_1^3 m_2^3}
      \bigg(192 C_F^2 (m_1 + m_2)^2 \Big(C_A^2 m_1^2 m_2^2 
      + C_F^2 (m_1 + m_2)^2 m_r^2 + 
      C_A C_F \big(2 m_1 m_2 m_r^2 
\nonumber
\\
&&       
      + m_2^2 m_r^2 + m_1^2 (2 m_2^2 + m_r^2)\big)\Big)
      \times z^{22 C_A/3} \ln^2(2 - z^{11 C_A/3})\bigg)
      \bigg]\,.
\end{eqnarray}

Finally, in Fig. \ref{Fig:pot}, we give the numerical evaluation $\delta \tilde D_{d,p}^{(2)NLL}$ for different values of $n_f$. The contribution is sizable. 

\subsection{Potential running, spin-dependent delta potential}
 
Even though not relevant for this paper, we profit to present the potential RG equation of $\tilde D^{(2)}_{S^2}$ in the basis we use in this paper, which is different from the basis used in \cite{Penin:2004ay}. The final solution is nevertheless the same. 

\bea
{d \tilde D^{(2)}_{S^2} \over d\ln\nu}&=&
-2 m_r^3 \left( {1\over m_1^3} +
{1\over m_2^3} \right)  C_F^2\alpha_{V}^2 D^{(2)}_{S^2}
+{m_r^2 \over m_1m_2}\,
C_F^2\alpha_{V}\Bigg(2D^{(2)}_{d}D^{(2)}_{S^2}
\nn\\
&&
-{4 \over 3}
\left(D^{(2)}_{S^2}\right)^2
-8 D^{(2)}_{S^2} D_{1}^{(2)}
-{5 \over 12} \left(D_{S_{12}}^{(2)}\right)^2  \Bigg)
-C_AC_FD^{(1)} D^{(2)}_{S^2}
\nn\\
&&+{1 \over 2}\alpha^3C_F^2 m_r^2
\left({c_{pp'}^{(1)}\over m_1^2}{c_F^{(2)}}
+{c_F^{(1)}}{c_{pp'}^{(2)}\over m_2^2}\right)
-2\alpha^3C_F^2 {m_r^2 \over m_1m_2}c_F^{(1)}c_F^{(2)}
\nn\\
&&
-{1 \over 4}\alpha^3C_F^2 {m_r^2 \over m_1m_2}c_S^{(1)}c_S^{(2)}
-\alpha^3C_F^2 m_r^2
\left({c_F^{(1)}}{c_S^{(2)}\over m_2^2}
+{c_S^{(1)}\over m_1^2}{c_F^{(2)}}\right)
\nn\\
&&
+{1 \over 2}\alpha^3C_F(4C_F-C_A) m_r
\left({c_F^{(1)}}{c_S^{(2)}\over m_2}+{c_S^{(1)}\over m_1}{c_F^{(2)}}\right)
\nn\\
&&
+{1 \over 8}\alpha^3C_FC_Am_r
\left({c_F^{(1)}}{c_F^{(2)\,2}\over m_2}+{c_F^{(1)\,2}\over m_1}{c_F^{(2)}}\right)
-{C_A^2(C_A-2C_F)\alpha^3 \over 2}D^{(2)}_{S^2,1/r^3}\,.
\label{Dsnllp}
\eea
This equation has slightly changed with respect to Eq. (36) in \cite{Penin:2004ay} because of the change in the basis of potentials. In particular the term proportional to $D_{S^2}^{(2)}$ changes to compensate the fact that $ D_d^{(2)}$ is also different so that the result is the same. 

\section{N$^3$LL heavy quarkonium mass}

For the organization of the computation and presentation of the results we closely follow the notation of \cite{Pwave}. In particular we split the total RG improved potential in the following way:
\be
V_{s}^{N^iLL}(\nu_h,\nu)=V_{s}^{N^iLO}(\nu)+\delta V_{s}^{N^iLL}(\nu_h,\nu)
\,,
\ee
where $V_{s}^{N^iLO}(\nu) \equiv V_{s}^{N^iLL}(\nu_h=\nu,\nu)$.
We then split the total energy into the N$^3$LO result and the new contribution associated to the resummation of logarithms. The S-wave spectrum at N$^3$LO was obtained in Ref.~\cite{Penin:2002zv} for the ground state, in Refs.~\cite{Penin:2005eu,Beneke:2005hg} for S-wave states, and in \rcites{Kiyo:2013aea,Kiyo:2014uca} for general quantum numbers but for the equal mass case. The result for the nonequal mass case was obtained in \rcite{Peset:2015vvi}. 

From the RG improved potential one obtains the N$^i$LL shift in the energy levels
\begin{align}
E_{\rm N^iLL}(\nu_h,\nu) & = E_{\rm N^iLO}(\nu)+ \delta E_{\RG}(\nu_h,\nu)\Big|_{\rm N^iLL}.
\end{align}
where the explicit expression for $E_{\rm N^iLO}(\nu)$ can be found in \rcite{Peset:2015vvi}, and in a different spin basis in Appendix B of \cite{Pwave}.

The LO and NLO energy levels are unaffected by the RG improvement, i.e.
\begin{align}
\delta E_{\RG}\Big|_{\rm LL}=\delta E_{\RG}\Big|_{\rm NLL}=0.
\end{align}
We now determine the variations with respect to the NNLO and N$^3$LO results.
 We are here interested in the corrections associated to the resummation of logarithms.
In order to obtain the spectrum at NNLL and N$^3$LL we need to add the following energy shift to the NNLO and N$^3$LO spectrum:
\begin{align}
&\delta E_{\RG}\Big|_{\rm NNLL}=
\langle nl|
\delta V_{s}^{NNLL}(\nu_h,\nu)|nl\rangle
\,,
\end{align}
which was computed in \rcite{Pineda:2001ra}, and 
\begin{align}
&\delta E_{nl,\RG}\Big|_{\rm N^3LL}=\label{deltaEN3LLdef}
\langle nl|
\delta V_{s}^{N^3LL}(\nu_h,\nu)|nl\rangle\\
&\hspace*{0.5cm}+
2\langle nl|V_1 \frac{1}{\left(E_n^C-h\right)'}\delta V_{s}^{NNLL}(\nu_h,\nu)|nl\rangle
+\left[\delta E_{US}(\nu,\nu_{us})-\delta E_{US}(\nu,\nu)\right]
\label{deltaEN3LL}
\end{align}
Note that $\langle nl|
\delta V_{s}^{N^3LL}(\nu_h,\nu)|nl\rangle$ includes $\langle nl|
\delta V_{s}^{NNLL}(\nu_h,\nu)|nl\rangle$. 

$\delta E_{nl,\RG}\Big|_{\rm N^3LL}$ was computed for $l\not=0$ in \rcite{Pwave}, and for $l=0$, $s=1$ in 
\rcites{Kniehl:2003ap,Penin:2004xi}.
To have the complete result for S-wave states, one needs to compute (and add) the new term for $l=0$
\be
\delta E^{\rm new}_{n0,\RG}\Big|_{\rm N^3LL}=
\langle n0|
[\delta V_{r}^{N^3LL}-\delta V_{r}^{NNLL}](\nu_h,\nu)|n0\rangle
+
2\langle n0|V_1 \frac{1}{\left(E_n^C-h\right)'}\delta V_{r}^{NNLL}(\nu_h,\nu)|n0\rangle
\,,
\ee
where 
\be
V_1=-\frac{C_F \al}{r}\frac{\al}{4\pi}(2\beta_0\ln(\nu r e^{\gamma_E})+a_1)
\,,
\ee
and 
$\delta V_{r}^{N^iLL}$ is the delta-related potential contribution 
to 
$\delta V_{s}^{N^iLL}$. The new term generated from $\tilde D_d^{(2)}$ then reads
\bea
\nn
\delta E^{\rm new}_{nl,\RG}\Big|_{\rm N^3LL}
&=& \frac{1}{m_1m_2}  \pi C_F [\delta \tilde D_d^{(2)NLL}(\nu_h,\nu)- \delta \tilde D_d^{(2)NLL}(\nu,\nu)] \frac{(m_r C_F \al)^3}{\pi n^3}\delta_{l0}
\\
\nn
&&
+2\frac{1}{m_1m_2} \pi C_F [\tilde D_d^{(2)LL}(\nu_h,\nu)- \tilde D_d^{(2)LL}(\nu,\nu)]\left[-\frac{\al}{4\pi}\right]
\frac{(m_r C_F \al)^3}{\pi n^3}\delta_{l0}
\\
&&\times
\left\{
2\beta_0
\left[
\frac{1}{2}+\frac{\pi^2 n}{6}-n\Sigma_2^{(k)}(n,0)-\frac{3}{2}\ln\left(\frac{n a \nu}{2}\right)-\frac{3}{2}S_1(n)
\right]
-\frac{3}{2}a_1
\right\}
\nn
\\
&&
+\frac{\pi C_F}{m_1 m_2}
\left[- \frac{1}{4\pi} \right] \frac{2(m_r C_F \alpha)^3}{n^3}\left(\ln\frac{na\nu}{2}-S_1(n)-\frac{n-1}{2n}\right)2\delta_{l0}
\nn
\\
&&
\times
\left[
\left( k {d\over d k}\tilde D_{d}^{(2)}\right)\Bigg|_{k=\nu}^{LL}(\nu_h;\nu)
-
\left( k {d\over d k}\tilde D_{d}^{(2)}\right)\Bigg|_{k=\nu}^{LL}(\nu;\nu)
\right]
\,,
\eea
where $ \delta \tilde D_d^{(2)NLL}$ is defined in \eq{Dsum}. 
The first three lines are generated by the term proportional to $\delta^{(3)}({\bf r})$. The last two lines are the contribution to the S-wave energy ($l=0$) from the last term of \eq{eq:Vr} (the contribution to the P-wave energy, proportional to $1-\delta_{l0}$ term, is already included in \rcite{Pwave}. Therefore, we do not include it in the expression above). To this contribution we have explicitly subtracted the fixed order contribution already included in the N$^3$LO result. 

By adding $\delta E^{\rm new}_{nl,\RG}\Big|_{\rm N^3LL}$ to the results  computed in these references\footnote{Note though that one should change $2=S(S+1)$ by $S(S+1)-3/2$ in the result obtained in \cite{Kniehl:2003ap,Penin:2004xi} to account for the change of basis of operators to the one we use here. One should also change from the on-shell to the Coulomb basis of potentials in \cite{Pwave} (this is very easy to do, as the ultrasoft running is not affected by this transformation).} one obtains the complete result.


\section{Conclusions}

In this paper we have computed the $\al/m^4$, and the $\al^2/m^3$, spin-independent potentials (in the Coulomb gauge), and an extra ultrasoft correction that contributes to the S-wave spin-average NNNLL spectrum. We have also obtained the potential RG equation of the delta potential with NLL accuracy (the first nonzero contribution). Combined with the previous results we solve this equation and obtain the 
complete (potential and ultrasoft) NLL running of the delta potential. 

We have also computed the bare and renormalized (soft-)$\al^3/m^2$ contribution to the spin-independent delta-like potential proportional to $[c_F^{(1)}]^2$, $[c_F^{(2)}]^2$, $\bar c_1^{(1)hl}$ and $\bar c_1^{(2)hl}$ and obtained (and solved) the RG equation. 

Combining all these results with the results in \cite{Pwave} and \rcites{Kniehl:2003ap,Penin:2004xi} allows us to obtain the S-wave mass with N$^3$LL accuracy. The missing terms to obtain the full results are to have the NLL running of $\bar c_1^{hl}$ (the associated missing term is of ${\cal O}(T_fn_f m \al^6\ln\al)$ and is expected to be quite small. Its computation will be carried out elsewhere), and a piece of the soft running of the delta potential. This computation will be performed in a separate paper. The magnitude of this contribution is estimated to be smaller compared with the potential running computed in this paper. It is also expected to be smaller than the complete running of the heavy quarkonium potential. Nevertheless, a detailed phenomenological analysis is postponed to future publications.

Finally, we remark that significant parts of the computations above are necessary building blocks for a future N$^4$LO evaluation of the heavy quarkonium spectrum. 

\bigskip

{\bf Acknowledgments}\\
We thank discussions with A.A. Penin, C. Peset, M. Stahlhofen and M. Steinhauser. We also thank A.A. Penin for partial checks of the computations of this paper. This work was supported in part by the Spanish grants FPA2014-55613-P, FPA2017-86989-P and SEV-2016-0588 from the ministerio de Ciencia, Innovaci\'on y Universidades, and the grant 2017SGR1069 from the Generalitat de Catalunya.


\appendix

\section{Matching scheme (in)dependence}

The potentials obtained in Sec. \ref{sec::matchSI} were computed in the Coulomb gauge. On the other hand, the potential RG equation obtained in Sec. \ref{Sec:RGpot} is generated by potential loops, which are independent of the gauge/matching scheme. The dependence on the matching scheme of \eq{eq:RGpot} is implicitly generated by the Wilson coefficients used for the running, such as $D_d^{(2)}$ or $D^{(1)}$, and explicitly, since we put the explicit expressions for the $1/m^3$ and $1/m^4$ potentials obtained in the Coulomb gauge. This last point makes that \eq{eq:RGpot} can only be used in the Coulomb gauge matching scheme, though with not much effort it could be written in terms of general structures of the $1/m^3$ and $1/m^4$ potentials that would make it also useful for a computation in a general matching scheme. Nevertheless, since we do not know the $1/m^3$ and $1/m^4$ potentials in other matching schemes, we refrain from doing so in this paper.  Still it is worth to study how the differences between different matching schemes show up in the terms where the entire matching scheme dependence is encoded in the $D's$ (the first four lines in \eq{eq:RGpot}). We do so in the following.
 
At ${\cal O}(m \al^4)$ the Coulomb and Feynman matching schemes produce the same potential but the on-shell scheme does not. At this order, the relation between the Wilson coefficients of the delta-like and the $1/m$ potentials in the off-shell Coulomb gauge (equal to the Feynman gauge at this order) and in the on-shell scheme are given by:
\begin{equation}
 D_{d,CG}^{(2)} = D_{d,ON}^{(2)} + \alpha(\nu)
\,,
\end{equation}
\begin{equation}
 D_{CG}^{(1)} = D_{ON}^{(1)} + \alpha^2(\nu) \frac{2C_F}{C_A}\frac{m_r^2}{m_1 m_2}
\,.\end{equation}
At the order we are working in this paper such differences produce the following difference between the RG equation for $\tilde D_d^{(2)}$ in the two 
schemes  (for the first four lines in \eq{eq:RGpot}):
\begin{equation}
\label{Ddscheme}
 \nu\frac{d}{d\nu}( \tilde D_{d,CG}^{(2)} - \tilde D_{d,ON}^{(2)}) 
 = C_F^2 \frac{m_r^2}{m_1 m_2}\left(  - 4\alpha^2 D_1^{(2)} + \alpha^3  
 - \alpha\frac{C_A}{C_F}\frac{m_1 m_2}{m_r^2}D_{CG}^{(1)}\right)
\,,
\end{equation}
which does not vanish. This difference can be understood through field redefinitions. 
The field redefinition that moves from the off-shell Coulomb to the on-shell scheme was already discussed in 
\cite{Brambilla:2000gk,Peset:2015vvi}. In the second reference the discussion was focused on effects to the spectrum up to ${\cal O}(m\al^5)$. We now need to see (the logarithmically enhanced) differences of  ${\cal O}(m\al^6)$. They can be traced back by using the following  Hamiltonian in the Coulomb (Feynman) gauge:
\begin{equation}
 h_{CG} = h^{(0)}+h_{CG}^{(2)}
\,,
\end{equation}
where $h^{(0)} \sim mv^2$ is the leading order Hamiltonian:
\be
h^{(0)}=\frac{ {\bf p}^2}{2m_r}+V^{(0)}(r)
\,,
\ee
and $h_{CG}^{(2)} \sim mv^4$ is the relativistic correction, with the explicit potentials:
\bea
h_{CG}^{(2)}&=&
-c_4{{\bf p}^4 \over 8m_1^3}-c_4{{\bf p}^4 \over 8m_2^3}-{\cf C_A
D^{(1)} \over 4m_rr^2}
\\
&&- { \cf D^{(2)}_{1} \over 2 m_1 m_2} \left\{ {1 \over r},{\bf p}^2 \right\}
+ { \cf D^{(2)}_{2} \over 2 m_1 m_2}{1 \over r^3}{\bf L}^2
+ {\pi \cf D^{(2)}_{d} \over m_1 m_2}\delta^{(3)}({\bf r})
\nn
\\
\nn
& & + {8\pi \cf D^{(2)}_{ S^2} \over 3m_1 m_2}{\bf S}_1\cdot {\bf S}_2 \delta^{(3)}({\bf r})
+ { 3 \cf  \over 2 m_1 m_2}{1 \over r^3}{\bf L} \cdot 
(D^{(2)}_{LS_1}{\bf S}_1+D^{(2)}_{LS_2}{\bf S}_2)
+ { \cf D^{(2)}_{S_{12}} \over 4 m_1 m_2}{1 \over r^3}S_{12}(\hat{\bf r})
\,.
\eea

$h_{CG}$ correctly reproduces the ${\cal O}(m\al^4)$ spectrum (for the purpose of the comparison we can neglect the ${\cal O}(\al)$ corrections to the static potential:  
$\cf {\alpha_{V_{s}} \over r} \simeq  -C_F\al\frac{1}{r}$). This will be enough 
for our purposes. We now consider the field redefinition that transforms $h_{CG}$ into $h_{OS}$, the on-shell Hamiltonian:
\begin{equation}
 U = \exp\left(-\frac{i}{m_r}\{{\bf W}(r),{\bf p}\}\right)
\,.
\end{equation}
${\bf W}$ can be determined from the equation:
\begin{equation}
 V^{(1)}_{ON} - V^{(1)}_{CG} = \frac{2}{m_r}{\bf W}\cdot ( {\bfnabla} V^{(0)})
\,.
\end{equation}
Since the only possible tensor structure of ${\bf W}$ is ${\bf W} = W(r^2){\bf r}^i$ the above equation can be written as:
\begin{equation}
 V^{(1)}_{ON} - V^{(1)}_{CG} = \frac{2}{m_r}W(r^2){\bf r}^i\cdot ( \bfnabla^i V^{(0)})
\,.
\end{equation}
We then obtain
\begin{equation}
   {\bf W}^i = \frac{\pi}{2g_B^2} C_A 
   (D^{(1)}_{CG} - D^{(1)}_{ON})  \frac{{\bf r}^i}{r^{1+2\epsilon}} 
\,,
\end{equation}
and
\begin{equation}
 h_{ON} = U^\dagger h_{CG} U=h_{CG}+\delta h=h^{(0)}+ h_{ON}^{(2)}+ h_{ON} ^{(4)}+\cdots
\,.\end{equation}
$h_{CG}$ and $h_{ON}$ obviously produce the same spectrum. Therefore, $\delta h$ cannot produce energy shifts, and any change in the RG equations has to be compensated among different terms. Let us see how it works. $ h_{ON}^{(2)}$ produces the differences reported in \eq{Ddscheme}. Such differences should be eliminated by $ h_{ON}^{(4)}$ (as the other contributions to the Hamiltonian are subleading), and indeed they do. In momentum space $ \tilde h_{ON}^{(4)}$ reads
\bea
 \tilde h_{ON}^{(4)} &=&  C_F^2 g_B^2 m_r  \frac{\pi D_1^{(2)}}{8m_1^2 m_2^2} 
\left( |{\bf k}|^{1+2\epsilon} + 4( {\bf p}\cdot{\bf p}' )|{\bf k}|^{-1+2\epsilon}
+ \frac{2({\bf p}\cdot{\bf k})({\bf p}'\cdot{\bf k})}{|{\bf k}|^{3-2\epsilon}}\right)
\nn
\\
&&
\nn
+C_F \frac{\pi}{m_1 m_2} \left(
 -\frac{1}{4}C_F C_A D_{CG}^{(1)}
 + \frac{1}{4}\frac{m_r^2}{m_1 m_2} C_F^2 \frac{g_B^4}{16\pi^2}  \right)
  \frac{g_B^2}{4\pi}  \frac{1}{\epsilon}\frac{1}{|{\bf k}|^{-4\epsilon}} 
  \\
  &&+\cdots
\,.
\eea
Note that the term proportional to $|{\bf k}|^{1+2\epsilon}$ in the first line gives a contribution to the potential RG equation through potential loops. It is equivalent to generating a new $1/m^3$ potential. The other two terms in the first line do not contribute to the potential RG equation. Looking at the 2nd line, it is also interesting to see that there is a kind of soft contribution, which nevertheless, has ultrasoft in the on-shell scheme. The second term in the second line can also be interpreted as a pure soft contribution. This brings the interesting observation that even if the potential RG equation can be written in a matching scheme independent way, the implicit scheme dependence of the potentials allows for a mixing with the soft computation (at least in the on-shell scheme). Finally, for the dots in the 3rd line we refer to extra contributions to $ \tilde h_{ON}^{(4)}$, generated by the field redefinitions, which nevertheless do not contribute to the running. 


\end{document}